%% file: SubmissionArxiv_changjun2026_v3.tex
\documentclass{article}

\usepackage[margin=1in]{geometry}
\usepackage{cmap}
\usepackage{mathrsfs}
\usepackage{lmodern} 

\usepackage{multirow}
\usepackage{graphicx}
\usepackage{caption}
\usepackage{subcaption}
\usepackage{booktabs}
\usepackage{placeins}

\DeclareCaptionLabelFormat{addS}{#1 S#2}

\usepackage{amsmath}
\usepackage{amsthm}
\usepackage{amssymb}
\usepackage{physics}
\usepackage[scr=esstix,cal=boondox]{mathalfa}

\newtheorem{assumption}{Assumption}

\usepackage{enumitem}
\setlist[itemize]{topsep=0.5em, itemsep=0.5em, left=1.5em}
\setlist[enumerate]{topsep=0.5em, itemsep=0.5em, left=1.5em}

\usepackage{anyfontsize}
\newcommand{\ninept}{\fontsize{8.5}{13}\selectfont}

\usepackage{tikz}

\usepackage{authblk}

\usepackage{hyperref}
\usepackage[numbers,sort&compress]{natbib}
\usepackage{xcolor}
\hypersetup{
    colorlinks,
    linkcolor={red!50!black},
    citecolor={blue!50!black},
    urlcolor={blue!80!black}
}

\newcommand{\HtoRB}{\ensuremath{\text{H}_2\text{RB}}}

\newcommand{\keywords}[1]{\par\medskip\noindent\textbf{Keywords:} #1}

\title{A Bayesian approach to the
survivor average causal effect in cluster-randomized crossover trials}

\author[1,2]{Changjun Li}
\author[3]{Dane Isenberg}
\author[3]{Michael O. Harhay}
\author[4,5]{Andrew B. Forbes}
\author[5,6,7,8]{Paul J. Young}
\author[1,2]{Fan Li\thanks{fan.f.li@yale.edu}\thanks{co-senior authors}}
\author[3]{Nandita Mitra$^\dagger$}

\affil[1]{Department of Biostatistics, Yale School of Public Health, New Haven, CT, USA}
\affil[2]{Center for Methods in Implementation and Prevention Science, Yale School of Public Health, New Haven, CT, USA}
\affil[3]{Department of Biostatistics, Epidemiology and Informatics, University of Pennsylvania, Philadelphia, PA, USA}
\affil[4]{Division of Quantitative Research Methodology, School of Public Health and Preventive Medicine, Monash University, Melbourne, VIC, Australia}
\affil[5]{Australian and New Zealand Intensive Care Research Centre, Department of Epidemiology and Preventive Medicine, School of Public Health and Preventive Medicine, Monash University, Melbourne, VIC, Australia}
\affil[6]{Medical Research Institute of New Zealand, Wellington, New Zealand}
\affil[7]{Intensive Care Unit, Wellington Regional Hospital, Wellington, New Zealand}
\affil[8]{Department of Critical Care, University of Melbourne, Parkville, VIC, Australia}

\begin{document}

\maketitle

\begin{abstract}
    
In cluster-randomized crossover (CRXO) trials, groups of individuals are randomly assigned {to one of} two or more
sequences of alternating treatments. Since clusters serve as their own control, the CRXO design is typically more statistically efficient than the usual parallel-arm design. CRXO trials are increasingly popular in many areas of health research where the number of available clusters is limited. Further, in trials among severely ill patients, researchers often want to assess the effect of treatments on secondary non-terminal outcomes, but frequently in these studies, there are patients who do not survive to have these measurements fully recorded. In this paper, we provide a causal inference framework and treatment effect estimation method for addressing truncation by death in the setting of CRXO trials. We target the survivor average causal effect (SACE) estimand, a well-defined subgroup treatment effect obtained via principal stratification. {We propose causal structural assumptions and standard parametric modeling assumptions that together enable estimating the SACE within a Bayesian paradigm.} We evaluate the small-sample performance of our proposed Bayesian approach for estimation of the SACE in CRXO trial settings via simulation studies. We apply our method to a previously conducted two-period cross-sectional CRXO study examining the impact of proton pump inhibitors compared to histamine-2 receptor blockers on length of hospitalization among adults requiring invasive mechanical ventilation.

\end{abstract}


\section{Introduction}

\begin{figure}[t]
    \captionsetup{labelfont=bf,justification=raggedright,singlelinecheck=false}
    \centering
    \includegraphics[scale=.6]{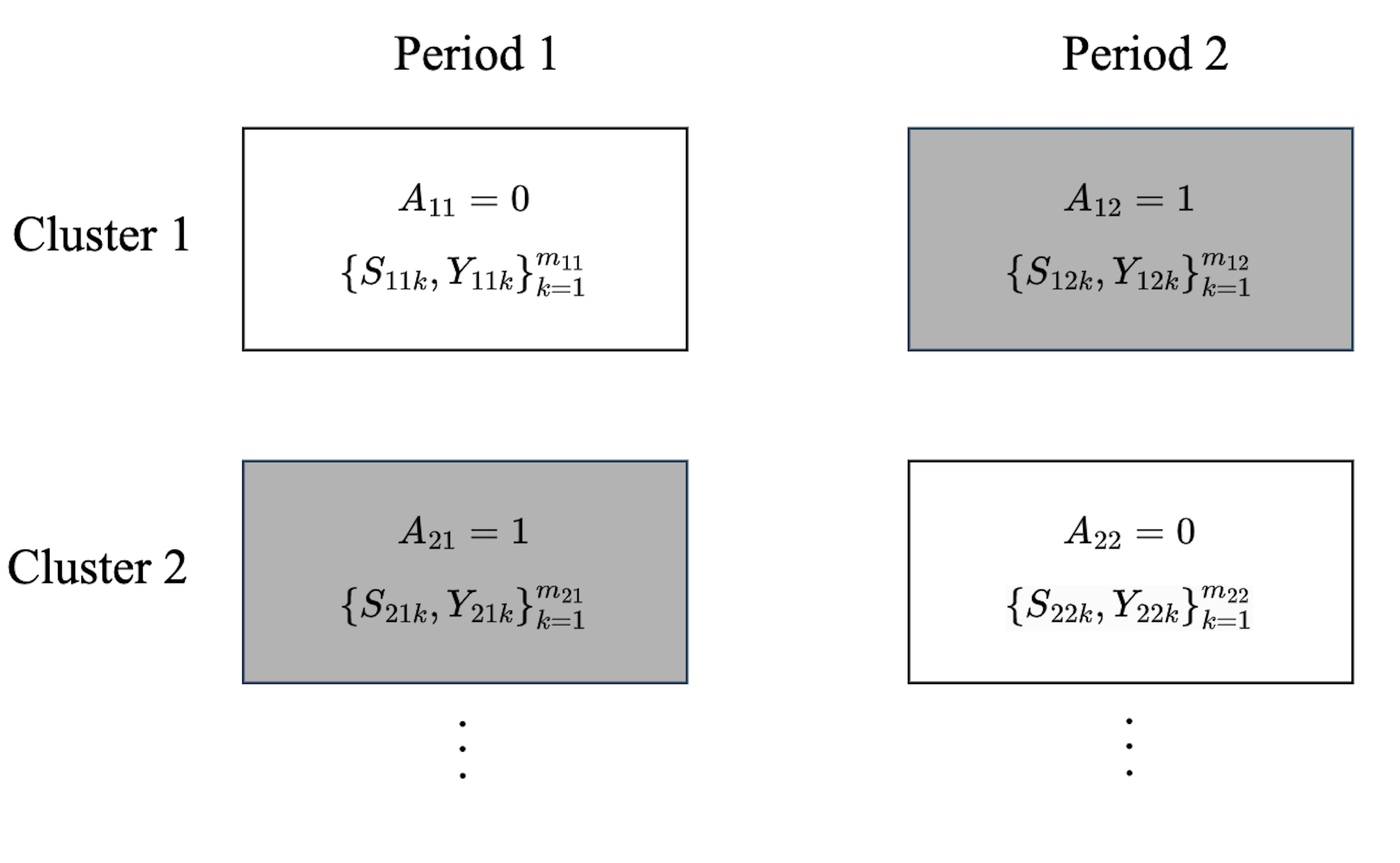}
    \caption{Illustration of a two-period cluster-randomized crossover (CRXO) design. Shaded cells indicate treatment condition and unshaded cells indicate the control condition. The first and second clusters have been randomized to the treatment sequences $A_1=(0,1)^T$ and $A_2=(1,0)^T$, respectively. The remaining clusters may be randomly assigned to either sequence. For each individual, the observed data shown are assigned treatment, survival status $S_{ijk}$, and non-mortality outcome $Y_{ijk}$. 
    We note that $Y_{ijk}$ is only fully observed when $S_{ijk}=1$, and $Y_{ijk}$ is truncated in the data when $S_{ijk}=0$.}
    \label{fig:crxodesign}
\end{figure}

In cluster-randomized trials (CRTs), {entire} groups of individuals, such as hospitals or schools, are randomly assigned to treatments 
{rather than individuals themselves} \citep{murray2004design}. CRTs are typically implemented when there are group-level therapies, concerns about contamination, or logistical challenges to individual random assignment \citep{turner2017review,hayes2017cluster}. However, when targeting an individual-level treatment effect, individually-randomized trials are generally more statistically efficient than CRTs, so they require smaller sample sizes \citep{donner1981randomization}. Therefore, if the number of available clusters or cluster recruitment is limited, a design of a CRT that augments statistical power is desirable. In particular, the cluster-randomized crossover (CRXO) design (Figure \ref{fig:crxodesign}) is an alternative to the standard parallel-arm CRT that can lead to sizable gains in precision in estimation \citep{hooper2025reporting,forbes2015cluster,arnup2017understanding,hayes2017cluster}.

In a CRXO design, clusters are randomly assigned to a sequence of alternating treatments of the trial duration, with each treatment administered for a designated time period before alternating to another treatment. These time periods are denoted as ``periods'', and considering a particular period for a particular cluster gives rise to the unit of a ``cluster-period''. In the simplest version, a two-period CRXO trial with binary treatment, clusters are randomly assigned to one treatment in the first time period and then are switched to the other treatment in the second time period, i.e., there are two cluster-periods associated with each cluster that receive different treatments. In this way, clusters are meant to serve as their own control, resulting in potential efficiency gains relative to a parallel-arm CRT \citep{arnup2017understanding}. As a result, the CRXO trial has emerged as an attractive option in critical care studies \citep{cook2021rationale}, where researchers typically only have access to a small number of clusters composed of many individuals within each cluster \citep{forbes2015cluster}.

In critical care studies, researchers often want to understand the effect of an intervention on a non-mortality outcome such as time to hospital discharge alive or time to requiring organ support. However, there may be considerable incomplete (or often times not well-defined) non-terminal outcome measurements as many critically or seriously ill patients do not survive to the end of a study's follow-up period. This dilemma, often termed ``truncation by death'', has implications for defining and estimating causal estimands for the effect of treatment. The method presented in this paper addresses the  challenge associated with truncation by death in the context of CRXO trials. Our work is principally motivated by the PEPTIC trial, a two-period CRXO trial among critical care patients requiring invasive mechanical ventilation \citep{young2018cluster,young2020effect}. As is common in critical care studies, each cluster (intensive care unit) in PEPTIC is randomized to a treatment sequence over time, and patients are assigned only the treatment allocated to their unit during the period of their admission {and appear in no subsequent periods per a cross-sectional design}. We perform an analysis of the effect of proton pump inhibitors (PPIs) versus histamine-2 receptor blockers (\HtoRB s) on time to hospital discharge alive, which we define as our hospital length of stay (LOS) outcome variable.

From the outset, it is necessary to determine a target causal estimand for an average effect of treatment on a non-mortality outcome that can be reasonably estimated from three-tiered multilevel data with informative truncation due to death.  For this setting, we choose to target the survivor average causal effect (SACE) \citep{rubin2000causal,zhang2003estimation}, a conditional treatment effect defined using the causal inference framework of principal stratification \citep{frangakis2002principal,vanderweele2011principal}. Specifically, the SACE is a causal contrast, often expressed in the literature as a difference in means, among the sub-population of ``always-survivors'' -- individuals that would survive regardless of assigned treatment such that their non-mortality outcomes under either treatment are potentially observable. With this estimand, we avoid having to condition on observed survival up to a particular time point, which would disrupt the balance in characteristics across treatment groups afforded by the initial randomization. Moreover, unlike a composite strategy, where the intercurrent event of death is included in the endpoint definition (generally as the most severe) along with the non-mortality outcome \citep{ICH_E9R1_2019,kahan2024estimands}, the SACE allows one to target a treatment effect of the specific non-mortality outcome of interest in the always-survivor subpopulation. {We establish causal structural assumptions and parametric modeling assumptions that render the SACE estimable using data from cross-sectional CRXO trials. These causal structural assumptions, such as randomization and consistency, are often referred to as identification assumptions in causal inference; however, because they do not suffice to point identify the SACE on their own, we avoid that terminology.}
Our approach follows a likelihood-based strategy \citep{zhang2009likelihood} within a Bayesian paradigm \citep{imbens1997bayesian,hirano2000assessing} initially developed for principal stratification estimands with independent and identically distributed (iid) data, which requires modeling the outcome and principal stratum membership jointly given covariates. Our work is further informed by the burgeoning body of literature for estimating SACE in parallel-arm CRTs (or their observational counterpart), which has been articulated from both frequentist \citep{wang2024mixed} and Bayesian perspectives \citep{tong2023bayesian,he2023bayesian}, where the requisite assumptions and modeling choices have accounted for cluster-level treatment assignment and potential clustering effects. {Our approach to targeting the SACE in CRXO trials is also informed by the potential-outcome formulation developed by Chen and Li \citep{chen2023model} for cross-sectional multilevel trial data. Their work focuses on stepped-wedge cluster-randomized trials, whose treatment assignment mechanism differs from the crossover randomization considered here. Appendix D describes how our Bayesian framework can be adapted to that setting.}

We have organized the remaining sections as follows. In Section~\ref{sec:notation}, we define notation and present causal {structural} assumptions for the SACE estimand in CRXO trials. In Section~\ref{sec:estim}, we describe a straightforward parametric mixed-effects modeling framework and detail a Bayesian approach for estimating the SACE. In Section~\ref{sec:sim}, we conduct a simulation study to assess the finite-sample operating characteristics of our estimation technique and demonstrate the necessity in incorporating and accounting for crossover via cluster-period random effects in the non-mortality outcome modeling. In Section~\ref{sec:app}, we apply our method to the PEPTIC trial where we estimate the SACE comparing the effect of PPIs to \HtoRB s on hospital LOS for mechanically ventilated critical care patients. Finally, we provide a discussion and considerations for future work in Section~\ref{sec:discuss}.

\section{Notation and Setup} \label{sec:notation}

\subsection{Survivor Average Causal Effect (SACE) Estimand} \label{sec:notation1}

CRXO trials have three unit levels: the cluster, the cluster-period, and the individual. We present notation that accounts for this hierarchy, so that we can clearly represent the individual-level SACE estimand when there is a binary treatment. 

Let $i=1,\dots,I$ represent cluster, $j=1,\dots,J$ represent period, and let $k=1,\dots,m_{ij}$ represent individual $k$ in cluster-period $ij$ of size $m_{ij}$. Therefore, terms subscripted by $ijk$ refer to individuals,  $ij$ refer to cluster-periods,  and $i$ refer to clusters. 
In accordance with the PEPTIC trial, we suppose the CRXO design has a cross-sectional structure, where each cluster-period contains distinct sets of individuals; hence, the $ijk$ uniquely index each individual. {Under the cross-sectional design, the index $k$ labels distinct individuals within a cluster-period rather than tracking the same individual across periods; for example, $(1,1,1)$ and $(1,2,1)$ refer to two different patients rather than two repeated observations from the same individual.} {For English letters}, we use uppercase notation to indicate random variables and lowercase for realizations of random variables. {Greek letters are used for parameters, with $\pi$ referring broadly to probabilities. Bold terms denote vectors or matrices; non-bold terms denote scalars. Bold numerical values denote a vector of repeated values (e.g., $\boldsymbol{1}=(1,\dots,1)^T$).}

\begin{table}[t]
\captionsetup{labelfont=bf,justification=raggedright,singlelinecheck=false}
\caption{Possible principal stratum membership, $G_{ijk}=(S_{ijk}(1),S_{ijk}(0))$, for an individual, and the corresponding potential non-mortality outcomes that are defined for each stratum.}
\centering
\begin{tabular}{c|c|c}
\toprule
\multirow{2}{*}{Principal Stratum} & \multirow{2}{*}{Stratum Name} & Defined Non-mortality \\
& & Potential Outcome(s)\\
\midrule
$G_{ijk}=(1,1)$ & Always-survivor & $Y_{ijk}(1)$ and $Y_{ijk}(0)$ \\
$G_{ijk}=(1,0)$ & Protected-patient & $Y_{ijk}(1)$ only          \\
$G_{ijk}=(0,1)$ & Harmed-patient    & $Y_{ijk}(0)$ only          \\
$G_{ijk}=(0,0)$ & Never-survivor & None                       \\
\bottomrule
\end{tabular} 
\label{tab:pstab}
\end{table}

{

We let vector $\boldsymbol{A}_i=(A_{i1},\dots,A_{iJ})^T$ be the sequence of treatments assigned to cluster $i$, and $A_{ij}$ be the treatment assigned at period $j$ in cluster $i$. In this setting, we assume that $A_{ij}$ has two options at each period $j$ (i.e., a period-$j$ specific treatment 1 and a period-$j$ specific treatment 0), but that the options assigned must alternate each period. We denote the full treatment vector $\boldsymbol{A}=(A_1^T,\dots,A^T_I)^T$. {For clarity, Figure \ref{fig:crxodesign} illustrates the two-period CRXO design with alternating treatment sequences, as specified in Assumption 1 (Section \ref{sec:notation2}).} Under this {design}, we take $A_{ij} \in \{1,0\}$ for all $i,j$ such that $\boldsymbol{A}_i \in \{1,0\}^J$, with $\boldsymbol{A}_i=(1,0,\dots)^T$ or $\boldsymbol{A}_i=(0,1,\dots)^T$. {Although the motivating PEPTIC trial has $J=2$ periods, below we formulate the framework for a general $J$-period CRXO design, with $J=2$ as a special case. We retain this general formulation because multi-period CRXO trials are not uncommon in practice. A systematic review found that 26 of the 84 trials with a reported number of periods (31\%) had more than two periods, including 17 (20\%) with four or more periods \citep{arnup2016appropriate}. Multiperiod CRXO designs are also discussed in Hemming et al \citep{hemming2020use}. The numerical studies in this article nevertheless focus on the two-period setting corresponding to PEPTIC. Appendix D also describes how the Bayesian framework can be adapted to cross-sectional multi-period parallel-arm and complete stepped-wedge CRTs by modifying the treatment assignment assumptions and design matrices.}

We use the parenthetical notation, $S_{ijk}(\cdot)$ and $Y_{ijk}(\cdot)$, to denote an individual's potential survival status and potential non-mortality outcome, respectively, under an arbitrary treatment. If $S_{ijk}(\cdot)=1$, then $Y_{ijk}(\cdot) \in \mathbb{R}$, but if $S_{ijk}(\cdot)=0$, then $Y_{ijk}(\cdot)=*$ deterministically as the non-mortality outcome (discharge alive) is structurally undefined, following standard notation \citep{zhang2003estimation}. Observed $Y_{ijk} \in \mathbb{R} \cup \{*\}$ is notated analogously with respect to observed survival status $S_{ijk} \in \{1,0\}$. We aim to summarize impacts of just two possible treatments 
on individual-level outcomes. Therefore, we set out to define and ultimately estimate (through assumptions) an estimand with respect to an individual being under either treatment 1 or treatment 0, i.e., an estimand expressed using $S_{ijk}(a)$ and $Y_{ijk}(a)$ with $a \in \{1,0\}$. 

To define the target estimand, we use a principal stratification framework \citep{frangakis2002principal}. We partition} the population into four subgroups, called principal strata, that are represented by joint potential survival outcomes $G_{ijk}=(S_{ijk}(1),S_{ijk}(0)) \in \{1,0\} \times \{1,0\}$ as shown in Table \ref{tab:pstab} {(transpose notation is suppressed for readability)}. The benefit of this joint representation is that principal stratum membership does not depend on one's treatment assignment (that is, free of $a$) and can be conceptualized as a latent pre-treatment covariate for which conditional effects can be targeted. In particular, we are interested in the SACE estimand, $\mu_{h}$, given by
\begin{align} \label{eq:sace}
\mu_{h}&=h\left\{E\left[Y_{ijk}(1)|G_{ijk}=(1,1)\right],E\left[Y_{ijk}(0)|G_{ijk}=(1,1)\right]
\right\}
\end{align}
where $h$ is a function $h:\mathbb{R}^2 \rightarrow \mathbb{R}$ representing an effect measure of interest. For example, setting $h(x,y)=x-y$ generates the difference in means summary measure and setting $h(x,y)=x/y$ generates the ratio of means summary measure (i.e., the relative risk for a binary outcome). Although the SACE frequently refers to the difference in means, we will also focus on it as a ratio of means and hence we use the term SACE to broadly to represent any contrast in potential non-mortality outcomes within the subgroup of individuals who would survive regardless of treatment. This subgroup of always-survivors is the only principal stratum comprised of individuals whose death status under either treatment would not preclude the complete measurement of their non-terminal outcome.  In constructing the target estimand as functions of individual conditional expectations for each potential non-mortality outcome as in (\ref{eq:sace}), we should clarify that we are working within a super-population framework as opposed to a finite-sample framework. {In a super-population framework, each unit's potential outcomes 
are conceptualized as arising from an underlying infinite population, and the estimand of interest is defined as a specific contrast of population-level expectations of these potential outcomes.} As such, association parameters involved in defining the joint distribution of the potential non-mortality outcomes do not factor into our estimand definition, and because they also do not appear in the identified observed likelihood described later, we do not need to include them in our posterior specification (assuming a priori independence from marginal parameters) \citep{ding2018causal,imbens1997bayesian}.

{For the remaining key terms, we use the following notation. For individuals in cluster $i$, we denote the vector of survival statuses as $\boldsymbol{S}_i=(\boldsymbol{S}^T_{i1},\dots,\boldsymbol{S}^T_{iJ})^T$, where $\boldsymbol{S}_{ij}=(S_{ij1},\dots,S_{ijm_{ij}})^T$, and the vector of non-mortality outcomes as $\boldsymbol{Y}_i=(\boldsymbol{Y}^T_{i1},\dots,\boldsymbol{Y}^T_{iJ})^T$, where 
 $\boldsymbol{Y}_{ij}=(Y_{ij1},\dots,Y_{ijm_{ij}})^T$. 
We define the potential outcome vectors $\boldsymbol{S}_{i}(\cdot)$ and  $\boldsymbol{Y}_{i}(\cdot)$ as having components $S_{ijk}(\cdot)$ and $Y_{ijk}(\cdot)$, respectively, which are indexed analogously. We use $\boldsymbol{L}_i$ to denote vectors of all observed individual-level covariates (e.g., sex), cluster-period-level covariates (e.g., season), and cluster-level covariates (e.g., urban or not) for cluster $i$. $\boldsymbol{U}_i$ similarly refers to  vectors of unobserved covariates. We let $\boldsymbol{X}_i$ correspond to the main-effects design matrix constructed from $\boldsymbol{L}_i$ that is indexed at the individual-level (less the intercept and period indicator) with component vectors $\boldsymbol{X}_{ijk}$. Its column dimension will match the number of corresponding covariates in $\boldsymbol{L}_i$ and its row dimension will equal the number of individuals in cluster $i$. Characteristics that are not at the individual-level will be repeated accordingly; for example, a  characteristic for cluster $i$ will be the same across all $j,k$. In general, any bold terms without subscripts, such as $\boldsymbol{S}$ and $\boldsymbol{Y}$, denote vectors or collections of variables over all indices; for parameter vectors the dimension will be implied by our posterior specification, such as the number of regressors (regression parameter) or the number of clusters (cluster random intercept).}

\subsection{Causal {Structural} Assumptions} \label{sec:notation2}
The SACE estimand cannot be directly estimated using the observed data (as in Figure \ref{fig:crxodesign}) without assumptions. The estimand requires knowing an individual's pair of survival status potential outcomes to determine if they are an always-survivor, but in reality individuals are only assigned to one form of treatment. Moreover, for an always-survivor, we cannot observe both of their potential non-mortality outcomes. The CRXO trial design introduces additional complexity in formulating these assumptions because data arising from this design must be used to estimate an individual-level treatment effect. {In this section, we present causal structural assumptions that are a necessary step in conducting inference on the SACE for cross-sectional CRXO trials with alternating treatments across periods.}

\begin{assumption}
\textit{Multi-period crossover treatment assignment}: \textup{For each cluster, $a_{ij}=a \in \{1,0\}$ for odd $j \le J$ and  $a_{ij}=1-a$ for even $j \le J$, so that $\boldsymbol{a}_i \in \{(1,0,\dots)^T,(0,1,\dots)^T\}$.}
\end{assumption}

{Assumption 1 specifies the treatment sequences permitted by the CRXO design. Each cluster starts with one of the two treatments and alternates between them in successive periods. The next assumption specifies how an individual's potential outcomes depend on these assignments.}

\begin{assumption}
\textit{No interference between clusters and cluster-periods}: \textup{
For each cluster, $\boldsymbol{a}_i=(a_{i1},\dots,a_{iJ})^T$ denotes an arbitrary vector of binary terms representing a possible treatment sequence. $\boldsymbol{a}=(\boldsymbol{a}_1^T,\dots,\boldsymbol{a}_I^T)^T$ is the corresponding vector of treatment sequences across all clusters. For each individual, $S_{ijk}(\boldsymbol{a})=S_{ijk}(\boldsymbol{a}_i)=S_{ijk}(a_{ij})$ and $Y_{ijk}(\boldsymbol{a})=Y_{ijk}(\boldsymbol{a}_i)=Y_{ijk}(a_{ij})$.}
\end{assumption}

Assumption 2 states that an individual's potential outcomes depend on the treatment strategy assigned to their own cluster-period, rather than on assignments in other clusters or periods. To illustrate this coherently, we have followed common practice and have not changed potential outcomes notation when dimension of the treatment input changes (e.g., we let $S_{ijk}(\boldsymbol{a}_i)=S_{ijk}(a_{ij})$). Assumption 2 could be violated, for instance, if there were treatment arms (e.g., surgical options or health education programs) where a physician's experience administering any treatment to one patient alters how they provide treatments to other patients that affects these patients' outcomes positively or negatively. This interference can occur at the cluster-level if physicians are treating patients at multiple sites and at the cluster-period/individual-level if they are treating multiple patients within a site. In our study, physicians are administering standard medications to patients in ICUs, which are highly monitored environments, so we do not expect this type of interference to occur. Furthermore, in this setting, an individual-level treatment (i.e., a medication) is provided to patients in ICUs who are physically isolated, so the treatment does not inherently involve other patients, unlike group therapies or vaccines. {Importantly, the cross-sectional design means that different individuals are observed in different cluster-periods. It does not, however, rule out effects of earlier treatment assignments on later patients through accumulated physician experience. Assumption 2 excludes such cluster-level carryover, a limitation we discuss in Section \ref{sec:discuss}.}

     \begin{assumption} \textit{Consistency}: 
     \textup{For each individual, if $A_{ij}=a$ in $\boldsymbol{A}_i$, then $S_{ijk}=S_{ijk}(a)$ and $Y_{ijk}=Y_{ijk}(a)$.}
    \end{assumption}

    Assumption 3 proposes that each individual's potential outcomes are equal to their observed 
    outcomes under the treatment assigned to their cluster-period. 
    In other words, there are no hidden variations in the treatments, which might give rise to distinctions in individuals' potential and observed outcomes.
     
     {Let $\boldsymbol{a}^*_i = (1,0,\dots)^T$, where $i$ simply denotes that it's a cluster-level sequence.}
    \begin{assumption}
    \textit{Cluster-level independence}: \textup{The collections of cluster-defined variables $\{\boldsymbol{Y}_{i}(\boldsymbol{a}^*_i),$
    $\boldsymbol{Y}_{i}(\boldsymbol{1}-\boldsymbol{a}^*_i),\boldsymbol{S}_{i}(\boldsymbol{a}^*_{i}),\boldsymbol{S}_{i}(\boldsymbol{1}-\boldsymbol{a}^*_i),\boldsymbol{L}_i,\boldsymbol{U}_i\}$ are mutually independent with respect to the cluster index $i$.}
    \end{assumption}

    Assumption 4 requires statistical independence of the cluster-level vectors of potential outcomes and covariates across clusters. It does not require independence of individuals or periods within the same cluster and is distinct from the absence of causal interference in Assumption 2.

    \begin{assumption}
    \textit{Cluster-level randomization}: \textup{For each cluster, the collection of cluster-defined variables is jointly independent of treatment assignment,  
    $\{\boldsymbol{Y}_{i}(\boldsymbol{a}^*_i),\boldsymbol{Y}_{i}(\boldsymbol{1}-\boldsymbol{a}^*_i),\boldsymbol{S}_{i}(\boldsymbol{a}^*_i),\boldsymbol{S}_{i}(\boldsymbol{1}-\boldsymbol{a}^*_i),\boldsymbol{L}_i,\boldsymbol{U}_i\} \perp \boldsymbol{A}_{i}$, with $0 <P(\boldsymbol{A}_i=\boldsymbol{a}^*_i) < 1$.} 
    \end{assumption}

    {Assumption 5 states that the mechanism of randomization is independent of each cluster's vectors of potential outcomes and covariates.} Although $\boldsymbol{U}_i$ is unobserved, randomization applies to $\boldsymbol{U}_i$ as well as observed $\boldsymbol{L}_i$.
    Looking ahead to our modeling setup, which is required for {estimation} of the SACE, we will leverage the structural features of the unmeasured $\boldsymbol{U}_i$ to appropriately account for
    latent clustering effects within and across periods.

    \begin{assumption}
    \textit{Survival monotonicity}: \textup{For each individual, $S_{ijk}(1) \ge S_{ijk}(0)$.} 
    \end{assumption}

    Assumption 6 signifies that each individual's potential survival status under the active treatment (treatment 1) is no worse than their potential survival status under the control treatment (treatment 0).
    Survival monotonicity simplifies modeling assumptions by removing the stratum of harmed patients, thereby identifying individuals observed to have died under the active treatment or have survived under the control treatment with a principal stratum (never-survivors and always-survivors respectively). This assumption requires substantive knowledge and, when justified, may improve estimation stability in practice \citep{chen2024bayesian} by eliminating what would have otherwise constituted two additional mixture components in the likelihood, as presented in the next section. {We assess violations of this assumption through supplemental simulations in Section \ref{sec:violations} and consider its practical implications in our discussion, Section \ref{sec:discuss}.}

    {Assumptions 1 and 5 describe features of the CRXO trial design and are guaranteed by design. Assumption 1 specifies the alternating treatment sequences prescribed by the trial protocol, whereas Assumption 5 states the independence and positivity conditions implied by randomized allocation of clusters to these sequences for the prespecified trial population. Assumptions 2--4 and 6 are additional conditions that are not guaranteed by the design. Assumptions 2 and 3 state the no-interference and consistency conditions for the cross-sectional CRXO setting \citep{vanderweele2013causal}, allowing us to express individual potential outcomes as $S_{ijk}(a)$ and $Y_{ijk}(a)$ for $a\in\{0,1\}$ and relate them to the observed data. Assumption 4 specifies independence between clusters in the super-population, as also assumed by \citet{isenberg2025weighting} in the parallel-arm design setting. Assumption 6 is the survival monotonicity condition used in previous SACE analyses of CRTs \citep{tong2023bayesian,wang2024mixed,li2026uncovering}. The plausibility of these additional assumptions may not be guaranteed by design and would need be assessed given the context of the study.}

    {As suggested in the introduction, these six assumptions alone are not sufficient to point identify the SACE, but our goal remains to conduct inference on the SACE in this setting.  
    The SACE is an estimand that depends on the distribution of the non-mortality outcomes given membership to the always-survivor subgroup, but principal stratum membership is only identified for individuals with certain combinations of observed treatment and survival status under these assumptions, making only bounding possible without further assumptions \citep{zhang2003estimation,zhang2009likelihood,chen2024bayesian}. Moreover, the SACE is an individual-level estimand, so we {require assumptions about how to} leverage information from clusters (i.e., patient information as well as cluster and cluster-period characteristics) that enable targeting individual-level effects. To resolve these issues, we conduct hierarchical Bayesian inference for the SACE by making parametric modeling assumptions and by selecting a coherent prior specification that leads to a proper posterior distribution \citep{tong2023bayesian}. Our modeling choices, as presented in Section \ref{sec:modelspec}, align with the conventional mixed-effects modeling approaches for CRXO trials \citep{turner2007analysis,morgan2017choosing}, which notably incorporate normally distributed latent variables for cluster and (possibly) cluster-period random effects. 
    In Appendix \ref{sec:achSACE}, we provide greater insight into why the SACE is not point identified under Assumptions 1-6, but we outline how with these assumptions and specifying a proper posterior distribution, one can perform Bayesian inference for the SACE  \citep{imbens1997bayesian,hirano2000assessing}.} 

    We expect the randomization assumption to hold for PEPTIC since patients must be undergoing mechanical ventilation to be treated. However, in more general settings, we may not be able to rule out the possibility that individuals gain knowledge about treatment assignment after the first period and that such information impacts the cluster from which they seek treatment.
    In fact, it is conceivable at an even time period in a crossover design that individuals may choose a treatment site which gives the \textit{opposite} of their sought after treatment, i.e., the initial treatment. If we cannot assume a strict randomization assumption as per Assumption 5, which posits independence of all cluster information and treatment assignment, we may only be able to make a weaker ignorability assumption, {$\{\boldsymbol{Y}_{i}(\boldsymbol{a}^*_i),\boldsymbol{Y}_{i}(\boldsymbol{1}-\boldsymbol{a}^*_i),\boldsymbol{S}_{i}(\boldsymbol{a}^*_i),\boldsymbol{S}_{i}(\boldsymbol{1}-\boldsymbol{a}^*_i)\} \perp \boldsymbol{A}_{i}|\boldsymbol{L}_i,\boldsymbol{U}_i$ (where $0<P(\boldsymbol{A}_i=\boldsymbol{a}^*_i|\boldsymbol{L}_i,\boldsymbol{U}_i)<1$)}. This would require that we have access to a rich enough set of covariates about each cluster that once conditioned upon, individuals' potential outcomes within each cluster are independent of treatment assignment. While this would not change our approach to modeling for SACE {estimation} per se, it would restrict our modeling flexibility by necessitating inclusion of certain variables. Importantly, it would necessitate specifying that the latent $\boldsymbol{U}_i$ has a parametric form compatible with our modeling assumptions. Separately, our analysis assumes that cluster size is uninformative, an additional assumption not implied by Assumption 5. Informative cluster size (ICS), in which outcomes or treatment effects vary with cluster size, can arise even when treatment assignment is independent of cluster size \citep{kahan2023estimands,chen2023model}. Accommodating ICS would require further consideration of cluster size in the modeling and standardization used to estimate the individual-level SACE, which we leave for future work. 
    
\section{Bayesian Estimation of Causal Effects} \label{sec:estim}
 
Recall the observed data for each cluster $i$ is comprised of the treatment assignment sequence $\boldsymbol{A}_{i}$, survival status vector $\boldsymbol{S}_{i}$, non-mortality outcome vector $\boldsymbol{Y}_{i}$, 
and covariates $\boldsymbol{L}_{i}$. 
$\boldsymbol{\theta}$ denotes the vector of parameters (defined shortly). Under the causal assumptions, specifically survival monotonicity, an individual who survives, $S_{ijk}=1$, under the control treatment, $A_{ij}=0$, is classified as an always-survivor, $G_{ijk}=(1,1)$, and conversely, anyone who dies, $S_{ijk}=0$, under the active treatment, $A_{ij}=1$, is classified as a never-survivor, $G_{ijk}=(0,0)$. However, for the remaining observed survival-treatment combinations, an individual may be a member of one of two principal strata. For those that survive under the active treatment, they might be an always-survivor or a protected-patient, $G_{ijk}=(1,0)$. For those that die under the control treatment, they might be a never-survivor or a protected-patient. As a result, the contributions of the individuals with one of those survival-treatment combinations to an observed data likelihood will be a mixture model involving the two plausible principal strata \citep{zhang2009likelihood}. For example, for the first pairing ($A_{ij}=1$ and $S_{ijk}=1$), there is a mixture of outcome models for $G_{ijk}=(1,1)$ and $G_{ijk}=(1,0)$, where the former is used directly in SACE estimation. In contrast, for the second pairing, there is no outcome modeling due to death. 

There are two common, related approaches to tackle mixture modeling: the frequentist method of the expectation-maximization (EM) algorithm \citep{dempster1977maximum} and the Bayesian method of data augmentation \citep{tanner1987calculation,diebolt1994estimation}. Both require specifying a complete data likelihood, which incorporates the missing data to avoid the intractable sum of mixture components present in the observed data likelihood after marginalization. With our approach, we complete the observed-data likelihood by including each individual's principal stratum membership. Working with this new likelihood, we can use the EM-algorithm to perform maximum likelihood estimation or in the Bayesian framework, to define an augmented posterior distribution that is itself more tractable and suited to simpler Markov chain Monte Carlo (MCMC) methods.  
We opt for a Bayesian approach since it 
intrinsically permits uncertainty quantification of our parameters, including the latent principal stratum membership, without relying on bootstrapping \citep{zhang2009likelihood}. Moreover, priors can help regularize parameter estimates, such as through L2 regularization, which can be particularly beneficial with limited data and complex modeling \citep{gelman2013bayesian,oganisian2021practical}.

\subsection{Model Specification}
\label{sec:modelspec} 

We propose a model for the principal strata-augmented likelihood $p(\boldsymbol{y},\boldsymbol{g}|\boldsymbol{a},\boldsymbol{l},\boldsymbol{\theta})$, which, in conjunction with the causal assumptions of Section \ref{sec:notation}, enables the estimation of SACE (see Appendix \ref{sec:achSACE}). To model this conditional joint distribution, we extend the normal and multinomial regression approaches for the outcome and principal strata models respectively as done in the parametric model-based literature for both iid and cluster-randomized data \citep{zhang2009likelihood,tong2023bayesian,wang2024mixed}. {Note, we make the implicit (super-population) assumption of common individual-level expectations when defining the two components of the target SACE contrasts, which might be achieved, for example, under identically distributed data for always-survivors' non-mortality outcomes \citep{wang2024mixed}.}

In our data application since $Y_{ijk}$ is length of stay, which is positive and typically right skewed, we will model it in terms of its natural logarithmic value as is common practice \citep{marazzi1998fitting}. We use the notation $\log(Y_{ijk})=\mathcal{Y}_{ijk}$. {For outcome modeling, the superscripts $(1,1)$ and $(1,0)$ on the non-mortality outcomes indicate conditioning on membership in $G_{ijk}=g$, i.e., regression relationships differ by principal stratum. Subscripts $(1,1)$ and $(1,0)$ on regression terms indicate that they are stratum-specific; terms subscripted by $(1,1),1$ only appear under treatment 1. Recall that $\boldsymbol{X}_{ijk}$ corresponds to the covariates $\boldsymbol{L}_i$ with respect to the individual (as per a design matrix without intercept and period indicator). Our proposed random-intercept outcome modeling is:} 
\begin{equation}
\label{eq:outmod}
\begin{gathered}
\mathcal{Y}^{(1,1)}_{ijk}=\alpha_{(1,1)}+\alpha_{(1,1),1}A_{ij}+\boldsymbol{\beta}_{(1,1)}^T\boldsymbol{X}_{ijk} + \boldsymbol{\delta}_{(1,1)}^T\boldsymbol{R}_j+\boldsymbol{\beta}_{(1,1),1}^TA_{ij}\boldsymbol{X}_{ijk}+ \xi_{i,(1,1)} + \gamma_{ij,(1,1)} + \epsilon_{ijk,(1,1)}\\
\mathcal{Y}^{(1,0)}_{ijk}|\{A_{ij}=1\}=\alpha_{(1,0)}+\boldsymbol{\beta}_{(1,0)}^T\boldsymbol{X}_{ijk} + \boldsymbol{\delta}_{(1,0)}^T\boldsymbol{R}_j  + \xi_{i,(1,0)} + \gamma_{ij,(1,0)} + \epsilon_{ijk,(1,0)}
\end{gathered}
\end{equation}
where $\boldsymbol{R}_j$ is a dummy vector of dimension $J-1$ indicating the period to which an individual belongs 
(with reference being the zero vector). For each individual, $\xi_{i,g} \sim N(0,\sigma^2_{C,g})$, $\gamma_{ij,g} \sim N(0,\sigma^2_{CP,g})$, and $\epsilon_{ijk,g} \sim N(0,\sigma_g^2)$ for $g \in \{(1,1),(1,0)\}$; all random effects and error terms are mutually independent and independent of covariates. Random effects account for additional variation in the outcome due to individuals potentially being correlated within cluster and cluster-period. Strength of correlation is quantified by two levels, the within-cluster between-period correlation (BPC), $\sigma^2_{C,g}/(\sigma^2_{C,g}+\sigma^2_{CP,g}+\sigma^2_{g})$, and the within-cluster within-period correlation (WPC), $(\sigma^2_{C,g}+\sigma^2_{CP,g})/(\sigma^2_{C,g}+\sigma^2_{CP,g}+\sigma^2_{g})$ \citep{arnup2017understanding}. These correlations are defined on the log-outcome scale, conditional on covariates and principal stratum.
The efficiency advantage of a CRXO design versus a standard parallel-arm CRT design is driven by the magnitude of $\sigma^2_{CP,g}$. For smaller values of $\sigma^2_{CP,g}$ relative to the cluster and error variances, the cluster-periods are more comparable across treatment, which tends to increase power \citep{forbes2015cluster,arnup2017understanding}. 
{We focus on Bayesian estimation of the SACE in the two-period CRXO trial specifically} in which protected-patients with observed outcomes who are in the same cluster will necessarily also be in the same cluster-period (since their non-mortality outcome can only be observed under the active treatment). Nonetheless, we include the cluster-period random effect to allow for generalization beyond two periods and to signify notationally that the period, along with the cluster, induces variation in the outcome. In the two-period setting, the protected-patient outcome likelihood depends on $\sigma^2_{C,(1,0)}+\sigma^2_{CP,(1,0)}$, so the separation of these two variance components depends on their prior distributions.

For principal strata modeling, we use a multinomial logistic model with random intercepts for cluster and a fixed effect for period. {Stratum probabilities for individuals are obtained by applying the softmax function to the following linear predictors: $Z_{ijk}$, which corresponds to $G_{ijk}=(1,1)$, $W_{ijk}$, which corresponds to $G_{ijk}=(1,0)$, and $0$, which corresponds to the reference group $G_{ijk}=(0,0)$. Regression terms are stratum-specific and are subscripted by $z$ and $w$ accordingly. Explicitly, $Z_{ijk}$ and $W_{ijk}$ are defined as:}
\begin{align}
\label{eq:prelogit}
\begin{split}
Z_{ijk}&=\alpha_z+\boldsymbol{\beta}_{z}^T\boldsymbol{X}_{ijk} + \boldsymbol{\delta}_z^T\boldsymbol{R}_j + \eta_{i,z}\\
W_{ijk}&=\alpha_w+\boldsymbol{\beta}_{w}^T\boldsymbol{X}_{ijk} + \boldsymbol{\delta}_w^T\boldsymbol{R}_j + \eta_{i,w}    
\end{split}
\end{align}
\noindent where $G_{ijk}|\boldsymbol{X}_{ijk},\boldsymbol{R}_j,\eta_{i,z},\eta_{i,w}$ is distributed (in bijection with) $\text{Multi}(1;\pi^{(1,1)}_{ijk},\pi^{(1,0)}_{ijk},\pi^{(0,0)}_{ijk})$ such that
\begin{align}
\label{eq:multi}
\begin{split}
\pi^{(1,1)}_{ijk}&=\exp(Z_{ijk})/(\exp(Z_{ijk})+\exp(W_{ijk})+1)\\
\pi^{(1,0)}_{ijk}&=\exp(W_{ijk})/(\exp(Z_{ijk})+\exp(W_{ijk})+1)
\end{split}
\end{align}
and $\pi^{(0,0)}_{ijk}=1-\pi^{(1,1)}_{ijk}-\pi^{(1,0)}_{ijk}$, where the {stratum superscripts correspond to the probability of being in that stratum}. For each individual, we assume $\eta_{i,z} \sim N(0,\tau^2_{C,z})$ and $\eta_{i,w} \sim N(0,\tau^2_{C,w})$, where all random effects are mutually independent and independent of covariates. We choose to represent the intracluster correlation coefficient (ICC) (on the latent scale) for each non-reference stratum as $\tau_{C,p}^2/(\tau_{C,p}^2+\pi^2/3)$, which is derivable using a latent variable representation of multinomial logistic regression \citep{hedeker2003mixed,goldstein2002partitioning}. Here, $p\in\{z,w\}$ indexes the always-survivor and protected-patient strata, respectively. We recommend not including the cluster-period random effects $\nu_{ij,z} \sim N(0,\tau^2_{CP,z})$ and $\nu_{ij,w} \sim N(0,\tau^2_{CP,w})$ as they have been shown to be somewhat computationally unstable even in a much simpler binary logistic mixed-effects model setting, particularly when there are few clusters \citep{morgan2017choosing}. Additionally, within the SACE literature for parallel-arm CRTs, where the treatment assignment shares an index with cluster membership (analogous to the cluster-period in CRXO trials), there is often limited improvement and potential convergence issues for including cluster-level random effects in the principal strata model \citep{tong2023bayesian,wang2024mixed}, which we evaluate further in simulations in Section \ref{sec:sim}.
 
In general, we assume that all random effects, error terms, and their respective variances are stratum-specific for modeling, marking an extension of previous works in the CRT parallel-arm setting \citep{tong2023bayesian,wang2024mixed}. Jo (2022) \cite{jo2022handling} noted that over-constraining the variance terms, when not warranted, may considerably hinder the performance of principal strata based Gaussian mixture modeling, so our method aims to be as flexible as possible within that framework. 

\subsection{Inference With Augmented Posterior Distribution}
\label{sec:augment}

Let $O(a)$ be the set of individuals assigned treatment $A_{ij}=a$. {Let $\boldsymbol{\theta}_{\text{PS}}$ and $\boldsymbol{\theta}_{\text{Out}}$ denote the parameters involved in modeling principal strata and the non-mortality outcome, respectively. Since $\boldsymbol{\theta}^T=(\boldsymbol{\theta}_{\text{Out}}^T,\boldsymbol{\theta}_{\text{PS}}^T)$ includes all random effects}, 
 we write the posterior under our model specification as follows
\begin{align}
\label{eq:posterior}
\begin{split}
p(\boldsymbol{\theta}|\boldsymbol{y},\boldsymbol{g},\boldsymbol{a},\boldsymbol{s},\boldsymbol{l}) &\propto p(\boldsymbol{y},\boldsymbol{g}|\boldsymbol{a},\boldsymbol{l},\boldsymbol{\theta})p(\boldsymbol{\theta})\\
&=p(\boldsymbol{y}|\boldsymbol{g},\boldsymbol{a},\boldsymbol{l},\boldsymbol{\theta}_{\text{Out}})p(\boldsymbol{g}|\boldsymbol{l},\boldsymbol{\theta}_{\text{PS}})p(\boldsymbol{\theta})\\
& = \prod_{ijk \in O(1)} \left[f^{((1,1),1)}_{y_{ijk}}\pi_{ijk}^{(1,1)}\right]^{I(G_{ijk}=(1,1))}\left[f^{(1,0)}_{y_{ijk}}\pi_{ijk}^{(1,0)}\right]^{I(G_{ijk}=(1,0))} \left[\pi_{ijk}^{(0,0)}\right]^{I(G_{ijk}=(0,0))} \\
& \times \prod_{ijk \in O(0)} \left[f^{((1,1),0)}_{y_{ijk}}\pi_{ijk}^{(1,1)}\right]^{I(G_{ijk}=(1,1))} \left[\pi_{ijk}^{(1,0)}\right]^{I(G_{ijk}=(1,0))}\left[\pi_{ijk}^{(0,0)}\right]^{I(G_{ijk}=(0,0))} \times p(\boldsymbol{\theta})
\end{split}
\end{align}
where $f^{(1,1),a}_{y_{ijk}}$ denotes the log-normal density function of $Y_{ijk}$  ({as per (\ref{eq:outmod})}) conditional on the following: being an always-survivor, assigned treatment $A_{ij}=a$, observed covariates, and random effects. $f^{(1,0)}_{y_{ijk}}$ is defined similarly among protected-patients under treatment $A_{ij}=1$. Note that one's principal stratum membership, $G_{ijk}$, necessarily includes the value of $S_{ijk}(a)$, and thus their observed survival $S_{ijk}$ status under their assigned treatment $A_{ij}$ by the causal assumptions; hence, the observed survival status vector is absorbed in the RHS of the above posterior.
The parameter vector $\boldsymbol{\theta}$ is given by

\begin{align}
\begin{split}
    \boldsymbol{\theta}=&(\boldsymbol{\theta}_{(1,1)}^T,\boldsymbol{\theta}_{(1,0)}^T,\sigma_{(1,1)}^2,\sigma_{(1,0)}^2,\boldsymbol{\xi}_{(1,1)}^T,\boldsymbol{\xi}_{(1,0)}^T,\sigma^2_{C,(1,1)},\sigma^2_{C,(1,0)},\boldsymbol{\gamma}_{(1,1)}^T,\boldsymbol{\gamma}_{(1,0)}^T,\sigma^2_{CP,(1,1)},\sigma^2_{CP,(1,0)},
    \\&\boldsymbol{\theta}_z^T,\boldsymbol{\theta}_w^T,\boldsymbol{\eta}_{z}^T,\boldsymbol{\eta}_{w}^T,\tau^2_{C,z},\tau^2_{C,w})^T
\end{split}
    \end{align}    
where $\boldsymbol{\theta}_{(1,1)}$ and $\boldsymbol{\theta}_{(1,0)}$ represent the coefficients including the intercepts for the outcome models in strata $(1,1)$ and $(1,0)$ as in (\ref{eq:outmod}), and $\boldsymbol{\theta}_z$ and $\boldsymbol{\theta}_w$ are the coefficients of the multinomial logistic model for {linear predictors $Z_{ijk}$ and $W_{ijk}$} respectively as in (\ref{eq:prelogit}) and (\ref{eq:multi}). 
The complete description of the prior is contained in Section \ref{sec:priors}, and details for conditional posterior derivations are in Section \ref{sec:condpost} of Appendix \ref{sec:gibbssample} {for a two-period cross-sectional CRXO design.} Although the principal strata cluster-period terms $\{\boldsymbol{\nu}_{z},\boldsymbol{\nu}_{w},\tau^2_{CP,z},\tau^2_{CP,w}\}$ are not incorporated in our proposed model, we include these parameters in Appendix \ref{sec:gibbssample} for full generality, and we evaluate their impact on modeling in additional simulations in Appendix \ref{sec:supptables}. For elements of the augmentation term $\boldsymbol{g}$, which are independent across individuals conditional on random effects and observed covariates, the \textit{form} of the conditional posterior will depend on an individual's observed survival and treatment. Patients with $A_{ij}=0$ and $S_{ijk}=1$ have $G_{ijk}=(1,1)$, and patients with $A_{ij}=1$ and $S_{ijk}=0$ have $G_{ijk}=(0,0)$. As noted above, the remaining survival-treatment combinations comprise mixtures of two strata; thus, they will be (in bijection with) Bernoulli-distributed random variables.  For those with $A_{ij}=1$ and $S_{ijk}=1$, their probability of being $G_{ijk}=(1,1)$ as opposed to $G_{ijk}=(1,0)$ will depend on their relative weights as determined from the principal strata and outcome models. For those with $A_{ij}=0$ and $S_{ijk}=0$, their probability of being $G_{ijk}=(1,0)$ as opposed to $G_{ijk}=(0,0)$ will depend only on their relative weights from the principal strata modeling. 

We now outline a mechanism for generating a closed form for a Gibbs sampler for the parameters of the principal strata multinomial logistic model, that is, without requiring a Metropolis-Hastings step. Full details are written in Section \ref{sec:condpost}. As demonstrated by Polson et al. (2013) \citep{polson2013bayesian} in the context of iid data, the likelihood contribution for an individual under logistic regression can be represented as a scale mixture of Gaussian kernels with mixture weights according to a P\'{o}lya-Gamma distribution $\text{PG}(1,0)$. Specifically, for each observation, 
\begin{equation}
\label{eq:pgmix}
\frac{[\exp(\varphi_i)]^{b_i}}{1+\exp(\varphi_i)}=2^{-1}\exp\left([b_i-1/2]\varphi_i\right)\int_0^\infty \exp(-\omega_i \varphi_i^2/2)p(\omega_i)d\omega_i
\end{equation}
where $\omega_i \sim \text{PG}(1,0)$, {with $\varphi_i$ and $b_i$ generic inputs.} By including $\boldsymbol{\omega}$, a vector of independent realizations $\omega_i$ from $\text{PG}(1,0)$, the likelihood can be written as (proportional to) the product of independent normal and $\text{PG}(1,0)$ densities. Therefore, under normal priors for the regression coefficients, the conditional posteriors for the regression coefficients are also normally distributed. Moreover, the conditional posteriors for the P\'{o}lya-Gamma variables are themselves P\'{o}lya-Gamma, a phenomenon that occurs since any $\text{PG}(b,c)$ is related to its base distribution $\text{PG}(b,0)$ through an exponential tilting. For more information regarding ergodicity and computational efficiency of this data augmentation derived Gibbs sampler for logistic regression, consult Choi and Hobert (2013)
\citep{choi2013polya} and Polson et al. (2013) \citep{polson2013bayesian}. We extend the P\'{o}lya-Gamma augmentation technique to our model for the principal strata with hierarchical data, which is a three-category multinomial logistic regression with cluster and possibly cluster-period random intercepts. Since there are three categories, we will use the augmented conditional posterior $p(\boldsymbol{\theta}_{\text{PS}}|\boldsymbol{\omega}_z,\boldsymbol{\omega}_w,\boldsymbol{g},\boldsymbol{l})$  with two independent P\'{o}lya-Gamma vectors containing elements $\omega_{ijk,z} \stackrel{\text{iid}}{\sim} \text{PG}(1,0)$ and $\omega_{ijk,w} \stackrel{\text{iid}}{\sim} \text{PG}(1,0)$ to represent the contributions $[\pi_{ijk}^g]^{I(G_{ijk}=g)}$ for complete Gibbs sampling (assuming $p(\boldsymbol{\theta})=p(\boldsymbol{\theta}_{\text{Out}})p(\boldsymbol{\theta}_{\text{PS}})$ and $\boldsymbol{\theta}_{\text{Out}} \perp \{\boldsymbol{\omega}_z,\boldsymbol{\omega}_w\}$) \citep{polson2013bayesian,allen2023bayesian}. Explicitly, the likelihood contribution of each individual relevant to $\boldsymbol{\theta}_{\text{PS}}$ is:
\begin{equation}\frac{\left[\exp(z_{ijk})\right]^{I(G_{ijk}=(1,1))}\left[\exp(w_{ijk})\right]^{I(G_{ijk}=(1,0))}}{1+\exp(z_{ijk})+\exp(w_{ijk})}\end{equation}
To illustrate how posterior augmentation with $\boldsymbol{\omega}_z$ works for parameters regarding the terms $Z_{ijk}$, we factor out $1+\exp(w_{ijk})$ from the denominator, leaving an expression proportional to
\begin{equation} \frac{\left[\exp(z_{ijk}-\log\{1+\exp(w_{ijk})\})\right]^{I(G_{ijk}=(1,1))}}{1+\exp(z_{ijk}-\log\{1+\exp(w_{ijk})\})}\end{equation}
 Setting $\varphi_{ijk}=z_{ijk}-\log\{1+\exp(w_{ijk})\}$ and $b_{ijk}=I(G_{ijk}=(1,1))$, we see that this expression, indexed by $ijk$, can be written according to the integral identity in Equation (\ref{eq:pgmix}) with weighting by $p(\omega_{ijk,z})$; an analogous construction can be done for parameters pertaining to the $W_{ijk}$ using $\omega_{ijk,w}$.  

For $\boldsymbol{\theta}$, we make normal and inverse gamma prior specifications on its components, which enable draws from closed-form conditional posteriors for all parameters (including those used in augmentation) per Gibbs sampling \citep{gelman2013bayesian,polson2013bayesian}. We assume that all the fixed effect regression parameter vectors involved in the non-mortality outcome and principal strata modeling are mean zero multivariate normally distributed with diffuse diagonal variances. Our derivations provided in the appendices are generalized to allow for any covariance structure on the fixed effects parameters if the context warrants it. All random effect vectors are mean zero multivariate normally distributed with diagonal variances per the design of the hierarchical model. All variance terms are ascribed diffuse inverse gamma priors. Non-diagonal covariance matrices could be specified for the error and/or random effects parameters, 
but we have avoided this complexity in adherence with our modeling assumptions and to prevent the risk of over-parametrization given our setting with small samples of clusters. As indicated by the term diffuse, we intend for our prior specifications to be weakly informative, allowing the data to principally drive inference. Nonetheless, as we later demonstrate in our simulation study, we would expect some, albeit minimal, shrinkage to 0 as our priors on the fixed effects result in an L2 penalty on their magnitude \citep{van2019shrinkage}. Lastly, all priors on the distinct fixed effect and random effect vectors are assumed mutually independent. With such specification on $\boldsymbol{\theta}$ and the data augmentation steps, all component conditional posteriors will be conjugate to the prior distributions as written out in Appendix \ref{sec:gibbssample}. Note, conditional conjugacy holds with log-normal densities for the outcome modeling since they are proportional to normal densities with logged arguments (by factors of $1/y_{ijk}$).

For estimation of the SACE, we consider it expressed both in difference in means for $\mathcal{Y}_{ijk}$ (log-time scale) and  ratio of means for $Y_{ijk}$ (time scale). Given our model specifications, estimators for these will avoid having to average over the distribution of the random effects, i.e., collapsible over any latent clustering variables. We let $\boldsymbol{\psi}_{i,(1,1)}$ be all random effects conditioned on in the outcome model for cluster $i$ for always-survivors. Under the causal structural and modeling assumptions, we have for the difference in means (LDiff){\begin{align}
\begin{split}
    \mu_{\text{LDiff}}&=E\left\{E\left(\mathcal{Y}_{ijk}|G_{ijk}=(1,1),A_{ij}=1,\boldsymbol{X}_{ijk},\boldsymbol{R}_j,\boldsymbol{\psi}_{i,(1,1)}\right)|G_{ijk}=(1,1)\right\}\\&\hspace{.4cm}-E\left\{E\left(\mathcal{Y}_{ijk}|G_{ijk}=(1,1),A_{ij}=0,\boldsymbol{X}_{ijk},\boldsymbol{R}_j,\boldsymbol{\psi}_{i,(1,1)}\right)|G_{ijk}=(1,1)\right\}
    \\&=\alpha_{(1,1),1}+ \boldsymbol{\beta}_{(1,1),1}^TE\{\boldsymbol{X}_{ijk}|G_{ijk}=(1,1)\}
\end{split}
\end{align}
We estimate $\mu_{\text{LDiff}}$ with $\hat \mu_{\text{LDiff}}$ defined by 
\begin{align}
\begin{split}
\label{eq:uldiff}
E(\mu_{\text{LDiff}}|\text{data})&=\int\left[\alpha_{(1,1),1}+ \boldsymbol{\beta}_{(1,1),1}^TE\{\boldsymbol{X}_{ijk}|G_{ijk}=(1,1)\}\right]dP_{\text{post}}\\
&\approx \int \alpha_{(1,1),1}+  {\boldsymbol{\beta}}_{(1,1),1}^T\overline{\boldsymbol{X}}_{(1,1)}dP_{\text{post}}=\hat\mu_{\text{LDiff}}
\end{split}
\end{align}}
where $P_{\text{post}}$ is short for posterior distribution. 
{Here, we have approximated the distribution of the covariates in the always-survivor subgroup with the empirical distribution, which assigns equal weight $1/n_{(1,1)}$ to each observation where $n_{(1,1)}$ denotes the number of always-survivors; this produces the sample mean vector $\overline{\boldsymbol{X}}_{(1,1)}$ as an estimator of the conditional expectation \citep{chen2024bayesian,tong2023bayesian}.}
Alternatively, the Bayesian bootstrap \citep{rubin1981bayesian,oganisian2021practical,oganisian2024hierarchical} puts a (hierarchical) Dirichlet prior on those weights, which permits uncertainty in the empirical estimate as well. While this approach is more adherent to Bayesian methods, it is computationally more expensive and may require more data for stability. We proceed with averaging over the observed covariates among always-survivors, which performs well in our simulation study. 

Similarly, with outer expectations as defined above, for the ratio of means (ROM),
{
\begin{align}
\begin{split}
    \mu_{\text{ROM}}&=\frac{E\left\{E(Y_{ijk}|G_{ijk}=(1,1),A_{ij}=1,\boldsymbol{X}_{ijk},\boldsymbol{R}_j,\boldsymbol{\psi}_{i,(1,1)})|G_{ijk}=(1,1)\right\}}{E\left\{E(Y_{ijk}|G_{ijk}=(1,1),A_{ij}=0,\boldsymbol{X}_{ijk},\boldsymbol{R}_j,\boldsymbol{\psi}_{i,(1,1)})|G_{ijk}=(1,1)\right\}}
    \\&=\frac{\exp\left(\alpha_{(1,1),1}\right)E\left\{\exp\left(\boldsymbol{\beta}_{(1,1)}^T\boldsymbol{X}_{ijk} + \boldsymbol{\delta}_{(1,1)}^T\boldsymbol{R}_j +\boldsymbol{\beta}_{(1,1),1}^T\boldsymbol{X}_{ijk}\right)\big|G_{ijk}=(1,1)\right\}}{E\left\{\exp\left(\boldsymbol{\beta}_{(1,1)}^T\boldsymbol{X}_{ijk} + \boldsymbol{\delta}_{(1,1)}^T\boldsymbol{R}_j \right)\big|G_{ijk}=(1,1)\right\}}
\end{split}
\end{align}}
since the random effects and error terms are independent of covariates. However, we do not assume covariates are mutually independent. As above, we estimate $\mu_{\text{ROM}}$ with the posterior mean $E(\mu_{\text{ROM}}|\text{data})$, estimating the always-survivor covariate distribution with the empirical distribution:
{\begin{equation}\label{eq:urom} \hat \mu_{\text{ROM}}=\int{\frac{\exp\left(\alpha_{(1,1),1}\right)\overline{ \exp\left(\boldsymbol{X}_{(1,1)}\boldsymbol{\beta}_{(1,1)} + \boldsymbol{R}_{(1,1)}\boldsymbol{\delta}_{(1,1)}+ \boldsymbol{X}_{(1,1)}\boldsymbol{\beta}_{(1,1),1}\right)}}{\overline{ \exp\left(\boldsymbol{X}_{(1,1)}\boldsymbol{\beta}_{(1,1)} + \boldsymbol{R}_{(1,1)}\boldsymbol{\delta}_{(1,1)}\right)}}}dP_{\text{post}}\end{equation}
where the subscripts (1,1) in $\boldsymbol{X}_{(1,1)}$ and $\boldsymbol{R}_{(1,1)}$  are used to denote the matrices of covariates and period indicators for all individual always-survivors. Moreover, we define $\exp(\boldsymbol{V})=(\exp(V_1),\dots,\exp(V_n))^T$ such that 
$\overline{\exp(\boldsymbol{V})}=1/n \boldsymbol{1}^T\exp(\boldsymbol{V})$. Therefore, the overlined components of $\hat \mu_{\text{ROM}}$ represent the sample means of the exponentiated terms for always-survivors.} 
The expression for the ratio of means does not collapse in the same way as the difference in means. Due to this distinction between the two contrasts, we expect the small-sample performance of their estimation to differ, and we investigate both in simulation studies in the next section. 

In practice, each MCMC draw of the SACE contrasts (difference and ratio of means) is obtained by averaging the necessary functions of covariates among sampled always-survivors under each treatment as shown above, at the current parameters. After a sufficient burn-in and number of draws, we use the mean of those contrasts to obtain a posterior mean estimate. For credible intervals, we find the highest posterior density (HPD) intervals due to potential asymmetries or skewness in the posterior distribution. We employ the R package \texttt{coda}  \citep{plummer2006coda} for efficient computation of intervals using posterior draws. {R code for the simulation studies and PEPTIC analysis is publicly available at \url{https://github.com/abcdane1/sims_sace_crxo_bayes}. The repository also includes a tutorial with example data and R code illustrating how to fit the proposed model and obtain SACE estimates without access to the PEPTIC trial data.}

\section{Simulation Study} \label{sec:sim}

In our simulation study, we evaluate the small-sample frequentist performance of our Bayesian approach to SACE estimation for a two-period cross-sectional CRXO design. While Bayesian estimation is not intrinsically suited to fixed parameter data generation, we nonetheless consider, for practical recommendations, the empirical bias for estimating the SACE expressed in both the difference in means using the log scale and the ratio of means using the {untransformed} scale. We additionally look at whether HPD credible intervals achieve nominal 95\% coverage. As comparison models to our proposed model \citep{arnup2016appropriate}, we evaluate the effect of removing the cluster-period random effects in the outcome model and/or cluster random effects in the principal strata model with respect to bias, root mean squared error (RMSE), and coverage. For the data generating process described in detail below, we aim to maintain the ratio of number of clusters to the average cluster-period size of the PEPTIC trial ($\approx$ 0.18) whilst varying the cluster/cluster-period correlation values as these will depend on which covariates are chosen for adjustment in practice. 

\subsection{Data Generating Process}
\label{sec:dgp}

We simulate our data using $I=18$ clusters with $J=2$ time periods and generate cluster-period size $M_{ij}$ according to a discrete uniform $\mathcal{U}\{50,150\}$ such that $0.18E(M_{ij})=I$. The true coefficient of variation is $\text{CV}=\sigma_{M_{ij}}/\mu_{M_{ij}} \approx 0.29$.
For each cluster-period, there are three covariates, $\boldsymbol{X}_{ij,1} \sim \text{MVN}(\boldsymbol{0.75},0.94\mathbb{I}_{m_{ij}})$, $\boldsymbol{X}_{ij,2} \sim \text{MVN}(\boldsymbol{0.25},1.32\mathbb{I}_{m_{ij}})$, and $\boldsymbol{X}_{ij,3} \sim \text{MVN}(-\boldsymbol{0.75},1.66\mathbb{I}_{m_{ij}})$, where 
$\mathbb{I}_d$ is the identity matrix with dimensions $d$. We choose these covariates in conjunction with parameter values of the data generating models such that the composition of the principal strata is roughly $(\pi_{(0,0)},\pi_{(1,0)},\pi_{(1,1)}) \approx (0.35,0.25,0.40)$ to reflect relatively high mortality and a positive effect of the active treatment on survival, and to set a SACE ratio of means of $\approx 0.5$ to represent a meaningful treatment effect. The exact values of these target characteristics vary some and are included in the results Section \ref{sec:results}. We first generate the principal stratum membership for each individual according to a multinomial distribution with probabilities defined by applying the {softmax} function (\ref{eq:multi}) to
\begin{align}
\label{eq:dgmps}
\begin{split}
Z_{ijk}&=0.1+0.2X_{ijk,1}-0.4X_{ijk,2}+0.1X_{ijk,3}+0.05I(j=2)+ \eta_{i,z}\\
W_{ijk}&=-0.1-0.4X_{ijk,1}-0.3X_{ijk,2}-0.1X_{ijk,3}+0.025I(j=2)+\eta_{i,w}
\end{split}
\end{align}
where all random effects terms are mutually independent and normally distributed with mean zero and independent of covariates.
We then obtain non-mortality outcomes according to generating conditions for an individual's principal stratum membership (which are none for never-survivors). {For always-survivors, we generate}:
\begin{align}
\begin{split}
\label{eq:dgmout11}
\mathcal{Y}_{ijk}^{(1,1)}(1)&=0.25+ 0.15X_{ijk,1}-0.5X_{ijk,2}+0.7X_{ijk,3}+0.05I(j=2)+\xi_{i,(1,1)}+\gamma_{ij,(1,1)}+\epsilon_{ijk,(1,1)}\\
\mathcal{Y}_{ijk}^{(1,1)}(0)&=0.9+ 0.3X_{ijk,1}-0.15X_{ijk,2}+0.1X_{ijk,3}+0.05I(j=2)+\xi_{i,(1,1)}+\gamma_{ij,(1,1)}+\epsilon_{ijk,(1,1)}
\end{split}
\end{align}
{and for protected-patients, we generate}:
\begin{equation}
\label{eq:dgmout01}
\mathcal{Y}_{ijk}^{(1,0)}(1)=0.2+0.25X_{ijk,1}-0.3X_{ijk,2}+0.15X_{ijk,3}+0.075I(j=2)+\xi_{i,(1,0)}+\gamma_{ij,(1,0)}+\epsilon_{ijk,(1,0)}\end{equation}
where all error and random effects terms are mutually independent and normally distributed with mean zero and are independent of covariates. The variances for $\epsilon_{{ijk},(1,1)}$ and $\epsilon_{{ijk},(1,0)}$ are 1 and 1.25 respectively. Clusters are randomly assigned to one of two sequences of alternating treatments. When a cluster-period has $A_{ij}=a$, the observed data are then $S_{ijk}(a)$ and $Y_{ijk}(a)$ in addition to any covariates. Importantly, under randomization and consistency, our modeling assumptions adhere to this data generation, which notably includes heterogeneous treatment effects in the always-survivor subgroup. To find the true parameters, we generate all (possible) potential outcomes for a large number of clusters, $I=5000$, which we use to obtain stratum membership and, subsequently, the target SACE contrasts.

Variances of all the random effects are determined by values of the sources of within-cluster and cluster-period correlations. For outcome modeling, we explore the effect of varying not only the size but also the proximity of BPC to WPC. Using the error variances, we consider values of BPC of $(0.01,0.03,0.05)$ paired with corresponding values of WPC of $(0.02,0.035,0.1)$ to generate the cluster and cluster-period random effects' variances. Note, these variances will be distinct for strata $(1,1)$ and $(1,0)$ due to different error variances. The BPC, WPC pairing, $(0.03,0.035)$, is chosen specifically to reflect the approximate BPC and WPC values reported in the initial analysis of the PEPTIC trial \citep{young2020effect}. Although we are interested in a different effect, these values provide a helpful benchmark for simulating realistic data in the context of the PEPTIC trial. For the principal strata model, we select ICC values of $(0.02,0.035,0.1)$ using the formula above involving $\pi^2/3$ to obtain the cluster random effects' variances for $\eta_{i,z}$ and $\eta_{i,w}$. 

Further simulation results are included in Appendix \ref{sec:supptables} that explore a) only having 12 clusters, b) two additional compositions of strata $(\pi_{(0,0)},\pi_{(1,0)},\pi_{(1,1)})$: one which has approximately the same proportion of protected-patients but fewer never-survivors, $(0.25, 0.25, 0.5)$, and one with fewer protected-patients and fewer never-survivors, $(0.18, 0.1, 0.72)$, and c) sensitivity of our method to: c.i) additional cluster-period variation in generating the principal strata, {c.ii) non-mortality outcome distribution mis-specification, and c.iii) violations of survival monotonicity (presence of harmed-patients).}

\subsection{Results}
\label{sec:results}

Results from 1,000 simulations for MCMC with 10,000 iterations with 2,500 burn-in iterations are reported in Table \ref{tab:simresults}. In this table, we evaluate our proposed model described in Section \ref{sec:modelspec}, which we denote Model 1, and we compare it to three other reduced models, each fitting within our broader Bayesian framework for SACE estimation. Model 2 drops only the (i) cluster random effects from the principal strata modeling in Model 1. Model 3 drops only the (ii) cluster-period random effects from the outcome modeling in Model 1. Lastly, Model 4 drops random effects (i) and (ii) from Model 1.
 
\begin{table}[t]
    \centering
    \captionsetup{labelfont=bf,justification=raggedright,singlelinecheck=false}
    \caption{Simulation study for a two-period CRXO trial with 18 clusters and {cluster-period sizes drawn} uniformly from 50 to 150. Empirical bias and root mean square error (RMSE) for Bayesian estimation of the SACE expressed in both difference in means on the log-scale and the ratio of means are included. {Empirical coverage of nominal $95\%$ highest posterior density credible intervals is provided.} Side-by-side comparisons of our proposed model (Model 1) to reduced models (Models 2-4) are included. Results are from 1,000 simulated data sets with Bayesian inference on 10,000 MCMC iterations and 2,500 burn-ins. 0.000 is used to indicate less than 0.001}
\begin{tabular}{ccccc|cccc}
    \toprule
    & \multicolumn{4}{c|}{\textbf{SACE Difference in means (Log)}} & \multicolumn{4}{c}{\textbf{SACE Ratio of means}}\\
     & Truth & Bias & RMSE & Coverage & Truth & Bias & RMSE & Coverage\\
     \midrule
    \multicolumn{1}{l}{\textbf{Scenario 1}} & \multicolumn{8}{c}{$\text{BPC}_{\text{Out}}=0.010$ \hspace{.1cm} $\text{WPC}_{\text{Out}}=0.020$ \hspace{.1cm} $\text{ICC}_{\text{PS}}=0.020$}\\
    \midrule 
    Model 1 & -1.180 & 0.014 & 0.096  & 94.8\% & 0.508 & 0.000 & 0.054 & 93.6\% \\
    Model 2 & -1.180 & 0.019 & 0.106 & 93.5\% & 0.508 & 0.000 & 0.054 & 93.7\% \\
    Model 3 & -1.180 & 0.012 & 0.096  & 92.1\% & 0.508 & -0.000 & 0.054 & 91.4\% \\
    Model 4 & -1.180 & 0.016 & 0.105  & 90.9\% & 0.508 & -0.000 & 0.055 & 91.9\% \\
    \midrule
    \multicolumn{1}{l}{\textbf{Scenario 2}} & \multicolumn{8}{c}{$\text{BPC}_{\text{Out}}=0.030$ \hspace{.1cm} $\text{WPC}_{\text{Out}}=0.035$ \hspace{.1cm} $\text{ICC}_{\text{PS}}=0.035$}\\
    \midrule
    Model 1 & -1.182 & 0.009 & 0.089 & 96.0\% & 0.510 & -0.001 & 0.054 & 94.4\%  \\
    Model 2 & -1.182 & 0.014 & 0.101 & 95.1\% & 0.510 & -0.001 & 0.055 & 95.0\% \\
    Model 3 & -1.182 & 0.007 & 0.090 & 92.6\% & 0.510 & -0.002 & 0.054 & 93.0\%\\
    Model 4 & -1.182 & 0.011 & 0.099 & 92.3\% & 0.510 & -0.002 & 0.055 & 92.9\% \\
    \midrule
    \multicolumn{1}{l}{\textbf{Scenario 3}} & \multicolumn{8}{c}{$\text{BPC}_{\text{Out}}=0.050$ \hspace{.1cm} $\text{WPC}_{\text{Out}}=0.100$ \hspace{.1cm} $\text{ICC}_{\text{PS}}=0.100$}\\
    \midrule
    Model 1 & -1.182 & 0.004 & 0.120 & 94.1\% & 0.513 & -0.005 & 0.067 & 92.8\% \\
    Model 2 & -1.182 & 0.018 & 0.133 & 93.1\% & 0.513 & -0.006 & 0.068 & 93.1\% \\
    Model 3 & -1.182 & 0.004 & 0.127 & 78.1\% & 0.513 & -0.008 & 0.069 & 82.5\% \\
    Model 4 & -1.182 & 0.017 & 0.135 & 80.3\% & 0.513 & -0.009 & 0.069 & 83.9\% \\
    \bottomrule
\end{tabular}
\caption*{We use the following abbreviations: BPC-within-cluster between-period correlation, WPC-within-cluster within-period correlation, ICC-intracluster correlation coefficient, Out-outcome, PS-principal strata. Model 1: includes cluster random effects in PS modeling and cluster and cluster-period random effects in outcome modeling. Model 2: removes cluster random effects from Model 1 in PS modeling only. Model 3: removes cluster-period random effects from Model 1 in outcome modeling only. Model 4: removes cluster random effects in PS modeling and cluster-period random effects in outcome modeling from Model 1.}
\label{tab:simresults}
\end{table}

Model 1 has low bias and achieves close to nominal coverage for both the difference and ratio of means SACE estimands across all the scenarios considered (including for supplemental scenarios in {Sections \ref{sec:compsupptables} and \ref{sec:diffsupptables}}). Model 2 also performs well and sometimes demonstrates better coverage, but it displays greater bias and RSME relative to Model 1 for SACE expressed as a difference in means. A comparison of the results for Models 1 and 3 to Models 2 and 4 further indicates that including cluster random effects in the principal strata modeling may mitigate bias for the difference of means SACE estimates. Model 3 and Model 4, which omit cluster-period random effects in outcome modeling, generally lead to noticeable undercoverage. Predictably, they have the worst coverage when the values of BPC and WPC are the most separated and greatest in magnitude. Results showing the principal strata probabilities can be found in Table \ref{tab:simresultsprop}. Across the three scenarios and all models, the magnitudes of bias in estimating the principal strata probabilities are within about $0.005$ for the $(0,0)$ and $(1,1)$ strata and within about $0.007$ for the $(1,0)$ stratum, which has the smallest percentage of individuals.

Overall, there are some performance distinctions between SACE estimation for the difference in means versus the ratio of means. As demonstrated in Section \ref{sec:augment}, the ratio of means is a function of more terms on different scales as compared to the difference in means, which may impact the strength of the prior specification relative to the data in driving posterior inference. For instance, the normal prior on the coefficients, which induces an L2 penalty, may be shrinking the magnitude of coefficients differentially and the effect of shrinkage on bias is scale-dependent. This shrinkage seems to be reflected in the difference in means estimates, which are somewhat attenuated towards zero across all simulation scenarios. Table \ref{tab:simresults} and the supplementary tables suggest Model 1 and/or Model 2 may also give rise to some undercoverage. One possible explanation for this undercoverage is that assigning diffuse inverse gamma priors on random effect variance terms, while allowing for conditional conjugacy, are known to pull small variance estimates towards zero \citep{gelman2013bayesian}. A potential remedy is to employ a different family of priors, such as half-Cauchy priors (although these are only conditionally conjugate through parameter expansion), that can avoid this attenuation.

Further supplementary results demonstrate how Model 1 performs under a wider range of data generating mechanisms. As expected, when we reduce the number of clusters from $I=18$ to $I=12$ (Tables \ref{tab:simresults12} and \ref{tab:simresults12prop}), there is greater variability in our SACE and principal strata estimates with a tendency to show more empirical bias, most noticeably for SACE expressed in difference in means. However, frequentist coverage for the SACE estimands is maintained; this may mean that with fewer clusters, the inverse gamma prior on the random effects variance terms helps regularize estimates as opposed to over-shrinking them. 
When considering all metrics, our proposed model typically outperforms the other models for 12 clusters with patterns mirroring our observations with 18 clusters. Estimation using Model 1 proved robust to the addition of cluster-period random effects in the principal strata model in data generation (Tables \ref{tab:simresultscp} and \ref{tab:simresultscpprop}). As compared to its performance in Table \ref{tab:simresults}, Model 1 generally leads to slightly more bias as well as higher RMSE and lower coverage, which we would anticipate because we do not account for the full hierarchical structure in principal strata modeling. However, Model 1 performs just as well if not better than the ``correctly specified'' fuller model, Model A, which explicitly includes those cluster-period random effects. This observation, together with the results of Table \ref{tab:simresults}, suggests that dropping the cluster-period random effects in the principal strata modeling is less consequential for coverage for the SACE than dropping them in outcome modeling.

When the composition of our strata is $(0.25, 0.25, 0.5)$ as in Tables \ref{tab:simresultsnew} and \ref{tab:simresultsnewprop}, we notice similar patterns to those in Table \ref{tab:simresults}. The performance of the models appear to be more impacted when we reduce the number of never-survivors and protected-patients with composition $(0.18, 0.1, 0.72)$ in Tables \ref{tab:simresultsnewsmP10} and \ref{tab:simresultsnewsmP10prop}. While nominal coverage for the SACE estimands is generally maintained for Model 1, empirical bias increases while variability decreases to small degrees. The undercoverage of Models 3 and 4 is accentuated relative to Models 1 and 2. There is also increased negative bias in estimating the proportion of protected-patients across all models. This bias might be explained by that fact that protected-patients are never directly identified by combinations of observed survival and treatment status (i.e., they only contribute to the likelihood as mixture components), making this stratum's prevalence susceptible to underestimation when its true prevalence is small. Moreover, the MCMC algorithm encountered no errors using Model 1 (or any other model) under the first two principal strata compositions with 25\% protected-patients, but 
it failed at a rate between 2\% to 2.8\% in this last setting (as seen in Table \ref{tab:simresultsnewsmP10}), likely due to the absence of the protected-patients stratum. Although we expect that this issue resolves when there are larger cluster sizes, we see that Model 2, which drops the principal strata cluster random effects, had slightly lower failure rates without demonstrably impacting performance; thus, Model 2 could be a good alternative for data applications if this issue arises. 

{Finally, we assess the sensitivity of our method for estimating the SACE under two important violations of assumptions in simulations. First, given that the log-normal distribution may not be flexible enough to model length-of-stay outcomes, we consider the impact of outcome model mis-specification on performance of our method. To this end, we generate the non-mortality outcome according to a mixture of two (conditional) log-normal distributions with a mixing parameter that decreases from 0.9 to 0.6 by 0.1 \citep{zhang2019lognormal}; when the mixing parameter is 1, outcome generation is as written in (\ref{eq:dgmout11}) and (\ref{eq:dgmout01}). Specifics of this data generating process and further explanation of results (Tables \ref{tab:simresultsmixsame},  \ref{tab:simresultsmixdiff}) are in Section \ref{sec:violations}.  
In general, we find that our method, under this data generating process, is robust to mixing parameters that are 0.9, 0.8, and occasionally 0.7, but performance wanes noticeably when these values are (sometimes) 0.7 and 0.6. Performance varies depending on cluster and cluster-period correlations as well as whether the mixing components have the same or different error variances. Second, we separately consider violations of the monotonicity assumption by generating some target percentage of harmed-patients: 2.5\%, 5\%, and 10\%. We also vary the extent to which the non-mortality outcomes of these harmed-patients differ from always-survivors under treatment 0, since harmed-patients under our method would be falsely classified as those individuals. Details of the data generation  for 2.5\%, 5\%, and 10\% harmed-patients and results (Tables \ref{tab:simresultsharm2p5}, \ref{tab:simresultsharm5}, \ref{tab:simresultsharm10} resp.) are in Section \ref{sec:violations}. 
Overall, we see that our method naturally declines in performance as the percentage of harmed-patients increases. However, at 2.5\% and 5\% harmed-patients, our method still exhibits relatively low bias and decent coverage in small samples across all settings, where bias increases as the harmed-patient outcomes look less like always-survivors under treatment 0. At 10\% harmed-patients, our method generally performs poorly and can even result in fairly frequent failures of the MCMC algorithm altogether (with rates worsening as harmed-patients are more distinct). That said, if an active treatment is suspected to harm at least 10\% of patients, such a treatment may be too toxic or deleterious to warrant use in a clinical trial, let alone justify the monotonicity assumption.} 

\section{Analysis of the PEPTIC trial} \label{sec:app}

The PEPTIC (Proton Pump Inhibitors versus Histamine-2 Receptor Blockers for Ulcer Prophylaxis Treatment in the Intensive Care Unit) trial, a two-period cross-sectional CRXO study, was conducted to compare the effect of two stress ulcer treatments, proton pump inhibitors (PPIs) and histamine-2 receptor blockers (\HtoRB s) for ICU patients undergoing invasive mechanical ventilation on 90-day survival \citep{young2018cluster,young2020effect}.  For patients undergoing mechanical ventilation, hospitalization duration is a key secondary outcome that is not only associated with the severity of illness and convalescence but also has major implications for resource utilization in health care \citep{hill2017long}. While the initial study explored the impacts of these two treatments on hospital LOS, we newly analyze these effects within our proposed SACE framework to allow for a causal principal stratification interpretation. {Descriptive summaries of the PEPTIC trial population and outcomes are provided in the original trial report and its supplementary material \citep{young2020effect}.}

In this trial, $I=50$ clusters were assigned to alternating treatments over $J=2$ periods with each period lasting 6 months; 25 ICUs were randomized to the sequence PPIs then \HtoRB s and the other 25 ICUs were randomized to the sequence \HtoRB s then PPIs. We label PPI as the active treatment, coded as 1, and $\HtoRB$ as the control treatment, coded as 0, in accordance with the original analysis of PEPTIC, which had been informed by a meta-analysis of randomized trials that showed PPIs are possibly better at preventing gastrointestinal (GI) bleeding than \HtoRB s \citep{alhazzani2013proton}. {Because the rationale for using these stress ulcer prophylaxis medications is to prevent GI bleeding and its associated complications \citep{young2018cluster}, this evidence \textit{might} suggest that PPIs are no worse than \HtoRB s with respect to survival, so we did not alter these designations and the implied survival monotonicity assumption. However, this (causal structural) assumption is untestable in truth, and we discuss the potential limitations of making this 
assumption in Section \ref{sec:discuss}.}

\begin{table}[t]
    \centering
    \captionsetup{labelfont=bf,justification=raggedright,singlelinecheck=false}
    
    \caption{Summary of parameter estimates on the PEPTIC data set for time to hospital discharge alive in days. For posterior inference related to SACE, posterior means and $95\%$ highest posterior density credible intervals are reported using the proposed method. Among observed survivors, standardization estimators are provided with 95\% non-parametric cluster bootstrap confidence intervals where the outcome model is a linear mixed-effects model (LMM) with treatment, covariate, period, and covariate by treatment fixed effects and cluster and cluster-period random effects.}
    
    \begin{tabular}{lcc}
        \toprule
        \textbf{Parameter} & \textbf{Point Estimate} & \textbf{95\% Interval} \\
        \midrule
        \multicolumn{3}{c}{\textit{Posterior Inference SACE}}\\
        \addlinespace
        $\mu_{\text{LDiff}}$ & 0.068 & (0.016, 0.118) \\
        $\mu_{\text{ROM}}$ & 1.063 & (1.008, 1.118) \\
        $\pi_{(0,0)}$ & 0.176 & (0.175, 0.177) \\
        $\pi_{(1,0)}$ & 0.063 & (0.060, 0.067) \\
        $\pi_{(1,1)}$ & 0.761 & (0.757, 0.764) \\
        $\sigma^2_{(1,1)}$ & 0.952 & (0.932, 0.973) \\
        $\sigma^2_{\text{C},(1,1)}$ & 0.113 & (0.066, 0.166) \\
        $\sigma^2_{\text{CP},(1,1)}$ & 0.009 & (0.003, 0.016) \\
        \midrule
        \multicolumn{3}{c}{\textit{Observed Survivor Estimates (LMM)}} \\
        \addlinespace
        LDiff & 0.012 & {(-0.022, 0.046)} \\
        ROM & 1.009 & {(0.974, 1.045)}  \\
        \bottomrule
    \end{tabular}
    \label{tab:truedataparm}
    \caption*{$\mu_{h}$ - survivor average causal effect for contrast $h$, LDiff - difference in means on log time scale, ROM - ratio of means on time scale, $\pi_{(\cdot)}$ - proportions of principal strata, $\sigma^2_{(1,1)}$ - error variance among always-survivors, $\sigma^2_{\text{C},(1,1)}$ - cluster random effect variance among always-survivors, $\sigma^2_{\text{CP},(1,1)}$ - cluster-period random effect variance among always-survivors.}
\end{table}

We apply our method to estimate the SACE for the difference in mean log times and ratio of mean times, where time is measured as days to hospital discharge alive. Key parameter estimates are presented in Table \ref{tab:truedataparm}. As proposed, random intercepts for cluster are included in principal strata modeling, and random intercepts for both cluster and cluster-period are included in outcome modeling; both models also adjust for period as a fixed effect. We include baseline covariates for age, sex, and APACHE II score \citep{knaus1985apache}, an aggregate measure for classifying disease severity. In our final analysis dataset, which includes only complete cases (in accordance with the initial study), the average cluster-period size is approximately 269 patients, with a total of 26,673 individuals and an overall mortality rate of 17.8\%. We run 4 chains of 10,000 draws with 2,500 burn-ins under random initial values (including per chain) for a total of 30,000 draws used for inference as outlined in the algorithmic framework provided in Section \ref{sec:pseudoalgorithm}. Trace plots for key parameters are shown in Figure \ref{fig:trace} and demonstrate sufficient mixing. As a comparison to the SACE estimates, we fit a naive linear mixed-effects model (LMM) for the outcome log LOS adjusted for treatment, covariates, treatment by covariate interactions, and period with two levels of random effects among observed survivors. We marginalize over the necessary fixed effects variables in this model to obtain the two point estimates and use the non-parametric cluster bootstrap with 1,000 iterations for confidence intervals \citep{field2007bootstrapping}. We emphasize that all these approaches are estimated as if they involved measurements at a single time, assuming complete outcome measurement upon release from the hospital.
A survival analysis framework for SACE would enable estimating treatment effect changes over time, whilst accounting for censoring mechanisms (if present), but there is no principal strata approach for cluster-randomized time-to-event data that exists to our knowledge, and this direction would necessitate future research. 

For these data, we obtain estimates and 95\% HPD credible intervals for the SACE in difference in means in log-days of 0.068 (0.016, 0.118) and ratio of means in days of 1.063 (1.008, 1.118). These credible intervals (at three decimal places) are slightly above 0 and 1, respectively, suggesting a statistically significant increase in
hospitalization LOS for those assigned to PPIs versus \HtoRB s among always-survivors. However, the effect may be quite small from a clinical point of view. For example, looking at the ratio of means, an expected LOS of one day among individuals assigned to \HtoRB s is estimated to be increased by about 0.063 days or 1.512 hours for those assigned to PPIs. The naive estimates obtained from the LMM among observed survivors are similar in direction and magnitude to the SACE estimates (see Table \ref{tab:truedataparm}), despite being derived from methodologically distinct approaches.  However, the naive approach lacks a causal interpretation. The proximity of these estimates might be explained by the fact that the classified always-survivors are comprised of similar individuals as the observed survivors -- a conjecture reinforced by the relatively small percentage of protected-patients. On the other hand, the minor attenuation in the survivor estimates relative to the SACE could be explained by the slightly greater \textit{observed} mortality among those taking PPIs (18.2\%) than among those taking \HtoRB s (17.3\%) since the deceased individuals, who are presumably the sickest patients, would otherwise have been hospitalized longer had they survived. 

\section{Discussion} \label{sec:discuss}

For estimating the effect of interventions on non-mortality outcomes truncated-by-death in CRXO trials, we extended the principal stratification framework beyond individually-randomized and simple cluster-randomized parallel-arm studies. We targeted SACE estimands among the always-survivor principal stratum where potential non-mortality outcomes under both treatments are completely defined. Under specified causal structural and modeling assumptions, we demonstrated how to estimate SACE using Bayesian methodology, where we augment the posterior with latent principal stratum membership so that conditional posteriors are more tractable for MCMC sampling. With further data augmentation by P\'{o}lya-Gamma variables, we avoided a Metropolis-Hastings step facilitating a complete Gibbs sampler.

In simulation studies, we illustrated that our method generally shows minimal bias and achieves near nominal coverage in studies with a small number of clusters with relatively large numbers of individuals per cluster, mirroring realistic critical care settings. While our proposed approach, which includes cluster random intercepts in principal strata modeling and both cluster and cluster-period random intercepts in non-terminal outcome modeling, performed the best overall relative to simpler comparator models in terms of bias, RMSE, and coverage in aggregate, removing all principal strata random effects did not result in any appreciable worsening of coverage as has been reported in the past \citep{tong2023bayesian,wang2024mixed}. We showed that including cluster-period random effects in the outcome model was necessary to maintain coverage across all data generating processes. Some small decline in the performance of our proposed model appeared as we reduced the number of clusters, but coverage rates remained acceptable. Our model displayed robustness to data generation that includes additional within-cluster within-period correlation in the principal strata modeling, often outperforming the more complex modeling that explicitly accounts for that additional level of clustering. Overall, we found that while it is imperative to account for the full random effects hierarchy in the non-mortality outcome modeling, we can reduce such complexity in the principal strata modeling, dropping say cluster-period or even cluster random effects, to ensure stable estimation in small samples without measurably sacrificing performance.

There are limitations of our method that warrant discussion. For practical implementation of our approach, we employed a parametric mixed-effects modeling framework that falls within standard methodology for analyzing CRXO trials \citep{turner2007analysis,morgan2017choosing}. However, parametric modeling inevitably comes with the risk of model mis-specification, which can result in {flawed} estimation of the SACE. 
For instance, length of stay is notoriously difficult to model parametrically and a log-normal based model may not suffice \citep{faddy2009modeling}.  {Our proposed method demonstrated some robustness to model mis-specification, when the non-mortality outcome (representing LOS) was generated using a mixture of two log-normal models. But, as expected, small-sample performance deteriorated as the mixing parameter approached 0.5, and this form of mis-specification represents just one of numerous possibilities.} A non-parametric Bayesian methodology using Bayesian additive regression trees (BART) has been proposed as a flexible alternative to estimating the SACE for individually-randomized trials \citep{chen2024bayesian}, and we intend to explore a hierarchical extension of this approach for the cluster-randomized setting \citep{teh2010hierarchical}. 
Additionally, since the observed likelihood contains mixture components, we can run into identifiability issues that are widely documented in the finite mixture modeling literature \citep{gelman2013bayesian}. 
In particular, if there is insufficient distance between the mixture components involved in the modeling of the outcome and principal strata, maximum likelihood estimation for the SACE may be biased due to a weakly identified or flat joint distribution even if conditioning on covariates \citep{feller2016principal, jo2022handling}. As noted earlier, priors can help regularize estimates that may otherwise have been unstable under maximum likelihood estimation in this setting. Moreover, from a Bayesian perspective, as long as posterior distributions are proper, Bayesian inference remains valid and will simply reflect greater uncertainty in estimation \citep{imbens1997bayesian}. While we did not see a major impact of weak identification as measured by the performance of our method in simulations or divergent MCMC chains in the data application, there is no guarantee that under different sets of conditions  the proposed SACE estimation remains stable. There are strategies that may mitigate Bayesian instability associated with principal strata mixture modeling that are worth exploring in future research, such as incorporating more informative or different families of priors, allowing for treatment arm-specific variance parameters \citep{jo2022handling}, and using auxiliary variables to help smooth the (mode of) the likelihood and associated posterior \citep{mercatanti2015improving}.

{A further limitation concerns treatment effects that may change over time. Although our outcome model includes period effects, it assumes that the conditional treatment effect on the log-outcome scale is the same across periods for a given covariate profile. Assumption 2 also excludes effects of earlier treatment assignments on patients enrolled in later periods. For example, physicians may become more proficient with an intervention as they gain experience, and its effect may therefore improve for patients treated later. If this improvement depends on experience gained under earlier treatment assignments, the current-period outcome would depend on the treatment history, contrary to Assumption 2. In stepped-wedge trials, a series of recent work \citep{kenny2022analysis,maleyeff2023assessing,wang2024achieve} have shown that ignoring variation in treatment effects by exposure time can lead to biased estimation. Extending the SACE framework to allow treatment effects to vary by period or cluster-level treatment history is an important direction for future work.}

There are also possible paths to {target} the SACE in CRXO trials that avoid fully specifying a conditional joint distribution for non-mortality outcomes and principal strata,
which can be extended from methods in the individually-randomized data context. In general, for {point} identification of principal strata estimands, 
a set of cross-world assumptions that propose a version of ``principal stratification ignorability'' would need to be adapted to tease apart survival potential outcomes across treatments \citep{hayden2005estimator,ding2017principal,jiang2022multiply,zehavi2023match,tong2025semiparametric}. These assumptions have been reframed for the context of parallel-arm CRTs using cluster-level covariates, where only correct specification of a survival status model is required for consistency and valid inference \citep{isenberg2025weighting}. However, general concerns with cross-world assumptions aside, the design of CRXO trials requires obscure formulations of principal ignorability-type assumptions that may be hard to interpret; this alternative type of identification assumptions and results for CRXO designs require further investigation.

In keeping the labeling of treatments (1=PPIs, 0=\HtoRB s) in the initial study \citep{young2020effect}, we maintained the survival monotonicity assumption that each individual's potential survival status under PPIs is no worse than that under \HtoRB s -- an assumption that is untestable and could be challenged in this and other comparative effectiveness studies with two (or more) active interventions. {However, through selected simulations for CRXO trials, we showed that our method for SACE estimation is robust to certain violations of survival monotonicity when the proportion of harmed-patients is modest, such as 2.5\% and even 5\%, consistent with prior observations for estimators achieved through point identification
\citep{ding2017principal,isenberg2025weighting}. While these simulations also indicate that our method is not recommended when there is a substantial proportion of harmed-patients (e.g., 10\%), we expect this occurrence to be rare for clinical trials and in particular, not feasible for PEPTIC (i.e., only a small monotonicity departure is conceivable).}
Moreover, the results of our analysis, which estimated that there are only 6.3\% protected-patients, suggests that at the very least a \textit{strict} survival status inequality may be quite uncommon in our application. 
Therefore, while the survival monotonicity assumption can promote estimation stability,
it may also leave the protected-patients stratum most vulnerable to underestimation since this subgroup only ever appears as a mixture in the observed data likelihood (i.e., unlike the other strata, it is never deterministically assigned from the observed data). {In the critical care setting, there are generally large cluster sizes (and associated cluster-period sizes) so that there should be enough protected-patients to avoid this underestimation or prevent it from being problematic; for PEPTIC, we did not encounter any failed MCMC chains related to the absence of protected-patients. 
Moreover, large clusters should reduce the extent to which posterior inference is driven by the prior specification associated with our modeling assumption that each cluster and (possibly) cluster-period contains protected-patients (and all other strata). 
 If these issues are present}, we saw in simulations that under realistic levels of ICCs, the model that removes cluster random effects from the principal strata model may reasonably be applied instead. 
{Overall, the development of a formal sensitivity analysis framework for the survival monotonicity assumption in CRXO trials is warranted. There are existing sensitivity analysis approaches for this assumption under simpler dependence structures, such as in Ding and Lu (2017) \cite{ding2017principal} and Jiang et al. (2022) \cite{jiang2022multiply}, that may offer a fruitful starting point for the CRXO design.}

In the PEPTIC trial, there were considerable estimated rates of non-compliance, where individuals received the opposite treatment to which they were assigned \citep{young2020effect}. Therefore, we caution that the SACE estimates described in this paper can only be interpreted from an intention-to-treat perspective. A future study possibly incorporating an additional layer of principal stratification \citep{frangakis2002principal,vanderweele2011principal} could be conducted to estimate always-survivor complier effects \citep{kennedy2019survivor}. In addition, while we only targeted SACE at a single point in time for PEPTIC, we are currently tackling the extension to continuous time SACE employing survival analysis techniques. Causal methods for SACE estimation using a semi-competing risks framework have been developed for iid time-to-event data but remain to be expanded to clustered designs \citep{comment2025survivor,xu2022bayesian,nevo2022causal}. Lastly, there is large variability in cluster-period sizes in the observed data ($\text{CV} \approx 0.76$). We assumed, based on the study protocol, that the associated cluster sizes are uninformative. However, if informative cluster size were present, our causal assumptions and corresponding modeling choices may not suffice to capture individual-level treatment effects \citep{kahan2023estimands}. Through this discussion of limitations, we hope to facilitate an informed interpretation of our SACE analysis and promote further research in the context of truncation by death for cluster-randomized trial designs.

\vspace{.5cm}

\noindent{\textbf{Research funding:}} Research in this article was supported by the Patient-Centered Outcomes Research Institute\textsuperscript{\textregistered} (PCORI\textsuperscript{\textregistered} Award ME-2020C1-19220), and the United States National Institutes of Health (NIH), National Heart, Lung, and Blood Institute (grant number R01-HL168202). All statements in this report, including its findings and conclusions, are solely those of the authors and do not necessarily represent the views of the NIH or PCORI\textsuperscript{\textregistered} or its Board of Governors or Methodology Committee.\par

\noindent{\textbf{Data availability:} R code for the simulation studies and PEPTIC analysis, together with a tutorial and example data, is available at \url{https://github.com/abcdane1/sims_sace_crxo_bayes}.}\par

\bibliographystyle{unsrtnat}
\bibliography{Bibliography_revised}

\appendix
\input{SubmissionSuppArxiv_changjun2026_v3}

\end{document}

%% file: SubmissionSuppArxiv_changjun2026_v3.tex
\section{Gibbs Sampler} \label{sec:gibbssample}

\subsection{Prior Specification} \label{sec:priors}
Recall our parameter vector is:
\begin{align*}
    \boldsymbol{\theta}=&(\boldsymbol{\theta}_{(1,1)}^T,\boldsymbol{\theta}_{(1,0)}^T,\sigma_{(1,1)}^2,\sigma_{(1,0)}^2,\boldsymbol{\xi}_{(1,1)}^T,\boldsymbol{\xi}_{(1,0)}^T,\sigma^2_{C,(1,1)},\sigma^2_{C,(1,0)},\boldsymbol{\gamma}_{(1,1)}^T,\boldsymbol{\gamma}_{(1,0)}^T,\sigma^2_{CP,(1,1)},\sigma^2_{CP,(1,0)},
    \\&\boldsymbol{\theta}_z^T,\boldsymbol{\theta}_w^T,\boldsymbol{\eta}_{z}^T,\boldsymbol{\eta}_{w}^T,\tau^2_{C,z},\tau^2_{C,w})^T
\end{align*}  

For the non-mortality outcome modeling $\boldsymbol{\theta}_{(1,1)}$ and $\boldsymbol{\theta}_{(1,0)}$ denote (fixed-effects) regression coefficient vectors (including intercept) corresponding to principal strata $(1,1)$ and $(1,0)$. The cluster random effects vectors are $\boldsymbol{\xi}_{(1,1)}$ and $\boldsymbol{\xi}_{(1,0)}$ with variance terms $\sigma^2_{C,(1,1)}$ and $\sigma^2_{C,(1,0)}$, and the cluster-period random effects are $\boldsymbol{\gamma}_{(1,1)}$ and $\boldsymbol{\gamma}_{(1,0)}$ with variance terms $\sigma^2_{CP,(1,1)}$ and $\sigma^2_{CP,(1,0)}$. The respective error variances are denoted $\sigma^2_{(1,1)}$ and $\sigma^2_{(1,0)}$. For the principal strata modeling, $\boldsymbol{\theta}_{z}$ and $\boldsymbol{\theta}_{w}$ denote regression coefficient vectors (including intercept). The cluster random effects vectors  are $\boldsymbol{\eta}_{z}$ and $\boldsymbol{\eta}_{w}$ with variance terms $\tau^2_{C,z}$ and $\tau^2_{C,w}$. 

The fullest model we consider additionally incorporates cluster-period random intercepts in the principal strata model, so we will be more general and include cluster-period random effect vectors $\boldsymbol{\nu}_z$ and $\boldsymbol{\nu}_w$ with variance terms $\tau^2_{CP,z}$, and $\tau^2_{CP,w}$
$$\boldsymbol{\theta}_{\text{ext}}=(\boldsymbol{\theta}^T,\boldsymbol{\nu}_z^T,\boldsymbol{\nu}_w^T,\tau^2_{CP,z},\tau^2_{CP,w})^T$$ Note, for all simpler models (including our proposed model), random effects parameters are removed accordingly. 

Let MVN and IG stand for multivariate normal and inverse gamma respectively. Let $\mathbb{I}_d$ represent the $d \times d$ identity matrix. The subscript $d$ is suppressed when the dimensions are implied. 

\begin{enumerate}
    \item $\boldsymbol{\theta}_g \sim \text{MVN}(\boldsymbol{\mu}_g, \boldsymbol{\Sigma}_g)$ for $g \in \{(1,1),(1,0)\}$
        \begin{itemize}
            \item Specification (diffuse): $\boldsymbol{\mu}_g=\boldsymbol{0}$, $\boldsymbol{\Sigma}_g=1000\mathbb{I}$
        \end{itemize}
    \item $\sigma_g^2 \sim \text{IG}(a_{g},b_{g})$  for $g \in \{(1,1),(1,0)\}$
    
    \begin{itemize}
            \item Specification (diffuse): $a_g=0.001,b_g=0.001$
        \end{itemize}
    \item $\boldsymbol{\xi}_g|\sigma^2_{C,g} \sim \text{MVN}(\boldsymbol{0},\sigma^2_{C,g}\mathbb{I}_{I})$ for $g \in \{(1,1),(1,0)\}$
    \item $\sigma^2_{C,g} \sim \text{IG}(a_{1,g},b_{1,g})$ for $g \in \{(1,1),(1,0)\}$
        \begin{itemize}
            \item Specification (diffuse): $a_{1,g}=0.001$, $b_{1,g}=0.001$
        \end{itemize}
    \item $\boldsymbol{\gamma}_{(1,1)}|\sigma^2_{CP,(1,1)} \sim \text{MVN}(\boldsymbol{0},\sigma^2_{CP,(1,1)}\mathbb{I}_{IJ})$; $\boldsymbol{\gamma}_{(1,0)}|\sigma^2_{CP,(1,0)} \sim \text{MVN}(\boldsymbol{0},\sigma^2_{CP,{(1,0)}}\mathbb{I}_{IJ/2})$
    \item $\sigma^2_{CP,g} \sim \text{IG}(a_{2,g},b_{2,g})$ for $g \in \{(1,1),(1,0)\}$
         \begin{itemize}
            \item Specification (diffuse): $a_{2,g}=0.001$, $b_{2,g}=0.001$
        \end{itemize}
    \item $\boldsymbol{\theta}_z \sim \text{MVN}(\boldsymbol{\mu}_z,\boldsymbol{\Sigma}_z)$ (note that this will correspond to regression parameters for always-survivors, (1,1), in multinomial logistic model)
         \begin{itemize}
            \item Specification (diffuse): $\boldsymbol{\mu}_z=\boldsymbol{0}$, $\boldsymbol{\Sigma}_z=1000\mathbb{I}$
        \end{itemize}
    
    \item $\boldsymbol{\theta}_w \sim \text{MVN}(\boldsymbol{\mu}_w,\boldsymbol{\Sigma}_w)$ (note that this will correspond to regression parameters for protected-patients, (1,0), in multinomial logistic model)
         \begin{itemize}
            \item Specification (diffuse): $\boldsymbol{\mu}_w=\boldsymbol{0}$, $\boldsymbol{\Sigma}_w=1000\mathbb{I}$
        \end{itemize}
    \item $\boldsymbol{\eta}_z|\tau^2_{C,z} \sim \text{MVN}(\boldsymbol{0},\tau^2_{C,z}\mathbb{I}_{I})$  
    \item $\tau^2_{C,z} \sim \text{IG}(k_{1,z},l_{1,z})$
         \begin{itemize}
            \item Specification (diffuse): $k_{1,z}=0.001$, $l_{1,z}=0.001$
        \end{itemize}
    \item* $\boldsymbol{\nu}_z|\tau^2_{CP,z} \sim \text{MVN}(\boldsymbol{0},\tau^2_{CP,z}\mathbb{I}_{IJ})$
    \item* $\tau^2_{CP,z} \sim \text{IG}(k_{2,z},l_{2,z})$
         \begin{itemize}
            \item Specification (diffuse): $k_{2,z}=0.001$, $l_{2,z}=0.001$
        \end{itemize}
    \item $\boldsymbol{\eta}_w|\tau^2_{C,w} \sim \text{MVN}(\boldsymbol{0},\tau^2_{C,w}\mathbb{I}_{I})$
    \item $\tau^2_{C,w} \sim \text{IG}(k_{1,w},l_{1,w})$
         \begin{itemize}
            \item Specification (diffuse): $k_{1,w}=0.001$, $l_{1,w}=0.001$
        \end{itemize}
    \item* $\boldsymbol{\nu}_w|\tau^2_{CP,w} \sim \text{MVN}(\boldsymbol{0},\tau^2_{CP,w}\mathbb{I}_{IJ})$
    \item* $\tau^2_{CP,w} \sim \text{IG}(k_{2,w},l_{2,w})$
         \begin{itemize}
            \item Specification (diffuse): $k_{2,w}=0.001$, $l_{2,w}=0.001$
        \end{itemize}
\end{enumerate}

* indicates only for model with cluster-period random effects in principal strata modeling, which is referred to as Model A.

\subsection{Conditional Posteriors} \label{sec:condpost}

{Let $\boldsymbol{D}_{y,(1,1)}$ be the submatrix of the complete design matrix associated with fixed effects in outcome modeling (includes intercept, covariates $\boldsymbol{x}_{ijk}$, period, and treatment [and its interactions]) and $\boldsymbol{y}_{(1,1)}$ the corresponding outcomes for individuals in principal stratum $(1,1)$. Let $\boldsymbol{D}_{y,(1,0)}$ be the submatrix of the design matrix and $\boldsymbol{y}_{(1,0)}$ the corresponding outcomes for individuals in principal stratum $(1,0)$ who have survived; since these individuals always have $A_{ij}=1$, columns for treatment terms are dropped in $\boldsymbol{D}_{y,(1,0)}$. Lastly, let unscripted $\boldsymbol{D}$ refer to the submatrix of the design matrix associated with fixed effects in principal strata modeling that excludes the treatment terms. We will use $N$ to denote the number of all individuals (e.g., number of rows for the complete design matrix and for $\boldsymbol{D}$) and the notation $N_{y,g}$ for the number of rows in $\boldsymbol{D}_{y,g}$ for each $g$.}

We represent the clusters numerically, $i=1,\dots,I$, and cluster-period identifiers according to the formula $i+I(j-1)$ for $j=1,\dots,J$. Recall that we will focus on $J=2$. Let $\boldsymbol{P}$ be the $N \times I$ matrix, where each row contains a 1 in the $i$-th column if the individual is in cluster $i$, and 0 otherwise, corresponding to the ordering of individuals in the design matrix. Let $\boldsymbol{P}_{y,g}$ be the $N_{y,g} \times I$ matrix, where each row contains a 1 in the $i$-th column if the individual is in cluster $i$, and 0 otherwise, subset to individuals who would survive under their treatment and are in stratum $g$ and maintaining the ordering of individuals in the respective design submatrices.

Let $\boldsymbol{Q}$ be the $N \times IJ$ matrix, where each row contains a 1 in the $i + I(j-1)$-th column if the individual is in cluster $i$ and period $j$, and 0 otherwise, corresponding to the ordering of individuals in the design matrix. Let $\boldsymbol{Q}_{y,g}$ be the $N_{y,g} \times IJ$ matrix, where each row contains a 1 in the $i + I(j-1)$-th column if the individual is in cluster $i$ and period $j$, and 0 otherwise, subset to individuals who would survive under their treatment and are in stratum $g$ and maintaining the ordering of individuals in the respective design submatrices.

For Gibbs sampling, we must derive the conditional posteriors for the component parameters of $\boldsymbol{\theta}$ and the augmentation parameters \citep{diebolt1994estimation}; denote an arbitrary such parameter by $\boldsymbol{\varphi}$ (vector or scalar). The conditional posteriors are defined by $p(\boldsymbol{\varphi}|\boldsymbol{\varphi_{-}},\boldsymbol{y},\boldsymbol{a},\boldsymbol{s},\boldsymbol{l})$ where $\boldsymbol{\varphi_{-}}$ denotes all remaining {$\boldsymbol{\theta}$ and augmentation parameters} except $\boldsymbol{\varphi}$. For these distributions below, we will use the $\cdots$ notation to indicate conditioning on the remaining unlisted terms so we can highlight any key terms involved in defining the form of conditional posteriors. The full covariate and incidence matrices are fixed, but their stratum-specific row subsets are updated according to the current sampled principal-stratum memberships. 

Before proceeding, we clarify that the multivariate normal and inverse gamma (or inverse-Wishart if non-diagonal variance) prior specification is in fact a conjugate prior for the multivariate log-normal distribution which characterizes our regression models. For illustration, suppose $$\boldsymbol{Y}=(\exp(\mathcal{Y}_1),\dots,\exp(\mathcal{Y}_n))^T$$ where
$$\boldsymbol{\mathcal{Y}}=(\mathcal{Y}_1,\dots,\mathcal{Y}_n)^T \sim \text{MVN}(\boldsymbol{\mu},\boldsymbol{\Sigma})$$ To invert the above transformation, we apply the multivariate function, $\log(\boldsymbol{Y})$, which takes the natural log of each element of the vector such that $\log(\boldsymbol{Y})=\boldsymbol{\mathcal{Y}}$. The determinant of the Jacobian of this element-wise log function is $1/(y_1y_2\cdots y_n)$. Therefore, we have the following relationship between densities, $f_{\boldsymbol{Y}}$ and $f_{\boldsymbol{\mathcal{Y}}}$:
$$f_{\boldsymbol{Y}}(\boldsymbol{y}|\boldsymbol{\mu},\boldsymbol{\Sigma})=\frac{1}{y_1y_2 \cdots y_n}f_{\boldsymbol{\mathcal{Y}}}(\log(\boldsymbol{y})|\boldsymbol{\mu},\boldsymbol{\Sigma}) \propto f_{\boldsymbol{\mathcal{Y}}}(\log(\boldsymbol{y})|\boldsymbol{\mu},\boldsymbol{\Sigma}) = f_{\text{MVN}}(\log(\boldsymbol{y})|\boldsymbol{\mu},\boldsymbol{\Sigma})$$
where the rightmost expression is shorthand for the multivariate normal density with parameters $\boldsymbol{\mu}$ and $\boldsymbol{\Sigma}$ where we have plugged in $\log(\boldsymbol{y})$. Proportionality is with respect to the parameters. 

Henceforth, we will define $\log(\boldsymbol{y})=\boldsymbol{\mathcal{y}}$ ($\log()$ taken element wise). Moreover, we will use the shorthand notation $\text{FN}(\text{arg}|\text{pars})$ to represent a named distribution, FN (e.g., multivariate normal), with respect to variable ``arg'' with parameters ``par''.

\begin{enumerate}
    \item $\boldsymbol{\theta}_{g}|\sigma_g^2,\dots \sim$ for $g \in \{(1,1),(1,0)\}$    
    \begin{align*}
    &\text{MVN}\left(\boldsymbol{\theta}_{g}|\left[\boldsymbol{D}_{y,g}^T\boldsymbol{D}_{y,g}+\sigma^2_g\boldsymbol{\Sigma}_{g}^{-1}\right]^{-1}\left[\boldsymbol{D}
    _{y,g}^T(\boldsymbol{\mathcal{y}}_g-\boldsymbol{P}_{y,g}\boldsymbol{\xi}_g-\boldsymbol{Q}_{y,g}\boldsymbol{\gamma}_g)+\sigma^2_g\boldsymbol{\Sigma}_{g}^{-1}\boldsymbol{\mu}_g\right]\right.,
    \\&\left. \sigma^2_g\left [\boldsymbol{D}_{y,g}^T\boldsymbol{D}_{y,g}+\sigma^2_g\boldsymbol{\Sigma}_{g}^{-1}\right]^{-1}
    \right)
    \end{align*}

    \item $\sigma_g^2|\dots \sim$  for $g \in \{(1,1),(1,0)\}$
    \begin{align*}
    \text{IG} \left(\sigma_g^2|a_g+\frac{N_{y,g}}{2}, b_g + \frac{1}{2}(\boldsymbol{\mathcal{y}}_g-\boldsymbol{D}_{y,g}\boldsymbol{\theta}_g-\boldsymbol{P}_{y,g}\boldsymbol{\xi}_g-\boldsymbol{Q}_{y,g}\boldsymbol{\gamma}_g)^T(\boldsymbol{\mathcal{y}}_g-\boldsymbol{D}_{y,g}\boldsymbol{\theta}_g-\boldsymbol{P}_{y,g}\boldsymbol{\xi}_g-\boldsymbol{Q}_{y,g}\boldsymbol{\gamma}_g)\right)
    \end{align*}
    where recall $N_{y,g}$ is the number of observed survivors in stratum $g$, i.e, those who have a non-mortality outcome defined. 

    \item $\boldsymbol{\xi}_g|\sigma_{C,g}^2,\dots \sim$  for $g \in \{(1,1),(1,0)\}$ 
    \begin{align*}
    &\text{MVN}\left(\boldsymbol{\xi}_g|\left[\sigma_g^2\sigma^{-2}_{C,g}\mathbb{I}_I+\boldsymbol{P}_{y,g}^T\boldsymbol{P}_{y,g}\right]^{-1}\boldsymbol{P}_{y,g}^T\left[\boldsymbol{\mathcal{y}}_g-\boldsymbol{D}_{y,g}\boldsymbol{\theta}_g-\boldsymbol{Q}_{y,g}\boldsymbol{\gamma}_g\right],\left [\sigma^{-2}_{C,g}\mathbb{I}_I+\sigma^{-2}_{g}\boldsymbol{P}_{y,g}^T\boldsymbol{P}_{y,g}\right]^{-1}
    \right)
    \end{align*}

    \item $\sigma^2_{C,g}|\dots \sim$ for $g \in \{(1,1),(1,0)\}$
    \begin{align*}
    &\text{IG} \left (\sigma_{C,g}^2|a_{1,g}+\frac{I}{2}, b_{1,g}+ \frac{1}{2}\boldsymbol{\xi}_g^T\boldsymbol{\xi}_g
    \right) 
    \end{align*} 

    This is appropriate since our random effects prior assumes that always-survivors and protected-patients appear in each cluster. The same will be true in possible cluster-periods as well.
    
    \item $\boldsymbol{\gamma}_{g}|\sigma_{CP,g}^2,\dots \sim$ 
    \begin{align*}
    \text{MVN}&\left(\boldsymbol{\gamma}_{(1,1)}|\left[\sigma_{(1,1)}^2\sigma^{-2}_{CP,(1,1)}\mathbb{I}_{IJ}+\boldsymbol{Q}_{y,(1,1)}^T\boldsymbol{Q}_{y,(1,1)}\right]^{-1}\boldsymbol{Q}_{y,(1,1)}^T\big[\boldsymbol{\mathcal{y}}_{(1,1)}-\boldsymbol{D}_{y,(1,1)}\boldsymbol{\theta}_{(1,1)}\right.
    \\&\left. -\boldsymbol{P}_{y,(1,1)}\boldsymbol{\xi}_{(1,1)}\big],\left [\sigma^{-2}_{CP,(1,1)}\mathbb{I}_{IJ}+\sigma_{(1,1)}^{-2}\boldsymbol{Q}_{y,(1,1)}^T\boldsymbol{Q}_{y,(1,1)}\right]^{-1}
    \right)\\ \text{MVN}&\left(\boldsymbol{\gamma}_{(1,0)}|\left[\sigma_{(1,0)}^2\sigma^{-2}_{CP,(1,0)}\mathbb{I}_{IJ/2}+\boldsymbol{Q}_{y,(1,0),1}^T\boldsymbol{Q}_{y,(1,0),1}\right]^{-1}\boldsymbol{Q}_{y,(1,0),1}^T\big[\boldsymbol{\mathcal{y}}_{(1,0)}-\boldsymbol{D}_{y,(1,0)}\boldsymbol{\theta}_{(1,0)}\right.\\
    &\left. -\boldsymbol{P}_{y,(1,0)}\boldsymbol{\xi}_{(1,0)}\big], \left [\sigma^{-2}_{CP,(1,0)}\mathbb{I}_{IJ/2}+\sigma_{(1,0)}^{-2}\boldsymbol{Q}_{y,(1,0),1}^T\boldsymbol{Q}_{y,(1,0),1}\right]^{-1}
    \right)
    \end{align*}
    where $\boldsymbol{Q}_{y,(1,0),1}$ just denotes the appropriately indexed submatrix of non-zero columns of $\boldsymbol{Q}_{y,(1,0)}$ (those under  treatment 1). We use this notation because in practice, to take care in indexing for the MCMC, we just set an $IJ$ random effect to 0 and only update the $IJ/2$ subvector of non-zero components (which is MVN). 

    \item $\sigma^2_{CP,g}|\dots \sim$
    \begin{align*}
    &\text{IG} \left (\sigma_{CP,(1,1)}^2|a_{2,(1,1)}+\frac{IJ}{2}, b_{2,(1,1)}+ \frac{1}{2}\boldsymbol{\gamma}_{(1,1)}^T\boldsymbol{\gamma}_{(1,1)}
    \right) \\
    &\text{IG} \left (\sigma_{CP,(1,0)}^2|a_{2,(1,0)}+\frac{IJ}{4}, b_{2,(1,0)}+ \frac{1}{2}\boldsymbol{\gamma}_{(1,0)}^T\boldsymbol{\gamma}_{(1,0)}
    \right) 
    \end{align*}


    The conditional posteriors for Steps 7-18 are written below for the multinomial logistic model employing a data augmentation strategy with P\'{o}lya-Gamma random variables following Polson et al. (2013)  \citep{polson2013bayesian} and Allen et al. (2023) \citep{allen2023bayesian}. A non-negative random variable has $\omega \sim \text{PG}(b,c)$ with $b>0$ and $c \in \mathbb{R}$ if 
    $$\omega \stackrel{d}{=} \frac{1}{2\pi^2} \sum_{k=1}^\infty \frac{G_k}{(k-1/2)^2+c^2/(4\pi^2)}$$
    where the $G_k \sim \text{Gamma}(b,1)$ are independent. $\stackrel{d}{=}$ denotes equal in distribution.

Let $\boldsymbol{D}_{ijk}$ denote the row of $\boldsymbol{D}$ corresponding to individual indexed by $ijk$ (includes intercept and all non-treatment regressors that are not random intercepts). Isolating the terms of the conditional posterior, $p(\boldsymbol{\theta}_{\text{PS}}|\boldsymbol{g},\boldsymbol{l})$, that pertain to strata membership, we have 
$$\prod_{ijk}\frac{\left[\exp(\boldsymbol{D}_{ijk}^T\boldsymbol{\theta}_z+\eta_{i,z}+\nu_{ij,z})\right]^{I(G_{ijk}=(1,1))}\left[\exp(\boldsymbol{D}_{ijk}^T\boldsymbol{\theta}_w+\eta_{i,w}+\nu_{ij,w})\right]^{I(G_{ijk}=(1,0))}}{1+\exp(\boldsymbol{D}_{ijk}^T\boldsymbol{\theta}_z+\eta_{i,z}+\nu_{ij,z})+\exp(\boldsymbol{D}_{ijk}^T\boldsymbol{\theta}_w+\eta_{i,w}+\nu_{ij,w})}\times p(\boldsymbol{\theta}_\text{PS})$$

which we'll represent in shorthand as
$$\prod_{ijk}\frac{\left[\exp(z_{ijk})\right]^{I(G_{ijk}=(1,1))}\left[\exp(w_{ijk})\right]^{I(G_{ijk}=(1,0))}}{1+\exp(z_{ijk})+\exp(w_{ijk})} \times p(\boldsymbol{\theta}_\text{PS})$$

For the purposes of determining conditional posterior distributions using data augmentation, we can represent the above in such a way that we isolate the parameters involving the always-survivor, (1,1), component of the principal strata model. We remove $p(\boldsymbol{\theta}_\text{PS})$ temporarily.
\begin{align*}
&\frac{\left[\exp(z_{ijk})\right]^{I(G_{ijk}=(1,1))}\left[\exp(w_{ijk})\right]^{I(G_{ijk}=(1,0))}}{1+\exp(z_{ijk})+\exp(w_{ijk})}\\& \quad =\frac{\left[\frac{\exp(z_{ijk})}{1+\exp(w_{ijk})}\right]^{I(G_{ijk}=(1,1))}\left[\exp(w_{ijk})\right]^{I(G_{ijk}=(1,0))}}{\left(1+\frac{\exp(z_{ijk})}{1+\exp(w_{ijk})}\right)(1+\exp(w_{ijk}))^{I(G_{ijk}\ne(1,1))}}\\
&\quad =  \frac{\left[\frac{\exp(z_{ijk})}{1+\exp(w_{ijk})}\right]^{I(G_{ijk}=(1,1))}}{1+\frac{\exp(z_{ijk})}{1+\exp(w_{ijk})}} \times \frac{\left[\exp(w_{ijk})\right]^{I(G_{ijk}=(1,0))}}{(1+\exp(w_{ijk}))^{I(G_{ijk}\ne(1,1))}}
\end{align*}

Therefore, for drawing from conditional posterior distributions involving parameters of $z_{ijk}$, it suffices to focus on the terms:

$$\frac{\left[\frac{\exp(z_{ijk})}{1+\exp(w_{ijk})}\right]^{I(G_{ijk}=(1,1))}}{1+\frac{\exp(z_{ijk})}{1+\exp(w_{ijk})}}=\frac{\left[\exp(z_{ijk}-\log\{1+\exp(w_{ijk})\})\right]^{I(G_{ijk}=(1,1))}}{1+\exp(z_{ijk}-\log\{1+\exp(w_{ijk})\})}$$
and analogously for $w_{ijk}$
$$\frac{\left[\frac{\exp(w_{ijk})}{1+\exp(z_{ijk})}\right]^{I(G_{ijk}=(1,0))}}{1+\frac{\exp(w_{ijk})}{1+\exp(z_{ijk})}}=\frac{\left[\exp(w_{ijk}-\log\{1+\exp(z_{ijk})\})\right]^{I(G_{ijk}=(1,0))}}{1+\exp(w_{ijk}-\log\{1+\exp(z_{ijk})\})}$$

For brevity, let $U_{ijk,11}=I(G_{ijk}=(1,1))$ and $U_{ijk,10}=I(G_{ijk}=(1,0))$ with observations denoted in lowercase. Each of these terms can be represented as a scale mixture of normals (as kernels) according to a P\'{o}lya-Gamma (PG) distribution as follows:

\begin{align*}
\frac{\left[\exp(z_{ijk}-\log\{1+\exp(w_{ijk})\})\right]^{u_{ijk,11}}}{1+\exp(z_{ijk}-\log\{1+\exp(w_{ijk})\})}&=\bigg\{2^{-1}\exp\left((u_{ijk,11}-1/2)(z_{ijk}-\log\{1+\exp(w_{ijk})\})\right)\\
&\hspace{-1.5cm}\times \int_0^\infty \exp\left(-\omega_{ijk,z}(z_{ijk}-\log\{1+\exp(w_{ijk})\})^2/2\right)p(\omega_{ijk,z})d\omega_{ijk,z}\bigg\}
\end{align*}
and 
\begin{align*}
\frac{\left[\exp(w_{ijk}-\log\{1+\exp(z_{ijk})\})\right]^{u_{ijk,10}}}{1+\exp(w_{ijk}-\log\{1+\exp(z_{ijk})\})}&=\bigg\{2^{-1}\exp\left((u_{ijk,10}-1/2)(w_{ijk}-\log\{1+\exp(z_{ijk})\})\right)\\
&\hspace{-1.5cm}\times \int_0^\infty \exp\left(-\omega_{ijk,w}(w_{ijk}-\log\{1+\exp(z_{ijk})\})^2/2\right)p(\omega_{ijk,w})d\omega_{ijk,w}\bigg\}
\end{align*}
where 
$\omega_{ijk,(\cdot)}$ are iid realizations of $\text{PG}(1,0)$. 

As a mixture of normals according to P\'{o}lya-Gamma distributions, we can use a data augmentation approach to represent the conditional posterior distributions of the relevant parameters by treating the $\boldsymbol{\omega}_z$ and $\boldsymbol{\omega}_w$ values as known in the augmented posterior $p(\boldsymbol{\theta}_{\text{PS}}|\boldsymbol{\omega}_z,\boldsymbol{\omega}_w,\boldsymbol{g},\boldsymbol{l})$.

Then conditional posterior of $\omega_{ijk,z}|u_{ijk,11},\dots$ is given by assuming the $\boldsymbol{\omega}$'s are independent of all other parameters in the prior,
\begin{align}
p(\omega_{ijk,z}|u_{ijk,11},\dots)&=\frac{p(u_{ijk,11}|\omega_{ijk,z},\dots)p(\omega_{ijk,z})}{\int_0^\infty p(u_{ijk,11}|\omega_{ijk,z},\dots)p(\omega_{ijk,z})d\omega_{ijk,z}} \nonumber \\
&=\frac{\exp\left(-\omega_{ijk,z}(z_{ijk}-\log\{1+\exp(w_{ijk})\})^2/2\right)p(\omega_{ijk,z})}{\int_0^\infty \exp\left(-\omega_{ijk,z}(z_{ijk}-\log\{1+\exp(w_{ijk})\})^2/2\right)p(\omega_{ijk,z})d\omega_{ijk,z}} \label{eq:tilt}
\end{align}

As demonstrated in Polson et al. (2013) \citep{polson2013bayesian}, if $\omega \sim \text{PG}(b,c)$, then it's related to the base distribution $\text{PG}(b,0)$ by an exponential tilting
$$p(\omega|b,c)=\frac{\exp\left(-\frac{c^2}{2}\omega\right)p(\omega|b,0)}{\int_0^\infty \exp\left(-\frac{c^2}{2}\omega\right)p(\omega|b,0)d\omega}$$

This is precisely what we have above in (\ref{eq:tilt}) where $\omega_{ijk,z} \sim \text{PG(1,0)}$ and the term $c$ corresponds to $(z_{ijk}-\log\{1+\exp(w_{ijk})\})$.

Thus, the conditional posteriors for the P\'{o}lya-Gamma variables used for augmentation are themselves P\'{o}lya-Gamma, where 

    \item (Augmentation) $\omega_{ijk,z}|\dots \sim$
    $$\text{PG}(\omega_{ijk,z}|1,z_{ijk}-\log\{1+\exp(w_{ijk})\})$$
    
and by the same steps, we would obtain
    \item (Augmentation)
    $\omega_{ijk,w}|\dots \sim$
    $$\text{PG}(\omega_{ijk,w}|1,w_{ijk}-\log\{1+\exp(z_{ijk})\})$$

Because of the mixture of normals representation, the normal priors on all the regression coefficients for $\boldsymbol{\theta}_z$ and $\boldsymbol{\theta}_w$ lead to closed form conditional posteriors. 

Let $\boldsymbol{u}_{11}$, $\boldsymbol{u}_{10}$, $\boldsymbol{\omega}_{z}$, $\boldsymbol{\omega}_{w}$ be the corresponding vector quantities. Let $\boldsymbol{\Omega}_z$ and $\boldsymbol{\Omega}_w$ be diagonal matrices with entries from the corresponding vectors $\boldsymbol{\omega}_z$ and $\boldsymbol{\omega}_w$. Assuming independence of $\boldsymbol{\omega}_z$ and $\boldsymbol{\theta}_z$, 
\begin{align*}
& p(\boldsymbol{\theta}_z|\boldsymbol{\omega}_z,\boldsymbol{u}_{11},\dots) \\
&\propto p(\boldsymbol{u}_{11}|\boldsymbol{\omega}_z,\boldsymbol{\theta}_z,\dots)p(\boldsymbol{\theta}_z)\\
&\propto \prod_{ijk} \bigg [\exp\left((u_{ijk,11}-1/2)(z_{ijk}-\log\{1+\exp(w_{ijk})\})\right)\\
&\quad \times \exp \left(-\omega_{ijk,z}(z_{ijk}-\log\{1+\exp(w_{ijk})\})^2/2\right)\bigg]\exp(-1/2(\boldsymbol{\theta}_z-\boldsymbol{\mu}_z)^T\boldsymbol{\Sigma}_z^{-1}(\boldsymbol{\theta}_z-\boldsymbol{\mu}_z))\\
&=\exp\left\{(\boldsymbol{u}_{11}-\mathbf{1/2})^T(\boldsymbol{D} \boldsymbol{\theta}_z + \boldsymbol{P}\boldsymbol{\eta}_z+\boldsymbol{Q}\boldsymbol{\nu}_z-\log(\mathbf{1}+\exp(\boldsymbol{D} \boldsymbol{\theta}_w + \boldsymbol{P}\boldsymbol{\eta}_w+\boldsymbol{Q}\boldsymbol{\nu}_w)))\right\}\\
&\quad \times \exp \bigg \{-1/2(\boldsymbol{D} \boldsymbol{\theta}_z + \boldsymbol{P}\boldsymbol{\eta}_z+\boldsymbol{Q}\boldsymbol{\nu}_z-\log(\mathbf{1}+\exp(\boldsymbol{D} \boldsymbol{\theta}_w + \boldsymbol{P}\boldsymbol{\eta}_w+\boldsymbol{Q}\boldsymbol{\nu}_w)))^T\boldsymbol{\Omega}_z(\boldsymbol{D} \boldsymbol{\theta}_z + \boldsymbol{P}\boldsymbol{\eta}_z+\boldsymbol{Q}\boldsymbol{\nu}_z
\\&\quad -\log(\mathbf{1}+\exp(\boldsymbol{D} \boldsymbol{\theta}_w + \boldsymbol{P}\boldsymbol{\eta}_w+\boldsymbol{Q}\boldsymbol{\nu}_w)))\bigg \}\exp(-1/2(\boldsymbol{\theta}_z-\boldsymbol{\mu}_z)^T\boldsymbol{\Sigma}_z^{-1}(\boldsymbol{\theta}_z-\boldsymbol{\mu}_z))
\end{align*}

If we isolate the below terms related to $\boldsymbol{\theta}_z$ in the above expression within the exponential
\begin{align*}&-1/2\left[\boldsymbol{\theta}_z^T(\boldsymbol{D}^T\boldsymbol{\Omega}_z\boldsymbol{D}+\boldsymbol{\Sigma}_z^{-1})\boldsymbol{\theta}_z-2\left\{(\log(\mathbf{1}+\exp(\boldsymbol{D} \boldsymbol{\theta}_w + \boldsymbol{P}\boldsymbol{\eta}_w+\boldsymbol{Q}\boldsymbol{\nu}_w))\right. \right.\\
&\left. \left. -(\boldsymbol{P} \boldsymbol{\eta}_z +\boldsymbol{Q}\boldsymbol{\nu}_z))^T\boldsymbol{\Omega}_z\boldsymbol{D}+(\boldsymbol{u}_{11}-\mathbf{1/2})^T\boldsymbol{D}+\boldsymbol{\mu}_z^T\boldsymbol{\Sigma}_z^{-1}\right\}\boldsymbol{\theta}_z\right]
\end{align*}

we can see by completing the square that this is the kernel of a multivariate Gaussian such that 

    \item $\boldsymbol{\theta}_z|\dots \sim$ 
    \begin{align*}
&\text{MVN}\left(\boldsymbol{\theta}_z|\left[\boldsymbol{D}^T\boldsymbol{\Omega}_z\boldsymbol{D}+\boldsymbol{\Sigma}_z^{-1}\right]^{-1} \big[\boldsymbol{D}^T\boldsymbol{\Omega}_z(\log(\mathbf{1}+\exp(\boldsymbol{D} \boldsymbol{\theta}_w + \boldsymbol{P}\boldsymbol{\eta}_w+\boldsymbol{Q}\boldsymbol{\nu}_w))-(\boldsymbol{P} \boldsymbol{\eta}_z +\boldsymbol{Q}\boldsymbol{\nu}_z))\right.\\
    & \left. +\boldsymbol{D}^T(\boldsymbol{u}_{11}-\mathbf{1/2})+\boldsymbol{\Sigma}_z^{-1}\boldsymbol{\mu}_z\big],\left[\boldsymbol{D}^T\boldsymbol{\Omega}_z\boldsymbol{D}+\boldsymbol{\Sigma}_z^{-1}\right]^{-1}
    \right)
    \end{align*}

similarly under independence of $\boldsymbol{\omega}_w$ and $\boldsymbol{\theta}_w$, we have

    \item $\boldsymbol{\theta}_w|\dots \sim$ 
    \begin{align*}
&\text{MVN}\left(\boldsymbol{\theta}_w|\left[\boldsymbol{D}^T\boldsymbol{\Omega}_w\boldsymbol{D}+\boldsymbol{\Sigma}_w^{-1}\right]^{-1} \big[\boldsymbol{D}^T\boldsymbol{\Omega}_w(\log(\mathbf{1}+\exp(\boldsymbol{D} \boldsymbol{\theta}_z + \boldsymbol{P}\boldsymbol{\eta}_z+\boldsymbol{Q}\boldsymbol{\nu}_z))-(\boldsymbol{P} \boldsymbol{\eta}_w+\boldsymbol{Q}\boldsymbol{\nu}_w))\right.\\
    & \left. +\boldsymbol{D}^T(\boldsymbol{u}_{10}-\mathbf{1/2})+\boldsymbol{\Sigma}_w^{-1}\boldsymbol{\mu}_w\big],\left[\boldsymbol{D}^T\boldsymbol{\Omega}_w\boldsymbol{D}+\boldsymbol{\Sigma}_w^{-1}\right]^{-1}
    \right)
    \end{align*}
If we proceed as we did with $\boldsymbol{\theta}_z$ and now isolate the below terms related to $\boldsymbol{\eta}_z$ in the above expression within the exponential (with its corresponding prior), we have
\begin{align*}&-1/2\big[\boldsymbol{\eta}_z^T(\boldsymbol{P}^T\boldsymbol{\Omega}_z\boldsymbol{P}+\tau_{C,z}^{-2}\mathbb{I}_I)\boldsymbol{\eta}_z-2\left\{(\log(\mathbf{1}+\exp(\boldsymbol{D} \boldsymbol{\theta}_w + \boldsymbol{P}\boldsymbol{\eta}_w+\boldsymbol{Q}\boldsymbol{\nu}_w))\right.\\
    &\left. -(\boldsymbol{D} \boldsymbol{\theta}_z +\boldsymbol{Q}\boldsymbol{\nu}_z))^T\boldsymbol{\Omega}_z\boldsymbol{P}+(\boldsymbol{u}_{11}-\mathbf{1/2})^T\boldsymbol{P}\right\}\boldsymbol{\eta}_z\big]
    \end{align*}

    \item $\boldsymbol{\eta}_z|\tau_{C,z}^2\dots \sim$ 
    \begin{align*}
&\text{MVN}\left(\boldsymbol{\eta}_z|\left[\boldsymbol{P}^T\boldsymbol{\Omega}_z\boldsymbol{P}+\tau_{C,z}^{-2}\mathbb{I}_I\right]^{-1} \big[\boldsymbol{P}^T\boldsymbol{\Omega}_z(\log(\mathbf{1}+\exp(\boldsymbol{D} \boldsymbol{\theta}_w + \boldsymbol{P}\boldsymbol{\eta}_w+\boldsymbol{Q}\boldsymbol{\nu}_w))-(\boldsymbol{D} \boldsymbol{\theta}_z +\boldsymbol{Q}\boldsymbol{\nu}_z))\right.\\
    & \left. +\boldsymbol{P}^T(\boldsymbol{u}_{11}-\mathbf{1/2})\big],\left[\boldsymbol{P}^T\boldsymbol{\Omega}_z\boldsymbol{P}+\tau_{C,z}^{-2}\mathbb{I}_I\right]^{-1}\right)
    \end{align*}

    \item  $\tau^2_{C,z}|\dots \sim$ 
    $$\text{IG}\left(\tau^2_{C,z}|k_{1,z}+\frac{I}{2},l_{1,z}+\frac{1}{2}\boldsymbol{\eta}_z^T\boldsymbol{\eta}_z \right)$$

similarly

    \item $\boldsymbol{\eta}_w|\tau_{C,w}^2\dots \sim$ 
    \begin{align*}
&\text{MVN}\left(\boldsymbol{\eta}_w|\left[\boldsymbol{P}^T\boldsymbol{\Omega}_w\boldsymbol{P}+\tau_{C,w}^{-2}\mathbb{I}_I\right]^{-1} \big[\boldsymbol{P}^T\boldsymbol{\Omega}_w(\log(\mathbf{1}+\exp(\boldsymbol{D} \boldsymbol{\theta}_z + \boldsymbol{P}\boldsymbol{\eta}_z+\boldsymbol{Q}\boldsymbol{\nu}_z))-(\boldsymbol{D} \boldsymbol{\theta}_w+\boldsymbol{Q}\boldsymbol{\nu}_w))\right.\\
    & \left. +\boldsymbol{P}^T(\boldsymbol{u}_{10}-\mathbf{1/2})\big],\left[\boldsymbol{P}^T\boldsymbol{\Omega}_w\boldsymbol{P}+\tau_{C,w}^{-2}\mathbb{I}_I\right]^{-1}\right)
    \end{align*}

    \item  $\tau^2_{C,w}|\dots \sim$ 
    $$\text{IG}\left(\tau^2_{C,w}|k_{1,w}+\frac{I}{2},l_{1,w}+\frac{1}{2}\boldsymbol{\eta}_w^T\boldsymbol{\eta}_w \right)$$

and further for cluster-periods 

    \item * $\boldsymbol{\nu}_z|\tau_{CP,z}^2\dots \sim$ 
    \begin{align*}
&\text{MVN}\left(\boldsymbol{\nu}_z|\left[\boldsymbol{Q}^T\boldsymbol{\Omega}_z\boldsymbol{Q}+\tau_{CP,z}^{-2}\mathbb{I}_{IJ}\right]^{-1} \big[\boldsymbol{Q}^T\boldsymbol{\Omega}_z(\log(\mathbf{1}+\exp(\boldsymbol{D} \boldsymbol{\theta}_w + \boldsymbol{P}\boldsymbol{\eta}_w+\boldsymbol{Q}\boldsymbol{\nu}_w))-(\boldsymbol{D} \boldsymbol{\theta}_z +\boldsymbol{P}\boldsymbol{\eta}_z))\right.\\
    & \left. +\boldsymbol{Q}^T(\boldsymbol{u}_{11}-\mathbf{1/2})\big],\left[\boldsymbol{Q}^T\boldsymbol{\Omega}_z\boldsymbol{Q}+\tau_{CP,z}^{-2}\mathbb{I}_{IJ}\right]^{-1}\right)
    \end{align*}

    \item * $\tau^2_{CP,z}|\dots \sim$ 
    $$\text{IG}\left(\tau^2_{CP,z}|k_{2,z}+\frac{IJ}{2},l_{2,z}+\frac{1}{2}\boldsymbol{\nu}_z^T\boldsymbol{\nu}_z \right)$$

    \item * $\boldsymbol{\nu}_w|\tau_{CP,w}^2\dots \sim$ 
    \begin{align*}
&\text{MVN}\left(\boldsymbol{\nu}_w|\left[\boldsymbol{Q}^T\boldsymbol{\Omega}_w\boldsymbol{Q}+\tau_{CP,w}^{-2}\mathbb{I}_{IJ}\right]^{-1} \big[\boldsymbol{Q}^T\boldsymbol{\Omega}_w(\log(\mathbf{1}+\exp(\boldsymbol{D} \boldsymbol{\theta}_z + \boldsymbol{P}\boldsymbol{\eta}_z+\boldsymbol{Q}\boldsymbol{\nu}_z))-(\boldsymbol{D} \boldsymbol{\theta}_w +\boldsymbol{P}\boldsymbol{\eta}_w))\right.\\
    & \left. +\boldsymbol{Q}^T(\boldsymbol{u}_{10}-\mathbf{1/2})\big],\left[\boldsymbol{Q}^T\boldsymbol{\Omega}_w\boldsymbol{Q}+\tau_{CP,w}^{-2}\mathbb{I}_{IJ}\right]^{-1}\right)
    \end{align*}

    \item * $\tau^2_{CP,w}|\dots \sim$ 
    $$\text{IG}\left(\tau^2_{CP,w}|k_{2,w}+\frac{IJ}{2},l_{2,w}+\frac{1}{2}\boldsymbol{\nu}_w^T\boldsymbol{\nu}_w \right)$$

\item (Augmentation) $G_{ijk}|S_{ijk},A_{ij},\dots$ drawn independently due to conditioning. In a slight abuse of notation, where we allow $\sim$ to represent in bijection with a Bernoulli distributed random variable
    $$G_{ijk}|S_{ijk}=1,A_{ij}=1, \dots \sim \text{Bern}(p^1_{ijk})$$
     where 
     $$p^1_{ijk} = \frac{f_{ijk}^{(1,1)}(y_{ijk})\exp(z_{ijk})}{f_{ijk}^{(1,1)}(y_{ijk})\exp(z_{ijk})+f_{ijk}^{(1,0)}(y_{ijk})\exp(w_{ijk})}$$
    $$z_{ijk}=\boldsymbol{D}_{ijk}^T\boldsymbol{\theta}_z+\eta_{i,z}+\nu_{i,z} \hspace{.3cm} w_{ijk}=\boldsymbol{D}_{ijk}^T\boldsymbol{\theta}_w+\eta_{i,w}+\nu_{i,w}$$
    and
    $$f_{ijk}^{(1,1)} = \text{pdf of } \text{Lognormal}(\boldsymbol{D}_{y,ijk,(1,1)}^T \boldsymbol{\theta}_{(1,1)}+\xi_{i,(1,1)}+\gamma_{ij,(1,1)},\sigma_{(1,1)}^2)$$
    $$f_{ijk}^{(1,0)}= \text{pdf of } \text{Lognormal}(\boldsymbol{D}_{y,ijk,(1,0)}^T \boldsymbol{\theta}_{(1,0)}+\xi_{i,(1,0)}+\gamma_{ij,(1,0)},\sigma_{(1,0)}^2)$$
    where $P(G_{ijk}=(1,1)|S_{ijk}=1,A_{ij}=1,\dots)=p^1_{ijk}=1-P(G_{ijk}=(1,0)|S_{ijk}=1,A_{ij}=1,\dots)$ and $\boldsymbol{D}_{ijk,y,g}$ are the covariate values for individual $ijk$ aligned with the parameters of $\boldsymbol{\theta}_g$ (so as to match dimension). Note that these draws are among individuals that survive under treatment $A_{ij}=1$.
    
    Next, we have
    $$G_{ijk}|S_{ijk}=0,A_{ij}=0, \dots \sim \text{Bern}{(p^0_{ijk})}$$
     where 
     $$p^0_{ijk} = \frac{\exp(w_{ijk})}{\exp(w_{ijk})+1}$$
    where $P(G_{ijk}=(1,0)|S_{ijk}=0,A_{ij}=0,\dots)=p^0_{ijk}=1-P(G_{ijk}=(0,0)|S_{ijk}=0,A_{ij}=0,\dots)$. Note these draws are among individuals that do not survive under treatment $A_{ij}=0$.
    Lastly due to survival monotonicity, we have
    $$P(G_{ijk}=(1,1)|S_{ijk}=1,A_{ij}=0,\dots)=1$$
    and
    $$P(G_{ijk}=(0,0)|S_{ijk}=0,A_{ij}=1,\dots)=1$$
    so these strata are drawn deterministically.
\end{enumerate}

Important note: the above results for the conditional posterior distributions assume that the treatment sequence is completely followed by every cluster (and that the design is strictly alternating). This adherence may not be the case; in the PEPTIC data set, there is one cluster that either started late in its sequence or dropped out. If one or more clusters were not assigned both treatments, we simply adjust dimensions corresponding to cluster-periods and/or clusters in all relevant conditional posteriors to reflect this, assuming that missingness is completely at random. 

\subsection{MCMC Algorithmic Framework}\label{sec:pseudoalgorithm}

The algorithmic framework for drawing from the posterior distribution using the Gibbs sampler is written out below. 

\begin{enumerate}
    \item Initialization of necessary components of $\boldsymbol{\theta}_{\text{ext}}$: Set all random effect vectors at the zero vector. Draw random effects variance and error variances randomly from an appropriately centered and ranged distribution. One way to accomplish this:
    \begin{itemize}
    \item Fit mixed-effects adjusted models on the binary survival status (logit link) and observed outcome variables (identity link with logged outcome) with explanatory variables as per the principal strata and outcome modeling respectively but without any interactions using standard software, e.g., \texttt{lme4} in \texttt{R}.
    \item Obtain REML/MLE random effect variance estimates for both models as well as error variances for the outcome model.
    \item Use estimates from the survival-status model as benchmarks for the principal-strata variance parameters and estimates from the observed-outcome model as benchmarks for the outcome-model variance parameters. For example, for the cluster random effects variance of both the always-survivor and protected-patients outcome models, one would use the cluster random effect variance in observed survivor outcome models.
    \item Initialize the variances by drawing from a uniform distribution with some band determined by lower bound term $0<\epsilon_{pl}<1$ and upper bound term $\epsilon_{pu}>0$. For an arbitrary $\sigma^2_{p}$ in $\boldsymbol{\theta}$, draw initials $\sigma^2_{p0} \sim \mathcal{U}\left((1-\epsilon_{pl}) \hat \sigma^2_p,(1+\epsilon_{pu})\hat \sigma^2_p\right)$, where $\hat \sigma^2_p$ is the corresponding estimate. We set $\epsilon_{pl}=\epsilon_{pu}=0.5$ for all variances in the PEPTIC data. But different size (and parameter-specific) bands can be used depending on the amount of data.
    \end{itemize}
    \item Initialization of augmentation term $\boldsymbol{g}$ by individual:
    \[G_{ijk} =
\begin{cases} 
(1,1) & \text{if } S_{ijk} = 1 \text{ and } A_{ij} = 0 \\
(0,0) & \text{if } S_{ijk} = 0 \text{ and } A_{ij} = 1 \\
\text{Ber}(0.5) \in \{(1,1),(1,0)\} & \text{if } S_{ijk} = 1 \text{ and } A_{ij} = 1\\
\text{Ber}(0.5) \in \{(1,0),(0,0)\}  & \text{if } S_{ijk} = 0 \text{ and } A_{ij} = 0
\end{cases}
\]
where $\text{Ber}(0.5) \in \{g_1,g_2\}$ is a slight abuse of notation meaning that $g_1$ and thus $g_2$ are sampled with probability 1/2. 

Repeat Steps 3--10 for $K$ iterations, where $K$ includes burn-in. Discard the first $B$ iterations and retain iterations $B+1,\ldots,K$ for inference; if thinning is used, retain every $T$-th iteration after burn-in. All updates refer to draws from conditional posterior distributions conditioned on current draws (i.e., within the iteration or from the last iteration) of all requisite parameters. 

    \begin{itemize}
    
    \item[3.] Non-mortality regression and SACE parameters: Update non-mortality regression parameter $\boldsymbol{\theta}_{(1,1)}$ from multivariate normal conditional posterior. Update SACE in difference in means (log-time scale), LDiff, and ratio of means (time scale), ROM, which are contrasts of averages of functions among current always-survivors (as in (\ref{eq:uldiff}) and (\ref{eq:urom})). Update error variance $\sigma^2_{(1,1)}$ from inverse gamma distribution. Update $\boldsymbol{\theta}_{(1,0)}$ from multivariate normal distribution. Update error variance $\sigma^2_{(1,0)}$ from inverse gamma distribution.

    \item[4.] Non-mortality cluster random effects and variances: Update cluster random effect vector $\boldsymbol{\xi}_{(1,1)}$ from multivariate normal distribution and variance term $\sigma^2_{C,(1,1)}$ from inverse gamma distribution. Do so analogously for $\boldsymbol{\xi}_{(1,0)}$ and $\sigma^2_{C,(1,0)}$.

    \item[5.] Non-mortality cluster-period random effects and variances (Models 1,2,A only): Update cluster-period random effect vector $\boldsymbol{\gamma}_{(1,1)}$ from multivariate normal distribution and variance term $\sigma^2_{CP,(1,1)}$ from inverse gamma distribution. Do so analogously for $\boldsymbol{\gamma}_{(1,0)}$ and $\sigma^2_{CP,(1,0)}$.

    \item[6.] P\'{o}lya-Gamma augmentation terms: Update variables $\boldsymbol{\omega}_z$ and $\boldsymbol{\omega}_w$ by drawing $\omega_{ijk,z}$ and $\omega_{ijk,w}$ from their respective P\'{o}lya-Gamma conditional posterior distributions for all individuals. 

    \item[7.] Principal strata regression parameters: Update non-mortality regression parameters $\boldsymbol{\theta}_{z}$ and $\boldsymbol{\theta}_{w}$ from multivariate normal conditional posteriors. 

    \item[8.]  Principal strata cluster random effects and variances (Models 1,3,A only): Update cluster random effect vector $\boldsymbol{\eta}_{z}$ from multivariate normal distribution and variance term $\tau^2_{C,z}$ from inverse gamma distribution. Do so analogously for $\boldsymbol{\eta}_{w}$ and $\tau^2_{C,w}$.

    \item[9.]  Principal strata cluster-period random effects and variances (Model A only): Update cluster-period random effect vector $\boldsymbol{\nu}_z$ from multivariate normal distribution and variance term $\tau^2_{CP,z}$ from inverse gamma distribution. Do so analogously for $\boldsymbol{\nu}_w$ and $\tau^2_{CP,w}$.

    \item[10.] Principal strata membership: Update principal strata membership, $G_{ijk}$, for each individual according to either a Bernoulli or deterministic conditional posterior distribution.

    \end{itemize}
\end{enumerate}

{\section{Achieving SACE Estimation}

\label{sec:achSACE}

For succinctness in this section, we will use the abbreviations A1 through A6 to refer to Assumption 1 through Assumption 6 as presented in Section \ref{sec:notation2}.

The individual-level SACE is defined as a contrast of the terms
$$E\{Y_{ijk}(a)|G_{ijk}=(1,1)\}=E\{Y_{ijk}(a)|S_{ijk}(1)=1,S_{ijk}(0)=1\}$$
for $a=1,0$. 
At the start, we see that we need target the terms $S_{ijk}(a)$ and $Y_{ijk}(a)$ (or its transformed value) using CRXO data. The purpose of A1 and A2 is to allow for us to work with these terms. 
But, we still are left with a problem since the SACE relies on:
$$p(y_{ijk}(a)|S_{ijk}(1)=1,S_{ijk}(0)=1)$$
but $S_{ijk}(1)=1$ and $S_{ijk}(0)=1$ cannot be simultaneously identified under treatment $A_{ij}$ of the assigned treatment sequence. As such, A1-A5 are not sufficient for point identification. Including survival monotonicity, A6, says that when $A_{ij}=0$ 
$$p(y_{ijk}(0)|S_{ijk}(1)=1,S_{ijk}(0)=1)\stackrel{\text{A6}}{=}p(y_{ijk}(0)|S_{ijk}(0)=1)=p(y_{ijk}|A_{ij}=0,S_{ijk}=1)$$
under randomization and consistency. However, this simplification is not possible when treatment $A_{ij}=1$, which is a consequence of the fact that observed survivors under treatment 1 could be either always-survivors or protected-patients. The SACE is only partially or set identified, with bounds proposed in the iid setting by Zhang and Rubin (2003) \citep{zhang2003estimation}.

Let's consider then how we might leverage covariates to get at point identification. $\boldsymbol{X}_{ijk}$ is a notational choice to denote an individual's covariates. Note, for brevity of notation in this section, $\boldsymbol{X}_{ijk}$ will refer to all covariates. Then,
\begin{align*}
&E\{Y_{ijk}(a)|S_{ijk}(1)=1,S_{ijk}(0)=1\}\\
&\quad =E\left\{E(Y_{ijk}(a)|S_{ijk}(1)=1,S_{ijk}(0)=1,\boldsymbol{X}_{ijk})|S_{ijk}(1)=1,S_{ijk}(0)=1\right\} \\
& \quad = \frac{E_{\boldsymbol{X}{ijk}}\left\{E(Y_{ijk}(a)|S_{ijk}(1)=1,S_{ijk}(0)=1,\boldsymbol{X}_{ijk})\pi(S_{ijk}(1)=1,S_{ijk}(0)=1|\boldsymbol{X}_{ijk})\right\}}{E_{\boldsymbol{X}_{ijk}}\left\{\pi(S_{ijk}(1)=1,S_{ijk}(0)=1|\boldsymbol{X}_{ijk})\right\}}
\end{align*}
Here we see that one can make cross-world assumptions (including or not including survival monotonicity) that break away counterfactual $S_{ijk}(1-a)$ from $Y_{ijk}(a)$ and $S_{ijk}(a)$ that leverages covariates to identify the SACE \citep{hayden2005estimator,zehavi2023match,isenberg2025weighting}. Further assumptions would then be required for using the data arising from a CRXO trial to estimate this individual-level SACE and ensure proper inference. We ultimately felt that this approach, in totality, may be rather confusing to represent coherently in the setting when we have cluster-periods. 

Alternatively and more directly, one could make modeling assumptions that specify individual-level distributions
$$S_{ijk}(1),S_{ijk}(0)|\boldsymbol{X}_{ijk}\hspace{.2cm} \text{and} \hspace{.2cm} Y_{ijk}(a)|S_{ijk}(1),S_{ijk}(0),\boldsymbol{X}_{ijk}$$
which share common (marginalized) expectations $E(Y_{ijk}(a)|S_{ijk}(1)=1,S_{ijk}(0)=1)$ for $a \in \{0,1\}$, as per the super-population definition of SACE. $Y_{ijk}(a)|S_{ijk}(1),S_{ijk}(0),\boldsymbol{X}_{ijk}$ is expressible as $Y_{ijk}|A_{ij}=a,S_{ijk}(1),S_{ijk}(0),\boldsymbol{X}_{ijk}$ under A3 and A5 (with truncated values under death) such that one could use latent variable techniques to make inferences about SACE. This motivates our Bayesian \textit{parametric} modeling approach for which we define the conditional joint distribution of potential non-mortality and survival status outcomes given covariates, as well as specify priors. Critically, we make parametric assumptions, via parameter $\boldsymbol{\theta}$, that enable us to target the individual-level SACE using CRXO data. Also, in keeping with previous SACE literature and CRXO methods, we opt for interpretable mixed-effects modeling.  

We outline this approach here in a similar exposition as Imbens and Rubin (1997) \cite{imbens1997bayesian} and Hirano et al. (2000) \cite{hirano2000assessing}.  
Recall $\boldsymbol{a}^*_i=(1,0,\dots)^T$. A1 and A2 are such that $\boldsymbol{S}^{\text{full}}=(\boldsymbol{S}_1(\boldsymbol{a}^*_i)^T,\boldsymbol{S}_1(\boldsymbol{1}-\boldsymbol{a}^*_i)^T,\dots, \boldsymbol{S}_I(\boldsymbol{a}^*_i)^T,\boldsymbol{S}_I(\boldsymbol{1}-\boldsymbol{a}^*_i)^T)^T$ is comprised of terms $S_{ijk}(1)$ and $S_{ijk}(0)$. Similarly, \\$\boldsymbol{Y}^{\text{full}}=(\boldsymbol{Y}_1(\boldsymbol{a}^*_i)^T,\boldsymbol{Y}_1(\boldsymbol{1}-\boldsymbol{a}^*_i)^T,\dots, \boldsymbol{Y}_I(\boldsymbol{a}^*_i)^T,\boldsymbol{Y}_I(\boldsymbol{1}-\boldsymbol{a}^*_i)^T)^T$ contains terms $Y_{ijk}(1)$ and $Y_{ijk}(0)$. Note, $Y_{ijk}(a)=*$ deterministically, under truncation $S_{ijk}(a)=0$, and thus does not impact likelihood (i.e., contributes 1). $\boldsymbol{S}^{\text{obs}},\boldsymbol{S}^{\text{mis}}$ and $\boldsymbol{Y}^{\text{obs}},\boldsymbol{Y}^{\text{mis}}$ are the observed and missing components of $\boldsymbol{S}^{\text{full}}$ and $\boldsymbol{Y}^{\text{full}}$, respectively. Under A3, the observed values are the components $S_{ijk}(a)=S_{ijk}^{\text{obs}}$ and $Y_{ijk}(a)=Y_{ijk}^{\text{obs}}$ for the sequence $\boldsymbol{A}_i$ where $A_{ij}=a$, and thus the missing values are the potential outcomes at $1-a$. We assume that there is a parameter $\boldsymbol{\theta}$ (i.e., parametric modeling) for which we consider a posterior 
$$p(\boldsymbol{\theta}|\boldsymbol{y}^{\text{obs}},\boldsymbol{s}^{\text{obs}},\boldsymbol{a},\boldsymbol{l}) \propto p(\boldsymbol{y}^{\text{obs}},\boldsymbol{s}^{\text{obs}}|\boldsymbol{a},\boldsymbol{l},\boldsymbol{\theta})p(\boldsymbol{\theta})$$
For illustration of this approach, it suffices to focus on the likelihood: 
\begingroup
\begin{align*}
p(\boldsymbol{y}^{\text{obs}},\boldsymbol{s}^{\text{obs}}|\boldsymbol{a},\boldsymbol{l},\boldsymbol{\theta}) &= \sum_{\boldsymbol{s}^{\text{mis}}}\int p(\boldsymbol{y}^{\text{full}},\boldsymbol{s}^{\text{full}}|\boldsymbol{a},\boldsymbol{l},\boldsymbol{\theta})d\boldsymbol{y}^{\text{mis}}\\
             & = \sum_{\boldsymbol{s}^{\text{mis}}}\int \prod_{i=1}^{I}p_i(\boldsymbol{y}^{\text{full}}_i,\boldsymbol{s}^{\text{full}}_i|\boldsymbol{l}_i,\boldsymbol{\theta})d\boldsymbol{y}^{\text{mis}} \hspace{.3cm} (\text{A4, A5)} \\
            & = \sum_{\boldsymbol{s}^{\text{mis}}}\int \prod_{ijk}f(\boldsymbol{y}^{\text{full}}_{ijk}|\boldsymbol{s}^{\text{full}}_{ijk},\boldsymbol{x}_{ijk},\boldsymbol{\theta}_\text{Out},\boldsymbol{\theta}_{\text{Assoc}})\pi(\boldsymbol{s}^{\text{full}}_{ijk}|\boldsymbol{x}_{ijk},\boldsymbol{\theta}_{\text{PS}})d\boldsymbol{y}^{\text{mis}}\hspace{.3cm} \text{(key P.A.)}
\end{align*}
\endgroup
P.A. is shorthand for parametric assumptions. In adherence with the above, $\boldsymbol{S}^{\text{full}}_{ijk}$ and  $\boldsymbol{Y}^{\text{full}}_{ijk}$ refer to individual's $ijk$ missing, $S^{\text{mis}}_{ijk}$, $Y^{\text{mis}}_{ijk}$, and observed, $S^{\text{obs}}_{ijk}$, $Y^{\text{obs}}_{ijk}$, components. For example, for survival status,``full'' would refer to both terms $S_{ijk}(1),S_{ijk}(0)$ under our assumptions. 
A key modeling assumption is conditional independence across individuals given the relevant cluster and cluster-period random effects, the covariates, and the remaining model parameters. The individual conditional distributions need not be identical, and survival status and non-mortality outcomes are modeled jointly through principal-stratum membership. Additionally, we assume that $\boldsymbol{\theta}=(\boldsymbol{\theta}_{\text{Out}}^T,\boldsymbol{\theta}_{\text{Assoc}}^T,\boldsymbol{\theta}_{\text{PS}}^T)^T$ such that we can have non-mortality outcomes given stratum membership and stratum membership components with separate parameters. $\boldsymbol{\theta}_{\text{Assoc}}$ contains the association parameters between the potential non-mortality outcomes for each individual.  We now continue from above, which equals
\begingroup
\ninept
\begin{align*}
&=\sum_{\boldsymbol{s}^{\text{mis}}}\prod_{ijk}\int f(\boldsymbol{y}^{\text{full}}_{ijk}|\boldsymbol{s}^{\text{full}}_{ijk},\boldsymbol{x}_{ijk},\boldsymbol{\theta}_{\text{Out}},\boldsymbol{\theta}_{\text{Assoc}})\pi(\boldsymbol{s}^{\text{full}}_{ijk}|\boldsymbol{x}_{ijk},\boldsymbol{\theta}_{\text{PS}})dy_{ijk}^{\text{mis}}\\
            &  =\prod_{ijk}\sum_{s_{ijk}^{\text{mis}}} f(y_{ijk}(a_{ij})|\boldsymbol{a}_i,\boldsymbol{s}^{\text{full}}_{ijk},\boldsymbol{x}_{ijk},\boldsymbol{\theta}_{\text{Out}})\pi(\boldsymbol{s}^{\text{full}}_{ijk}|\boldsymbol{x}_{ijk},\boldsymbol{\theta}_{\text{PS}}) \hspace{.3cm} (\text{A3, A5})\\
            & = \prod_{ijk}\bigg\{ f(y^{\text{obs}}_{ijk}|a_{ij},S_{ijk}(a_{ij})=s_{ijk}^{\text{obs}},S_{ijk}(1-a_{ij})=1,\boldsymbol{x}_{ijk},\boldsymbol{\theta}_{\text{Out}})\pi(S_{ijk}(a_{ij})=s_{ijk}^{\text{obs}},S_{ijk}(1-a_{ij})=1|\boldsymbol{x}_{ijk},\boldsymbol{\theta}_{\text{PS}})\\
            & + f(y^{\text{obs}}_{ijk}|a_{ij},S_{ijk}(a_{ij})=s_{ijk}^{\text{obs}},S_{ijk}(1-a_{ij})=0,\boldsymbol{x}_{ijk},\boldsymbol{\theta}_{\text{Out}})\pi(S_{ijk}(a_{ij})=s_{ijk}^{\text{obs}},S_{ijk}(1-a_{ij})=0|\boldsymbol{x}_{ijk},\boldsymbol{\theta}_{\text{PS}})\bigg\} \hspace{.3cm} (\text{A3})
\end{align*}
\endgroup
At this point, we see that there are four possible two-component mixture contributions. 
If under $A_{ij}=1$, $S_{ijk}(1)=S_{ijk}^{\text{obs}}=1$, we would have
\begin{align*}
&f(y_{ijk}|A_{ij}=1,S_{ijk}(1)=1,S_{ijk}(0)=1,\boldsymbol{x}_{ijk},\boldsymbol{\theta}_{\text{Out}})\pi(S_{ijk}(1)=1,S_{ijk}(0)=1|\boldsymbol{x}_{ijk},\boldsymbol{\theta}_{\text{PS}})\\
& + f(y_{ijk}|A_{ij}=1,S_{ijk}(1)=1,S_{ijk}(0)=0,\boldsymbol{x}_{ijk},\boldsymbol{\theta}_{\text{Out}})\pi(S_{ijk}(1)=1,S_{ijk}(0)=0|\boldsymbol{x}_{ijk},\boldsymbol{\theta}_{\text{PS}})
\end{align*}
where an individual could be an always-survivor or a protected-patient. Similarly, if under $A_{ij}=1$,  $S_{ijk}(1)=S_{ijk}^{\text{obs}}=0$, we would have \begin{align*}
&\pi(S_{ijk}(1)=0,S_{ijk}(0)=1|\boldsymbol{x}_{ijk},\boldsymbol{\theta}_{\text{PS}})+ \pi(S_{ijk}(1)=0,S_{ijk}(0)=0|\boldsymbol{x}_{ijk},\boldsymbol{\theta}_{\text{PS}})
\end{align*}
where an individual could be a harmed-patient or a never-survivor. $f$ is suppressed for clarity here since outcomes are truncated (deterministically *). Under A6, survival monotonicity, the harmed-patient stratum is removed, reducing the previous expression to 
$$\pi(S_{ijk}(1)=0,S_{ijk}(0)=0|\boldsymbol{x}_{ijk},\boldsymbol{\theta}_{\text{PS}})$$

The remaining two components under treatment 0 follow similarly. When $A_{ij}=0$ with $S_{ijk}(0)=S_{ijk}^{\text{obs}}=1$, there's a mixture for always-survivors and harmed-patients (which is removed under A6) and for $A_{ij}=0$ with $S_{ijk}(0)=S_{ijk}^{\text{obs}}=0$, there's a mixture for protected-patients and never-survivors.

Importantly, we see that the likelihood does not depend on association parameters, $\boldsymbol{\theta}_{\text{Assoc}}$, between the terms $Y_{ijk}(1)$ and $Y_{ijk}(0)$. Therefore, estimands, such as the superpopulation SACE, that do not involve these association parameters (since the two means comprising the SACE only involve $Y_{ijk}(a)$ individually) are estimable once parametric assumptions and further prior specifications are made. To avoid confusion, we have excluded $\boldsymbol{\theta}_{\text{Assoc}}$ from $\boldsymbol{\theta}$ in the main paper. 

Our final parametric assumption is that we assume that the data generating mechanisms for the component terms  $G_{ijk}|\boldsymbol{X}_{ijk},\boldsymbol{\theta}_{\text{PS}}$ takes the form of (\ref{eq:dgmps}) and $Y_{ijk}(a)|G_{ijk},\boldsymbol{X}_{ijk},\boldsymbol{\theta}_{\text{Out}}$ (for survivors) takes the form of (\ref{eq:dgmout11}) and (\ref{eq:dgmout01}), with general parameter values. Under our causal structural assumptions, the latter can be expressed compatibly in terms of observed non-mortality outcomes $Y_{ijk}|A_{ij}=a,G_{ijk},\boldsymbol{X}_{ijk},\boldsymbol{\theta}_{\text{Out}}$ as in (\ref{eq:outmod}), LMMs with logged outcomes. Proper priors were specified according to Section \ref{sec:priors}. We keep outcome modeling in terms of observed outcomes in our paper as this is what is implemented in practice, from which the underlying parametric data generating mechanism is implied. 

Putting it all together, we see under certain parametric assumptions alongside A1-A6, the SACE can be represented as a function of parameters $\boldsymbol{\theta}_{\text{Out}}$ (per principal stratum membership) contained in a mixture model with observable data. Therefore, these assumptions allow us to construct a proper posterior distribution for performing Bayesian inference on the SACE, where we use data augmentation \citep{tanner1987calculation} to handle the latent principal strata. While we present a fully parametric approach, more flexible modeling assumptions have been considered in the iid setting \citep{chen2024bayesian} that warrant future research for application to CRXO trials.}

\section{Supplemental Tables and Figures} \label{sec:supptables}

\renewcommand{\thetable}{A.\arabic{table}}
\setcounter{table}{0}

\subsection{Companion Tables to Table 1} \label{sec:compsupptables}

\begin{enumerate}
    \item[A1] Bias and RMSE of principal strata proportions as per main data generating mechanism (Table~\ref{tab:simresultsprop})
    \item[A2] SACE estimation as per main data generating mechanism but with cluster-period random effects for the principal strata. Compares proposed model (Model 1) with model (Model A) that explicitly includes cluster-period random effects in principal strata modeling  (Table~\ref{tab:simresultscp})
        \begin{itemize}
            \item[A3] Bias and RMSE of principal strata proportions as per A2 data generating mechanism (Table~\ref{tab:simresultscpprop})
        \end{itemize}
    \item[A4] SACE estimation as per main data generating mechanism but with $I=12$ clusters (Table~\ref{tab:simresults12})
    \begin{itemize}
        \item[A5] Bias and RMSE of principal strata proportions as per A4 data generating mechanism (Table~\ref{tab:simresults12prop})
    \end{itemize}
\end{enumerate}

\begin{table}[ht]
    \centering
    \captionsetup{labelfont=bf,justification=raggedright,singlelinecheck=false}
    \caption{Simulation study for a two-period CRXO trial with 18 clusters and {cluster-period sizes drawn} uniformly from 50 to 150. Empirical bias and root mean square error (RMSE) for Bayesian estimation of the principal strata proportions are presented. Side-by-side comparisons of our proposed model (Model 1) to reduced models (Models 2-4) are included. Results are from 1,000 simulated data sets with Bayesian inference on 10,000 MCMC iterations and 2,500 burn-ins. 0.000 is used to indicate less than 0.001}
\begin{tabular}{cccc|ccc|ccc}
    \toprule
     Model & $\pi_{(0,0)}$ & Bias & RMSE & $\pi_{(1,0)}$ & Bias & RMSE & $\pi_{(1,1)}$ & Bias & RMSE \\
     \midrule
    \multicolumn{1}{l}{\textbf{Scenario 1}} & \multicolumn{9}{c}{$\text{BPC}_{\text{Out}}=0.010$ \hspace{.1cm} $\text{WPC}_{\text{Out}}=0.020$ \hspace{.1cm} $\text{ICC}_{\text{PS}}=0.020$}\\
    \midrule 
    Model 1 & 0.352 & 0.003 & 0.015 & 0.252 & -0.006 & 0.021 & 0.396 & 0.003 & 0.019 \\
    Model 2 & 0.352 & 0.003 & 0.015 & 0.252 & -0.005 & 0.021 & 0.396 & 0.002 & 0.019 \\
    Model 3 & 0.352 & 0.003 & 0.015 & 0.252 & -0.005 & 0.021 & 0.396 & 0.002 & 0.019 \\
    Model 4 & 0.352 & 0.002 & 0.015 & 0.252 & -0.004 & 0.021 & 0.396 & 0.002 & 0.019 \\
    \midrule
    \multicolumn{1}{l}{\textbf{Scenario 2}} & \multicolumn{9}{c}{$\text{BPC}_{\text{Out}}=0.030$ \hspace{.1cm} $\text{WPC}_{\text{Out}}=0.035$ \hspace{.1cm} $\text{ICC}_{\text{PS}}=0.035$}\\
    \midrule
    Model 1 & 0.351 & 0.003 & 0.018 & 0.255 & -0.007 & 0.024 & 0.394 & 0.005 & 0.023 \\
    Model 2 & 0.351 & 0.002 & 0.017 & 0.255 & -0.006 & 0.024 & 0.394 & 0.004 & 0.023 \\
    Model 3 & 0.351 & 0.002 & 0.018 & 0.255 & -0.007 & 0.024 & 0.394 & 0.004 & 0.023 \\
    Model 4 & 0.351 & 0.002 & 0.017 & 0.255 & -0.005 & 0.024 & 0.394 & 0.003 & 0.023 \\
    \midrule
    \multicolumn{1}{l}{\textbf{Scenario 3}} & \multicolumn{9}{c}{$\text{BPC}_{\text{Out}}=0.050$ \hspace{.1cm} $\text{WPC}_{\text{Out}}=0.100$ \hspace{.1cm} $\text{ICC}_{\text{PS}}=0.100$}\\
    \midrule
    Model 1 & 0.346 & 0.004 & 0.023 & 0.256 & -0.004 & 0.030 & 0.398 & -0.000 & 0.032 \\
    Model 2 & 0.346 & 0.003 & 0.023 & 0.256 & -0.001 & 0.031 & 0.398 & -0.001 & 0.032 \\
    Model 3 & 0.346 & 0.003 & 0.023 & 0.256 & -0.000 & 0.030 & 0.398 & -0.002 & 0.032 \\
    Model 4 & 0.346 & 0.002 & 0.023 & 0.256 & 0.002 & 0.031 & 0.398 & -0.004 & 0.032 \\
    \bottomrule
\end{tabular}
\label{tab:simresultsprop}
\caption*{We use the following abbreviations: BPC-within-cluster between-period correlation, WPC-within-cluster within-period correlation, ICC-intracluster correlation coefficient, Out-outcome, PS-principal strata. Model 1: includes cluster random effects in PS modeling and cluster and cluster-period random effects in outcome modeling. Model 2: removes cluster random effects from Model 1 in PS modeling only. Model 3: removes cluster-period random effects from Model 1 in outcome modeling only. Model 4: removes cluster random effects in PS modeling and cluster-period random effects in outcome modeling from Model 1.}
\end{table}

\FloatBarrier

\begin{table}[ht]
    \centering
    \captionsetup{labelfont=bf,justification=raggedright,singlelinecheck=false}
    
    \caption{Simulation study for a two-period CRXO trial with 18 clusters and {cluster-period sizes drawn} uniformly from 50 to 150, which includes cluster-period random effects in principal strata generation. Empirical bias and root mean square error (RMSE) for Bayesian estimation of the SACE expressed in both difference in means on the log-scale and the ratio of means are included. {Empirical coverage of nominal $95\%$ highest posterior density credible intervals is provided.} Side-by-side comparisons of our proposed model (Model 1) to fuller model  (Model A) are included. Results are from 1,000 simulated data sets with Bayesian inference on 10,000 MCMC iterations and 2,500 burn-ins. 0.000 is used to indicate less than 0.001.}
\begin{tabular}{ccccc|cccc}
    \toprule
    & \multicolumn{4}{c|}{\textbf{SACE Difference in means (Log)}} & \multicolumn{4}{c}{\textbf{SACE Ratio of means}}\\
     & Truth & Bias & RMSE & Coverage & Truth & Bias & RMSE & Coverage\\
     \midrule
    \multicolumn{1}{l}{\textbf{Scenario 1: }} & \multicolumn{8}{l}{$\text{BPC}_{\text{Out}}=0.010$ \hspace{.1cm} $\text{WPC}_{\text{Out}}=0.020$ \hspace{.1cm} $\text{BPC}_{\text{PS}}=0.010$ \hspace{.1cm}
    $\text{WPC}_{\text{PS}}=0.020$}\\
    \midrule 
    Model A & -1.176 & 0.020 & 0.111 & 92.0\% & 0.510 & -0.002 & 0.056 & 92.8\% \\
    Model 1 & -1.176 & 0.015 & 0.101 & 93.0\% & 0.510 & -0.000 & 0.056 & 93.2\% \\
    \midrule 
    \multicolumn{1}{l}{\textbf{Scenario 2: }} & \multicolumn{8}{l}{$\text{BPC}_{\text{Out}}=0.030$ \hspace{.1cm} $\text{WPC}_{\text{Out}}=0.035$ \hspace{.1cm} $\text{BPC}_{\text{PS}}=0.03$ \hspace{.1cm}
    $\text{WPC}_{\text{PS}}=0.035$}\\
    \midrule
    Model A & -1.181 & 0.017 & 0.101 & 94.8\% & 0.509 & -0.001 & 0.054 & 94.0\% \\
    Model 1 & -1.181 & 0.013 & 0.096 & 95.1\% & 0.509 & -0.000 & 0.054 & 94.1\% \\
    \midrule 
    \multicolumn{1}{l}{\textbf{Scenario 3: }} & \multicolumn{8}{l}{$\text{BPC}_{\text{Out}}=0.050$ \hspace{.1cm} $\text{WPC}_{\text{Out}}=0.100$ \hspace{.1cm} $\text{BPC}_{\text{PS}}=0.05$ \hspace{.1cm}
    $\text{WPC}_{\text{PS}}=0.100$}\\
    \midrule
    Model A & -1.185 & 0.026 & 0.132 & 92.9\% & 0.508 & 0.005 & 0.070 & 93.1\%\\
    Model 1 & -1.185 & 0.027 & 0.125 & 93.6\% & 0.508 & 0.006 & 0.071 & 93.3\%\\
    \bottomrule
\end{tabular}
\caption*{We use the following abbreviations: BPC-within-cluster between-period correlation, WPC-within-cluster within-period correlation, Out-outcome, PS-principal strata. Model 1: includes cluster random effects in PS modeling and cluster and cluster-period random effects in outcome modeling. Model A: adds cluster-period random effects to PS modeling in Model 1.}
\label{tab:simresultscp}
\end{table}

\begin{table}[ht]
    \centering
    \captionsetup{labelfont=bf,justification=raggedright,singlelinecheck=false}
    
    \caption{Simulation study for a two-period CRXO trial with 18 clusters and {cluster-period sizes drawn} uniformly from 50 to 150, which includes cluster-period random effects in principal strata generation. Empirical bias and root mean square error (RMSE) for Bayesian estimation of the principal strata proportions are presented. Side-by-side comparisons of our proposed model (Model 1) to fuller model  (Model A) are included. Results are from 1,000 simulated data sets with Bayesian inference on 10,000 MCMC iterations and 2,500 burn-ins. 0.000 is used to indicate less than 0.001.}
\begin{tabular}{cccc|ccc|ccc}
    \toprule
     Model & $\pi_{(0,0)}$ & Bias & RMSE & $\pi_{(1,0)}$ & Bias & RMSE & $\pi_{(1,1)}$ & Bias & RMSE \\
     \midrule
    \multicolumn{1}{l}{\textbf{Scenario 1}} & \multicolumn{9}{c}{$\text{BPC}_{\text{Out}}=0.010$ \hspace{.1cm} $\text{WPC}_{\text{Out}}=0.020$ \hspace{.1cm} $\text{BPC}_{\text{PS}}=0.010$ \hspace{.1cm} $\text{WPC}_{\text{PS}}=0.020$}\\
    \midrule 
    Model A & 0.353 & 0.002 & 0.016 & 0.252 & -0.008 & 0.025 & 0.395 & 0.005 & 0.020 \\
    Model 1 & 0.353 & 0.001 & 0.015 & 0.252 & -0.004 & 0.023 & 0.395 & 0.003 & 0.019 \\
    \midrule
    \multicolumn{1}{l}{\textbf{Scenario 2}} & \multicolumn{9}{c}{$\text{BPC}_{\text{Out}}=0.030$ \hspace{.1cm} $\text{WPC}_{\text{Out}}=0.035$ \hspace{.1cm} $\text{BPC}_{\text{PS}}=0.030$ \hspace{.1cm} $\text{WPC}_{\text{PS}}=0.035$}\\
    \midrule
    Model A & 0.350 & 0.004 & 0.018 & 0.254 & -0.009 & 0.027 & 0.396 & 0.005 & 0.024 \\
    Model 1 & 0.350 & 0.003 & 0.018 & 0.254 & -0.005 & 0.024 & 0.396 & 0.002 & 0.023 \\
    \midrule
    \multicolumn{1}{l}{\textbf{Scenario 3}} & \multicolumn{9}{c}{$\text{BPC}_{\text{Out}}=0.050$ \hspace{.1cm} $\text{WPC}_{\text{Out}}=0.100$ \hspace{.1cm} $\text{BPC}_{\text{PS}}=0.050$ \hspace{.1cm} $\text{WPC}_{\text{PS}}=0.100$}\\
    \midrule
    Model A & 0.346 & 0.003 & 0.025 & 0.257 & -0.011 & 0.040 & 0.397 & 0.008 & 0.034 \\
    Model 1 & 0.346 & -0.002 & 0.023 & 0.257 & 0.002 & 0.034 & 0.397 & -0.000 & 0.032 \\
    \bottomrule
\end{tabular}
\label{tab:simresultscpprop}
\caption*{We use the following abbreviations: BPC-within-cluster between-period correlation, WPC-within-cluster within-period correlation, Out-outcome, PS-principal strata. Model 1: includes cluster random effects in PS modeling and cluster and cluster-period random effects in outcome modeling. Model A: adds cluster-period random effects to PS modeling in Model 1.}
\end{table}

\FloatBarrier

\begin{table}[t]
    \centering
    \captionsetup{labelfont=bf,justification=raggedright,singlelinecheck=false}
    
    \caption{Simulation study for a two-period CRXO trial with 12 clusters and {cluster-period sizes drawn} uniformly from 50 to 150. Empirical bias and root mean square error (RMSE) for Bayesian estimation of the SACE expressed in both difference in means on the log-scale and the ratio of means are included. {Empirical coverage of nominal $95\%$ highest posterior density credible intervals is provided.} Side-by-side comparisons of our proposed model (Model 1) to reduced models (Models 2-4) are included. Results are from 1,000 simulated data sets with Bayesian inference on 10,000 MCMC iterations and 2,500 burn-ins. 0.000 is used to indicate less than 0.001}
\begin{tabular}{ccccc|cccc}
    \toprule
    & \multicolumn{4}{c|}{\textbf{SACE Difference in means (Log)}} & \multicolumn{4}{c}{\textbf{SACE Ratio of means}}\\
     & Truth & Bias & RMSE & Coverage & Truth & Bias & RMSE & Coverage\\
     \midrule
    \multicolumn{1}{l}{\textbf{Scenario 1}} & \multicolumn{8}{c}{$\text{BPC}_{\text{Out}}=0.010$ \hspace{.1cm} $\text{WPC}_{\text{Out}}=0.020$ \hspace{.1cm} $\text{ICC}_{\text{PS}}=0.020$}\\
    \midrule 
    Model 1 & -1.180 & 0.021 & 0.116 & 95.3\% & 0.508 & 0.000 & 0.065 & 95.2\% \\
    Model 2 & -1.180 & 0.024 & 0.123 & 95.4\% & 0.508 & 0.001 & 0.065 & 95.8\% \\
    Model 3 & -1.180 & 0.019 & 0.120 & 92.2\% & 0.508 & -0.000 & 0.065 & 93.1\% \\
    Model 4 & -1.180 & 0.028 & 0.136 & 91.6\% & 0.508 & 0.000 & 0.067 & 92.8\% \\
    \midrule
    \multicolumn{1}{l}{\textbf{Scenario 2}} & \multicolumn{8}{c}{$\text{BPC}_{\text{Out}}=0.030$ \hspace{.1cm} $\text{WPC}_{\text{Out}}=0.035$ \hspace{.1cm} $\text{ICC}_{\text{PS}}=0.035$}\\
    \midrule
    Model 1 & -1.182 & 0.025 & 0.111 & 96.4\% & 0.510 & 0.004 & 0.064 & 96.2\% \\
    Model 2 & -1.182 & 0.037 & 0.127 & 95.0\% & 0.510 & 0.005 & 0.065 & 96.3\% \\
    Model 3 & -1.182 & 0.023 & 0.114 & 92.7\% & 0.510 & 0.003 & 0.064 & 93.7\% \\
    Model 4 & -1.182 & 0.036 & 0.134 & 90.7\% & 0.510 & 0.003 & 0.066 & 93.6\% \\
    \midrule
    \multicolumn{1}{l}{\textbf{Scenario 3}} & \multicolumn{8}{c}{$\text{BPC}_{\text{Out}}=0.050$ \hspace{.1cm} $\text{WPC}_{\text{Out}}=0.100$ \hspace{.1cm} $\text{ICC}_{\text{PS}}=0.100$}\\
    \midrule
    Model 1 & -1.182 & 0.023 & 0.148 & 94.0\% & 0.513 & 0.005 & 0.083 & 94.2\% \\
    Model 2 & -1.182 & 0.048 & 0.164 & 93.1\% & 0.513 & 0.004 & 0.082 & 94.7\% \\
    Model 3 & -1.182 & 0.023 & 0.154 & 79.3\% & 0.513 & 0.001 & 0.086 & 83.8\% \\
    Model 4 & -1.182 & 0.050 & 0.168 & 81.6\% & 0.513 & -0.000 & 0.082 & 87.2\% \\
    \bottomrule
\end{tabular}
\caption*{We use the following abbreviations: BPC-within-cluster between-period correlation, WPC-within-cluster within-period correlation, ICC-intracluster correlation coefficient, Out-outcome, PS-principal strata. Model 1: includes cluster random effects in PS modeling and cluster and cluster-period random effects in outcome modeling. Model 2: removes cluster random effects from Model 1 in PS modeling only. Model 3: removes cluster-period random effects from Model 1 in outcome modeling only. Model 4: removes cluster random effects in PS modeling and cluster-period random effects in outcome modeling from Model 1.}
\label{tab:simresults12}
\end{table}

\FloatBarrier

\begin{table}[ht]
    \centering
    \captionsetup{labelfont=bf,justification=raggedright,singlelinecheck=false}
    
    \caption{Simulation study for a two-period CRXO trial with 12 clusters and {cluster-period sizes drawn} uniformly from 50 to 150. Empirical bias and root mean square error (RMSE) for Bayesian estimation of the principal strata proportions are presented. Side-by-side comparisons of our proposed model (Model 1) to reduced models (Models 2-4) are included. Results are from 1,000 simulated data sets with Bayesian inference on 10,000 MCMC iterations and 2,500 burn-ins. 0.000 is used to indicate less than 0.001}
\begin{tabular}{cccc|ccc|ccc}
    \toprule
     Model & $\pi_{(0,0)}$ & Bias & RMSE & $\pi_{(1,0)}$ & Bias & RMSE & $\pi_{(1,1)}$ & Bias & RMSE \\
     \midrule
    \multicolumn{1}{l}{\textbf{Scenario 1}} & \multicolumn{9}{c}{$\text{BPC}_{\text{Out}}=0.010$ \hspace{.1cm} $\text{WPC}_{\text{Out}}=0.020$ \hspace{.1cm} $\text{ICC}_{\text{PS}}=0.020$}\\
    \midrule 
    Model 1 & 0.352 & 0.004 & 0.019 & 0.252 & -0.006 & 0.026 & 0.396 & 0.002 & 0.023 \\
    Model 2 & 0.352 & 0.003 & 0.019 & 0.252 & -0.005 & 0.026 & 0.396 & 0.002 & 0.023 \\
    Model 3 & 0.352 & 0.003 & 0.019 & 0.252 & -0.006 & 0.025 & 0.396 & 0.002 & 0.023 \\
    Model 4 & 0.352 & 0.003 & 0.019 & 0.252 & -0.004 & 0.026 & 0.396 & 0.001 & 0.023 \\
    \midrule
    \multicolumn{1}{l}{\textbf{Scenario 2}} & \multicolumn{9}{c}{$\text{BPC}_{\text{Out}}=0.030$ \hspace{.1cm} $\text{WPC}_{\text{Out}}=0.035$ \hspace{.1cm} $\text{ICC}_{\text{PS}}=0.035$}\\
    \midrule
    Model 1 & 0.351 & 0.003 & 0.022 & 0.255 & -0.009 & 0.032 & 0.394 & 0.005 & 0.028 \\
    Model 2 & 0.351 & 0.003 & 0.022 & 0.255 & -0.006 & 0.031 & 0.394 & 0.004 & 0.028 \\
    Model 3 & 0.351 & 0.003 & 0.022 & 0.255 & -0.008 & 0.031 & 0.394 & 0.005 & 0.028 \\
    Model 4 & 0.351 & 0.003 & 0.022 & 0.255 & -0.006 & 0.031 & 0.394 & 0.003 & 0.028 \\
    \midrule
    \multicolumn{1}{l}{\textbf{Scenario 3}} & \multicolumn{9}{c}{$\text{BPC}_{\text{Out}}=0.050$ \hspace{.1cm} $\text{WPC}_{\text{Out}}=0.100$ \hspace{.1cm} $\text{ICC}_{\text{PS}}=0.100$}\\
    \midrule
    Model 1 & 0.346 & 0.004 & 0.028 & 0.256 & -0.007 & 0.036 & 0.398 & 0.003 & 0.038 \\
    Model 2 & 0.346 & 0.002 & 0.028 & 0.256 & -0.004 & 0.037 & 0.398 & 0.001 & 0.039 \\
    Model 3 & 0.346 & 0.003 & 0.028 & 0.256 & -0.004 & 0.035 & 0.398 & 0.000 & 0.038 \\
    Model 4 & 0.346 & 0.001 & 0.028 & 0.256 & -0.000 & 0.037 & 0.398 & -0.000 & 0.039 \\
    \bottomrule
\end{tabular}
\label{tab:simresults12prop}
\caption*{We use the following abbreviations: BPC-within-cluster between-period correlation, WPC-within-cluster within-period correlation, ICC-intracluster correlation coefficient, Out-outcome, PS-principal strata. Model 1: includes cluster random effects in PS modeling and cluster and cluster-period random effects in outcome modeling. Model 2: removes cluster random effects from Model 1 in PS modeling only. Model 3: removes cluster-period random effects from Model 1 in outcome modeling only. Model 4: removes cluster random effects in PS modeling and cluster-period random effects in outcome modeling from Model 1.}
\end{table}

\FloatBarrier

\subsection{Different Strata Composition Tables}
\label{sec:diffsupptables}

\begin{enumerate}
    \item[A6] Performance of SACE estimation with new principal strata proportions (Table~\ref{tab:simresultsnew})
        \begin{itemize}
            \item 25\% never-survivors, 25\% protected-patients, 50\% always-survivors
        \end{itemize}
    \item[A7] Bias and RMSE of principal strata proportions as per A6 data generating mechanism  (Table~\ref{tab:simresultsnewprop})
    
    \item[A8] Performance of SACE estimation with new principal strata proportions (Table~\ref{tab:simresultsnewsmP10})
        \begin{itemize}
            \item 18\% never-survivors, 10\% protected-patients, 72\% always-survivors
        \end{itemize}
    \item[A9] Bias and RMSE of principal strata proportions as per A8 data generating mechanism  (Table~\ref{tab:simresultsnewsmP10prop})
    
\end{enumerate}

\begin{table}[t]
    \centering

    \captionsetup{labelfont=bf,justification=raggedright,singlelinecheck=false}
    \caption{Simulation study for a two-period CRXO trial with 18 clusters and {cluster-period sizes drawn} uniformly from 50 to 150 with new principal strata proportions (25\%-NS, 25\%-PP, 50\%-AS). Empirical bias and root mean square error (RMSE) for Bayesian estimation of the SACE expressed in difference in means on the log-scale and the ratio of means are included. {Empirical coverage of nominal $95\%$ highest posterior density credible intervals is provided.} Side-by-side comparisons of our proposed model (Model 1) to reduced models (Models 2-4) are included. Results are from 1,000 simulated data sets with Bayesian inference on 10,000 MCMC iterations and 2,500 burn-ins. 0.000 is used to indicate less than 0.001.}
\begin{tabular}{ccccc|cccc}
    \toprule
    & \multicolumn{4}{c|}{\textbf{SACE Difference in means (Log)}} & \multicolumn{4}{c}{\textbf{SACE Ratio of means}}\\
     & Truth & Bias & RMSE & Coverage & Truth & Bias & RMSE & Coverage\\
     \midrule
    \multicolumn{1}{l}{\textbf{Scenario 1}} & \multicolumn{8}{c}{$\text{BPC}_{\text{Out}}=0.010$ \hspace{.1cm} $\text{WPC}_{\text{Out}}=0.020$ \hspace{.1cm} $\text{ICC}_{\text{PS}}=0.020$}\\
    \midrule 
        Model 1 & -1.186 & 0.011 & 0.094 & 93.6\% & 0.506 & -0.003 & 0.049 & 92.4\% \\
        Model 2 & -1.186 & 0.015 & 0.099 & 93.3\% & 0.506 & -0.003 & 0.049 & 93.2\% \\
        Model 3 & -1.186 & 0.009 & 0.094 & 89.3\% & 0.506 & -0.004 & 0.049 & 89.1\% \\
        Model 4 & -1.186 & 0.011 & 0.092 & 89.2\% & 0.506 & -0.004 & 0.049 & 89.7\% \\
    \midrule
    \multicolumn{1}{l}{\textbf{Scenario 2}} & \multicolumn{8}{c}{$\text{BPC}_{\text{Out}}=0.030$ \hspace{.1cm} $\text{WPC}_{\text{Out}}=0.035$ \hspace{.1cm} $\text{ICC}_{\text{PS}}=0.035$}\\
    \midrule
        Model 1 & -1.189 & 0.011 & 0.078 & 96.1\% & 0.505 & -0.003 & 0.043 & 96.0\% \\
        Model 2 & -1.189 & 0.016 & 0.087 & 95.7\% & 0.505 & -0.004 & 0.044 & 95.5\% \\
        Model 3 & -1.189 & 0.008 & 0.078 & 93.0\% & 0.505 & -0.004 & 0.043 & 93.3\% \\
        Model 4 & -1.189 & 0.014 & 0.092 & 92.4\% & 0.505 & -0.004 & 0.045 & 93.6\% \\
    \midrule
    \multicolumn{1}{l}{\textbf{Scenario 3}} & \multicolumn{8}{c}{$\text{BPC}_{\text{Out}}=0.050$ \hspace{.1cm} $\text{WPC}_{\text{Out}}=0.100$ \hspace{.1cm} $\text{ICC}_{\text{PS}}=0.100$}\\
    \midrule
        Model 1 & -1.194 & 0.013 & 0.114 & 94.3\% & 0.500 & 0.008 & 0.065 & 94.0\% \\
        Model 2 & -1.194 & 0.033 & 0.136 & 92.2\% & 0.500 & 0.007 & 0.065 & 93.3\% \\
        Model 3 & -1.194 & 0.015 & 0.120 & 74.9\% & 0.500 & 0.006 & 0.066 & 78.3\% \\
        Model 4 & -1.194 & 0.033 & 0.131 & 75.2\% & 0.500 & 0.004 & 0.065 & 80.0\% \\
    \bottomrule
\end{tabular}
\caption*{We use the following abbreviations: BPC-within-cluster between-period correlation, WPC-within-cluster within-period correlation, ICC-intracluster correlation coefficient, Out-outcome, PS-principal strata, NS-never-survivor, PP-protected-patient, AS-always-survivor. Model 1: includes cluster random effects in PS modeling and cluster and cluster-period random effects in outcome modeling. Model 2: removes cluster random effects from Model 1 in PS modeling only. Model 3: removes cluster-period random effects from Model 1 in outcome modeling only. Model 4: removes cluster random effects in PS modeling and cluster-period random effects in outcome modeling from Model 1.}
\label{tab:simresultsnew}
\end{table}

\FloatBarrier

\begin{table}[ht]
    \centering
    \captionsetup{labelfont=bf,justification=raggedright,singlelinecheck=false}
    
    \caption{Simulation study for a two-period CRXO trial with 18 clusters and {cluster-period sizes drawn} uniformly from 50 to 150 with new principal strata proportions (25\%-NS, 25\%-PP, 50\%-AS). Empirical bias and root mean square error (RMSE) for Bayesian estimation of the principal strata proportions are presented. Side-by-side comparisons of our proposed model (Model 1) to reduced models (Models 2-4) are included. Results are from 1,000 simulated data sets with Bayesian inference on 10,000 MCMC iterations and 2,500 burn-ins. 0.000 is used to indicate less than 0.001}
\begin{tabular}{cccc|ccc|ccc}
    \toprule
     Model & $\pi_{(0,0)}$ & Bias & RMSE & $\pi_{(1,0)}$ & Bias & RMSE & $\pi_{(1,1)}$ & Bias & RMSE \\
     \midrule
    \multicolumn{1}{l}{\textbf{Scenario 1}} & \multicolumn{9}{c}{$\text{BPC}_{\text{Out}}=0.010$ \hspace{.1cm} $\text{WPC}_{\text{Out}}=0.020$ \hspace{.1cm} $\text{ICC}_{\text{PS}}=0.020$}\\
    \midrule 
    Model 1 & 0.245 & 0.002 & 0.014 & 0.254 & -0.005 & 0.022 & 0.501 & 0.003 & 0.020 \\
    Model 2 & 0.245 & 0.002 & 0.014 & 0.254 & -0.005 & 0.022 & 0.501 & 0.003 & 0.020 \\
    Model 3 & 0.245 & 0.002 & 0.014 & 0.254 & -0.005 & 0.022 & 0.501 & 0.003 & 0.020 \\
    Model 4 & 0.245 & 0.002 & 0.014 & 0.254 & -0.004 & 0.022 & 0.501 & 0.003 & 0.020 \\
    \midrule
    \multicolumn{1}{l}{\textbf{Scenario 2}} & \multicolumn{9}{c}{$\text{BPC}_{\text{Out}}=0.030$ \hspace{.1cm} $\text{WPC}_{\text{Out}}=0.035$ \hspace{.1cm} $\text{ICC}_{\text{PS}}=0.035$}\\
    \midrule
    Model 1 & 0.246 & 0.000 & 0.015 & 0.254 & -0.005 & 0.025 & 0.500 & 0.004 & 0.024 \\
    Model 2 & 0.246 & 0.000 & 0.015 & 0.254 & -0.004 & 0.025 & 0.500 & 0.003 & 0.024 \\
    Model 3 & 0.246 & 0.000 & 0.015 & 0.254 & -0.004 & 0.025 & 0.500 & 0.003 & 0.024 \\
    Model 4 & 0.246 & 0.000 & 0.015 & 0.254 & -0.003 & 0.025 & 0.500 & 0.003 & 0.024 \\
    \midrule
    \multicolumn{1}{l}{\textbf{Scenario 3}} & \multicolumn{9}{c}{$\text{BPC}_{\text{Out}}=0.050$ \hspace{.1cm} $\text{WPC}_{\text{Out}}=0.100$ \hspace{.1cm} $\text{ICC}_{\text{PS}}=0.100$}\\
    \midrule
    Model 1 & 0.243 & 0.002 & 0.020 & 0.260 & -0.004 & 0.031 & 0.498 & 0.002 & 0.034 \\
    Model 2 & 0.243 & 0.000 & 0.020 & 0.260 & -0.002 & 0.033 & 0.498 & 0.001 & 0.035 \\
    Model 3 & 0.243 & 0.001 & 0.020 & 0.260 & -0.001 & 0.031 & 0.498 & -0.000 & 0.034 \\
    Model 4 & 0.243 & -0.000 & 0.020 & 0.260 & 0.001 & 0.032 & 0.498 & -0.001 & 0.035 \\
    \bottomrule
\end{tabular}
\label{tab:simresultsnewprop}
\caption*{We use the following abbreviations: BPC-within-cluster between-period correlation, WPC-within-cluster within-period correlation, ICC-intracluster correlation coefficient, Out-outcome, PS-principal strata, NS-never-survivor, PP-protected-patient, AS-always-survivor. Model 1: includes cluster random effects in PS modeling and cluster and cluster-period random effects in outcome modeling. Model 2: removes cluster random effects from Model 1 in PS modeling only. Model 3: removes cluster-period random effects from Model 1 in outcome modeling only. Model 4: removes cluster random effects in PS modeling and cluster-period random effects in outcome modeling from Model 1.}
\end{table}

\FloatBarrier

\begin{table}[t]
    \centering

    \captionsetup{labelfont=bf,justification=raggedright,singlelinecheck=false}
    
    \caption{Simulation study for a two-period CRXO trial with 18 clusters and {cluster-period sizes drawn} uniformly from 50 to 150 with new principal strata proportions (18\%-NS, 10\%-PP, 72\%-AS). Empirical bias and root mean square error (RMSE) for Bayesian estimation of the SACE expressed in both difference in means on the log-scale and the ratio of means are included. {Empirical coverage of nominal $95\%$ highest posterior density credible intervals is provided.} Side-by-side comparisons of our proposed model (Model 1) to reduced models (Models 2-4) are included. Results are from 1,000 simulated data sets with Bayesian inference on 10,000 MCMC iterations and 2,500 burn-ins. Failed runs by percentage are denoted as Failure Rate. 0.000 is used to indicate less than 0.001.}
\begin{tabular}{ccccc|cccc|c}
    \toprule
    & \multicolumn{4}{c|}{\textbf{SACE Difference in means (Log)}} & \multicolumn{4}{c|}{\textbf{SACE Ratio of means}} & \textbf{Failure} \\
     & Truth & Bias & RMSE & Coverage & Truth & Bias & RMSE & Coverage & \textbf{Rate} \\
     \midrule
    \multicolumn{1}{l}{\textbf{Scenario 1}} & \multicolumn{8}{c|}{$\text{BPC}_{\text{Out}}=0.010$ \hspace{.1cm} $\text{WPC}_{\text{Out}}=0.020$ \hspace{.1cm} $\text{ICC}_{\text{PS}}=0.020$} & \\
    \midrule 
    Model 1 & -1.122 & 0.020 & 0.065 & 92.2\% & 0.539 & 0.006 & 0.040 & 95.1\% & 2.1\% \\
    Model 2 & -1.122 & 0.020 & 0.065 & 92.0\% & 0.539 & 0.006 & 0.040 & 95.4\% & 1.8\% \\
    Model 3 & -1.122 & 0.020 & 0.065 & 85.9\% & 0.539 & 0.005 & 0.040 & 89.7\% & 2.7\% \\
    Model 4 & -1.122 & 0.019 & 0.065 & 85.4\% & 0.539 & 0.005 & 0.040 & 90.8\% & 1.2\% \\
    \midrule
    \multicolumn{1}{l}{\textbf{Scenario 2}} & \multicolumn{8}{c|}{$\text{BPC}_{\text{Out}}=0.030$ \hspace{.1cm} $\text{WPC}_{\text{Out}}=0.035$ \hspace{.1cm} $\text{ICC}_{\text{PS}}=0.035$} & \\
    \midrule
    Model 1 & -1.121 & 0.016 & 0.059 & 95.6\% & 0.540 & 0.004 & 0.038 & 95.6\% & 2.8\% \\
    Model 2 & -1.121 & 0.016 & 0.059 & 95.3\% & 0.540 & 0.004 & 0.038 & 96.0\% & 1.5\% \\
    Model 3 & -1.121 & 0.014 & 0.059 & 89.5\% & 0.540 & 0.003 & 0.038 & 91.5\% & 2.4\% \\
    Model 4 & -1.121 & 0.015 & 0.059 & 90.3\% & 0.540 & 0.004 & 0.038 & 91.9\% & 1.2\% \\
    \midrule
    \multicolumn{1}{l}{\textbf{Scenario 3}} & \multicolumn{8}{c|}{$\text{BPC}_{\text{Out}}=0.050$ \hspace{.1cm} $\text{WPC}_{\text{Out}}=0.100$ \hspace{.1cm} $\text{ICC}_{\text{PS}}=0.100$} & \\
    \midrule
    Model 1 & -1.117 & 0.013 & 0.097 & 94.8\% & 0.542 & 0.007 & 0.057 & 95.6\% & 2.0\% \\
    Model 2 & -1.117 & 0.015 & 0.098 & 94.3\% & 0.542 & 0.007 & 0.057 & 96.5\% & 0.8\% \\
    Model 3 & -1.117 & 0.010 & 0.101 & 66.6\% & 0.542 & 0.003 & 0.059 & 75.4\% & 1.7\% \\
    Model 4 & -1.117 & 0.014 & 0.099 & 67.7\% & 0.542 & 0.004 & 0.057 & 78.2\% & 0.5\% \\
    \bottomrule
\end{tabular}
\caption*{We use the following abbreviations: BPC-within-cluster between-period correlation, WPC-within-cluster within-period correlation, ICC-intracluster correlation coefficient, Out-outcome, PS-principal strata, NS-never-survivor, PP-protected-patient, AS-always-survivor. Model 1: includes cluster random effects in PS modeling and cluster and cluster-period random effects in outcome modeling. Model 2: removes cluster random effects from Model 1 in PS modeling only. Model 3: removes cluster-period random effects from Model 1 in outcome modeling only. Model 4: removes cluster random effects in PS modeling and cluster-period random effects in outcome modeling from Model 1.}
\label{tab:simresultsnewsmP10}
\end{table}

\FloatBarrier

\begin{table}[ht]
    \centering
    \captionsetup{labelfont=bf,justification=raggedright,singlelinecheck=false}
    
    \caption{Simulation study for a two-period CRXO trial with 18 clusters and {cluster-period sizes drawn} uniformly from 50 to 150 with new principal strata proportions (18\%-NS, 10\%-PP, 72\%-AS). Empirical bias and root mean square error (RMSE) for Bayesian estimation of the principal strata proportions are presented. Side-by-side comparisons of our proposed model (Model 1) to reduced models (Models 2-4) are included. Results are from 1,000 simulated data sets with Bayesian inference on 10,000 MCMC iterations and 2,500 burn-ins. 0.000 is used to indicate less than 0.001}
\begin{tabular}{cccc|ccc|ccc}
    \toprule
     Model & $\pi_{(0,0)}$ & Bias & RMSE & $\pi_{(1,0)}$ & Bias & RMSE & $\pi_{(1,1)}$ & Bias & RMSE \\
     \midrule
    \multicolumn{1}{l}{\textbf{Scenario 1}} & \multicolumn{9}{c}{$\text{BPC}_{\text{Out}}=0.010$ \hspace{.1cm} $\text{WPC}_{\text{Out}}=0.020$ \hspace{.1cm} $\text{ICC}_{\text{PS}}=0.020$}\\
    \midrule 
    Model 1 & 0.183 & 0.006 & 0.014 & 0.098 & -0.013 & 0.022 & 0.719 & 0.006 & 0.018 \\
    Model 2 & 0.183 & 0.006 & 0.014 & 0.098 & -0.011 & 0.021 & 0.719 & 0.005 & 0.017 \\
    Model 3 & 0.183 & 0.006 & 0.014 & 0.098 & -0.012 & 0.021 & 0.719 & 0.006 & 0.017 \\
    Model 4 & 0.183 & 0.005 & 0.013 & 0.098 & -0.010 & 0.021 & 0.719 & 0.005 & 0.017 \\
    \midrule
    \multicolumn{1}{l}{\textbf{Scenario 2}} & \multicolumn{9}{c}{$\text{BPC}_{\text{Out}}=0.030$ \hspace{.1cm} $\text{WPC}_{\text{Out}}=0.035$ \hspace{.1cm} $\text{ICC}_{\text{PS}}=0.035$}\\
    \midrule
    Model 1 & 0.185 & 0.006 & 0.015 & 0.099 & -0.011 & 0.022 & 0.717 & 0.006 & 0.020 \\
    Model 2 & 0.185 & 0.005 & 0.015 & 0.099 & -0.009 & 0.021 & 0.717 & 0.005 & 0.020 \\
    Model 3 & 0.185 & 0.005 & 0.015 & 0.099 & -0.011 & 0.021 & 0.717 & 0.005 & 0.020 \\
    Model 4 & 0.185 & 0.004 & 0.014 & 0.099 & -0.009 & 0.021 & 0.717 & 0.004 & 0.020 \\
    \midrule
    \multicolumn{1}{l}{\textbf{Scenario 3}} & \multicolumn{9}{c}{$\text{BPC}_{\text{Out}}=0.050$ \hspace{.1cm} $\text{WPC}_{\text{Out}}=0.100$ \hspace{.1cm} $\text{ICC}_{\text{PS}}=0.100$}\\
    \midrule
    Model 1 & 0.185 & 0.007 & 0.019 & 0.104 & -0.010 & 0.023 & 0.711 & 0.003 & 0.025 \\
    Model 2 & 0.185 & 0.006 & 0.019 & 0.104 & -0.007 & 0.023 & 0.711 & 0.001 & 0.026 \\
    Model 3 & 0.185 & 0.006 & 0.019 & 0.104 & -0.006 & 0.022 & 0.711 & -0.000 & 0.026 \\
    Model 4 & 0.185 & 0.004 & 0.019 & 0.104 & -0.004 & 0.023 & 0.711 & -0.000 & 0.026 \\
    \bottomrule
\end{tabular}
\label{tab:simresultsnewsmP10prop}
\caption*{We use the following abbreviations: BPC-within-cluster between-period correlation, WPC-within-cluster within-period correlation, ICC-intracluster correlation coefficient, Out-outcome, PS-principal strata, NS-never-survivor, PP-protected-patient, AS-always-survivor. Model 1: includes cluster random effects in PS modeling and cluster and cluster-period random effects in outcome modeling. Model 2: removes cluster random effects from Model 1 in PS modeling only. Model 3: removes cluster-period random effects from Model 1 in outcome modeling only. Model 4: removes cluster random effects in PS modeling and cluster-period random effects in outcome modeling from Model 1.}
\end{table}

\FloatBarrier

\newpage

\renewcommand{\thefigure}{A.\arabic{figure}}
\setcounter{figure}{0}

\subsection{Violations of Core Assumptions}
\label{sec:violations}

{In this section, we consider data generating mechanisms that violate one of two key assumptions to empirically assess the sensitivity of our estimation approach for SACE in CRXO trials under these departures. First, we generate the data according to non-mortality outcome modeling that does not match our log-normal modeling assumption. Second, we allow for the presence of harmed-patients, which violates the survival monotonicity assumption (Assumption 6). We recognize that there are numerous other ways to generate the data that violate these assumptions than what we have proposed below. However, our hope is that our choices hint at how our method realistically stands up to varying levels of departures from our assumptions. Throughout this section, we use the number of clusters, number of periods, sizes of cluster-periods, and covariates as in the main data generating mechanism of Section \ref{sec:dgp}. Specifically, we have 18 clusters, 2 periods, and cluster-period sizes $m_{ij}$ of 50-150 drawn uniformly, where $\boldsymbol{X}_{ij,1} \sim \text{MVN}(\boldsymbol{0.75},0.94\mathbb{I}_{m_{ij}})$, $\boldsymbol{X}_{ij,2} \sim \text{MVN}(\boldsymbol{0.25},1.32\mathbb{I}_{m_{ij}})$, and $\boldsymbol{X}_{ij,3} \sim \text{MVN}(-\boldsymbol{0.75},1.66\mathbb{I}_{m_{ij}})$. 
}

{

For the outcome model mis-specification scenarios, we generate the non-mortality outcomes for both protected-patients and always-survivors according to a two-component mixture of (conditional) log-normal distributions. This mixture has been proposed as a more flexible model for length of stay (LOS), which is capable of capturing two distinct groups of patients based on LOS, such as those with shorter and longer stays \citep{zhang2019lognormal}. Principal strata are generated according to (\ref{eq:dgmps}). For each principal stratum (with defined outcomes), one component of the outcome mixture matches our original data generating mechanism for outcomes (as in (\ref{eq:dgmout11}),(\ref{eq:dgmout01})), and the second component produces longer lengths of stay.

We obtain potential outcomes for always-survivors according to the mixture distribution, defined by
$$P_{ijk}\sim\text{Bern}(\pi_{\text{mix}})$$ 
$$\mathcal{Y}_{ijk}^{(1,1)}(1)|\{P_{ijk}=1\}=0.25+ 0.15X_{ijk,1}-0.5X_{ijk,2}+0.7X_{ijk,3}+0.05I(j=2)+\xi_{i,(1,1)}+\gamma_{ij,(1,1)}+\epsilon_{ijk,1,(1,1)}$$
$$\mathcal{Y}_{ijk}^{(1,1)}(1)|\{P_{ijk}=0\}=0.75+ 0.35X_{ijk,1}-0.3X_{ijk,2}+0.5X_{ijk,3}+0.05I(j=2)+\xi_{i,(1,1)}+\gamma_{ij,(1,1)}+\epsilon_{ijk,0,(1,1)}$$
$$\mathcal{Y}_{ijk}^{(1,1)}(0)|\{P_{ijk}=1\}=0.9+ 0.3X_{ijk,1}-0.15X_{ijk,2}+0.1X_{ijk,3}+0.05I(j=2)+\xi_{i,(1,1)}+\gamma_{ij,(1,1)}+\epsilon_{ijk,1,(1,1)}$$
$$\mathcal{Y}_{ijk}^{(1,1)}(0)|\{P_{ijk}=0\}=1.2+ 0.5X_{ijk,1}+0.05X_{ijk,2}-0.1X_{ijk,3}+0.05I(j=2)+\xi_{i,(1,1)}+\gamma_{ij,(1,1)}+\epsilon_{ijk,0,(1,1)}$$
and for protected-patients
$$P_{ijk}\sim\text{Bern}(\pi_{\text{mix}})$$
$$\mathcal{Y}_{ijk}^{(1,0)}(1)|\{P_{ijk}=1\}=0.2+0.25X_{ijk,1}-0.3X_{ijk,2}+0.15X_{ijk,3}+0.075I(j=2)+\xi_{i,(1,0)}+\gamma_{ij,(1,0)}+\epsilon_{ijk,1,(1,0)}$$
$$\mathcal{Y}_{ijk}^{(1,0)}(1)|\{P_{ijk}=0\}=0.5+0.45X_{ijk,1}-0.1X_{ijk,2}-0.05X_{ijk,3}+0.075I(j=2)+\xi_{i,(1,0)}+\gamma_{ij,(1,0)}+\epsilon_{ijk,0,(1,0)}$$
We consider two scenarios: a) when the error variances are the same across mixture components with results in Table \ref{tab:simresultsmixsame} and b) when the error variances are different across mixture components with results in Table \ref{tab:simresultsmixdiff}. For b), we make a sizable increase in the error variances of component 0 so that a (mis-specified) single component log-normal model should perform worse. For a), the variances for $\epsilon_{{ijk},1,(1,1)}$ and $\epsilon_{{ijk},0,(1,1)}$  are both 1, and the variances for $\epsilon_{{ijk},1,(1,0)}$ and $\epsilon_{{ijk},0,(1,0)}$ are both 1.25. For b), the variances for $\epsilon_{{ijk},1,(1,1)}$ and $\epsilon_{{ijk},0,(1,1)}$  are 1 and 2.17, respectively, and the variances for $\epsilon_{{ijk},1,(1,0)}$ and $\epsilon_{{ijk},0,(1,0)}$ are 1.25 and 3.42, respectively. The remaining specifications and independence assumptions are as in the main data generating process of Section \ref{sec:dgp}, where all within cluster and cluster-period correlations are based on the error variances of component 1.  

Under both a) and b), we then vary the mixing parameter, $\pi_{\text{mix}} \in \{0.9,0.8,0.7,0.6\}$, to reflect how each component contributes to the generated outcomes. We vary this parameter because as it gets closer to $0.5$ (i.e., the most ``mixed'' setting), we expect the single component log-normal models (as per our proposed approach) to be progressively less suitable.

The patterns in the results for Tables \ref{tab:simresultsmixsame}, \ref{tab:simresultsmixdiff} illustrate what we had anticipated. For values of the mixture parameters of 0.9 and 0.8 across both a) and b), performance of our method is fairly adequate, with the absolute value of relative bias never exceeding (and often well below) 5\%. The bias and coverage do generally deteriorate when the mixture parameters are 0.7 and 0.6. However, as expected, under a), the performance decline relative to the correctly specified settings is less pronounced than under b), and even at $\pi_{\text{mix}}$ of 0.7 or 0.6, the results under a) are not consistently problematic. In fact, the ratio of means appears quite robust to the mis-specification under a), which might be explained by the fact that the error variance terms factor out of the (unscaled) mean for the mixture of log-normals. In summary, it seems that our approach to SACE estimation in CRXO trials can withstand modest mis-specification of this form, but it remains to be tested against more complex non-mortality outcome distribution departures.}

\begin{table}[ht]
    \centering
    \captionsetup{labelfont=bf,justification=raggedright,singlelinecheck=false}
    \caption{{Simulation study for a two-period CRXO trial with 18 clusters and cluster-period sizes drawn uniformly from 50 to 150. Empirical bias and root mean square error (RMSE) for Bayesian estimation of the SACE expressed in both difference in means on the log-scale and the ratio of means are included. Empirical coverage of nominal $95\%$ highest posterior density credible intervals is provided. Proposed method is assessed under violation of log-normal outcome modeling assumptions. A mixture of short (as in main data generation) and long stays, with mixing parameter $\pi_{\text{mix}}$ for proportion of short stays, is used with the same error variances across mixture components. Results are from 1,000 simulated data sets with Bayesian inference on 10,000 MCMC iterations and 2,500 burn-ins. 0.000 is used to indicate less than 0.001}}
{
\begin{tabular}{ccccc|cccc}
    \toprule
    & \multicolumn{4}{c|}{\textbf{SACE Difference in means (Log)}} & \multicolumn{4}{c}{\textbf{SACE Ratio of means}}\\
     $\pi_{\text{mix}}$ & Truth & Bias & RMSE & Coverage & Truth & Bias & RMSE & Coverage\\
     \midrule
    \multicolumn{2}{l}{\textbf{Scenario 1}} & \multicolumn{7}{c}{$\text{BPC}_{\text{Out}}=0.010$ \hspace{.1cm} $\text{WPC}_{\text{Out}}=0.020$ \hspace{.1cm} $\text{ICC}_{\text{PS}}=0.020$}\\
    \midrule 
    0.9 & -1.155 & -0.022 & 0.114 & 92.5\% & 0.497 & -0.006 & 0.054 & 92.5\% \\
    0.8 & -1.135 & -0.034 & 0.118 & 90.6\% & 0.489 & -0.005 & 0.054 & 92.3\% \\
    0.7 & -1.114 & -0.053 & 0.138 & 87.0\% & 0.484 & -0.011 & 0.052 & 92.3\% \\
    0.6 & -1.094 & -0.056 & 0.147 & 85.0\% & 0.478 & -0.012 & 0.052 & 91.8\% \\
    \midrule
    \multicolumn{2}{l}{\textbf{Scenario 2}} & \multicolumn{7}{c}{$\text{BPC}_{\text{Out}}=0.030$ \hspace{.1cm} $\text{WPC}_{\text{Out}}=0.035$ \hspace{.1cm} $\text{ICC}_{\text{PS}}=0.035$}\\
    \midrule
    0.9 & -1.164 & -0.011 & 0.097 & 95.6\% & 0.495 &  0.000 & 0.050 & 96.0\% \\
    0.8 & -1.144 & -0.033 & 0.110 & 93.2\% & 0.486 & -0.003 & 0.050 & 95.5\% \\
    0.7 & -1.125 & -0.051 & 0.119 & 90.7\% & 0.479 & -0.009 & 0.050 & 94.0\% \\
    0.6 & -1.105 & -0.056 & 0.128 & 89.1\% & 0.473 & -0.006 & 0.050 & 93.6\% \\
    \midrule
    \multicolumn{2}{l}{\textbf{Scenario 3}} & \multicolumn{7}{c}{$\text{BPC}_{\text{Out}}=0.050$ \hspace{.1cm} $\text{WPC}_{\text{Out}}=0.100$ \hspace{.1cm} $\text{ICC}_{\text{PS}}=0.100$}\\
    \midrule
    0.9 & -1.163 & -0.017 & 0.126 & 93.2\% & 0.495 &  0.004 & 0.067 & 94.4\% \\
    0.8 & -1.142 & -0.040 & 0.125 & 93.5\% & 0.489 & -0.002 & 0.064 & 94.3\% \\
    0.7 & -1.122 & -0.059 & 0.138 & 90.8\% & 0.482 & -0.006 & 0.063 & 93.8\% \\
    0.6 & -1.103 & -0.066 & 0.140 & 89.6\% & 0.476 & -0.008 & 0.062 & 93.2\% \\
    \bottomrule
\end{tabular}}
\caption*{{We use the following abbreviations: BPC-within-cluster between-period correlation, WPC-within-cluster within-period correlation, ICC-intracluster correlation coefficient, Out-outcome, PS-principal strata.}}
\label{tab:simresultsmixsame}
\end{table}

\begin{table}[t]
    \centering
    \captionsetup{labelfont=bf,justification=raggedright,singlelinecheck=false}
    \caption{{Simulation study for a two-period CRXO trial with 18 clusters and cluster-period sizes drawn uniformly from 50 to 150. Empirical bias and root mean square error (RMSE) for Bayesian estimation of the SACE expressed in both difference in means on the log-scale and the ratio of means are included. Empirical coverage of nominal $95\%$ highest posterior density credible intervals is provided. Proposed method is assessed under violation of log-normal outcome modeling assumptions. A mixture of short (as in main data generation) and long stays, with mixing parameter $\pi_{\text{mix}}$ for proportion of short stays, is used with the different error variances across mixture components. Results are from 1,000 simulated data sets with Bayesian inference on 10,000 MCMC iterations and 2,500 burn-ins. 0.000 is used to indicate less than 0.001}}
{
\begin{tabular}{ccccc|cccc}
    \toprule
    & \multicolumn{4}{c|}{\textbf{SACE Difference in means (Log)}} & \multicolumn{4}{c}{\textbf{SACE Ratio of means}}\\
     $\pi_{\text{mix}}$ & Truth & Bias & RMSE & Coverage & Truth & Bias & RMSE & Coverage\\
     \midrule
    \multicolumn{2}{l}{\textbf{Scenario 1}} & \multicolumn{7}{c}{$\text{BPC}_{\text{Out}}=0.010$ \hspace{.1cm} $\text{WPC}_{\text{Out}}=0.020$ \hspace{.1cm} $\text{ICC}_{\text{PS}}=0.020$}\\
    \midrule 
    0.9 & -1.155 & -0.018 & 0.102 & 94.3\% & 0.492 & -0.012 & 0.051 & 92.4\% \\
    0.8 & -1.136 & -0.047 & 0.115 & 90.9\% & 0.481 & -0.021 & 0.054 & 90.6\% \\
    0.7 & -1.114 & -0.076 & 0.140 & 84.4\% & 0.478 & -0.032 & 0.063 & 83.7\% \\
    0.6 & -1.094 & -0.091 & 0.152 & 83.6\% & 0.471 & -0.033 & 0.064 & 85.4\% \\
    \midrule
    \multicolumn{2}{l}{\textbf{Scenario 2}} & \multicolumn{7}{c}{$\text{BPC}_{\text{Out}}=0.030$ \hspace{.1cm} $\text{WPC}_{\text{Out}}=0.035$ \hspace{.1cm} $\text{ICC}_{\text{PS}}=0.035$}\\
    \midrule
    0.9 & -1.164 & -0.013 & 0.090 & 96.4\% & 0.488 & -0.005 & 0.051 & 93.9\% \\
    0.8 & -1.144 & -0.046 & 0.103 & 94.2\% & 0.475 & -0.018 & 0.053 & 92.3\% \\
    0.7 & -1.125 & -0.067 & 0.119 & 90.7\% & 0.468 & -0.020 & 0.054 & 91.4\% \\
    0.6 & -1.105 & -0.086 & 0.138 & 87.0\% & 0.462 & -0.026 & 0.058 & 89.3\% \\
    \midrule
    \multicolumn{2}{l}{\textbf{Scenario 3}} & \multicolumn{7}{c}{$\text{BPC}_{\text{Out}}=0.050$ \hspace{.1cm} $\text{WPC}_{\text{Out}}=0.100$ \hspace{.1cm} $\text{ICC}_{\text{PS}}=0.100$}\\
    \midrule
    0.9 & -1.163 & -0.022 & 0.123 & 94.2\% & 0.496 & -0.007 & 0.066 & 93.2\% \\
    0.8 & -1.142 & -0.056 & 0.127 & 92.8\% & 0.488 & -0.024 & 0.066 & 90.7\% \\
    0.7 & -1.122 & -0.079 & 0.140 & 90.1\% & 0.478 & -0.029 & 0.066 & 90.8\% \\
    0.6 & -1.103 & -0.089 & 0.154 & 88.8\% & 0.475 & -0.032 & 0.070 & 88.2\% \\
    \bottomrule
\end{tabular}}
\caption*{{We use the following abbreviations: BPC-within-cluster between-period correlation, WPC-within-cluster within-period correlation, ICC-intracluster correlation coefficient, Out-outcome, PS-principal strata. Values of WPC and BPC are computed using Component 1 of mixture as in (\ref{eq:dgmout11}),(\ref{eq:dgmout01}).}}
\label{tab:simresultsmixdiff}
\end{table}

\FloatBarrier

{
 For the scenarios that violate survival monotonicity, we generate the principal stratum membership according to a multinomial logistic regression model with random intercepts for four possible categories:
\begin{align}
\label{eq:allogit}
\begin{split}
Z_{ijk}&=0.2+0.2X_{ijk,1}-0.4X_{ijk,2}+0.1X_{ijk,3}+0.05I(j=2)+\eta_{i,z}\\
W_{ijk}&=-0.1-0.4X_{ijk,1}-0.3X_{ijk,2}-0.1X_{ijk,3}+0.025I(j=2)+\eta_{i,w}\\
H_{ijk}&=\alpha_h-0.5X_{ijk,1}-0.6X_{ijk,2}-0.4X_{ijk,3}+0.05I(j=2)+\eta_{i,h}
\end{split}
\end{align}
\noindent where $G_{ijk}|\boldsymbol{X}_{ijk},R_j,\eta_{i,z},\eta_{i,w},\eta_{i,h}$ is distributed (in bijection with) $\text{Multi}(1;\pi^{(1,1)}_{ijk},\pi^{(1,0)}_{ijk},\pi^{(0,1)}_{ijk},\pi^{(0,0)}_{ijk})$ such that
\begin{align}
\label{eq:multi4}
\begin{split}
\pi^{(1,1)}_{ijk}&=\exp(Z_{ijk})/(\exp(Z_{ijk})+\exp(W_{ijk})+\exp(H_{ijk})+1)\\
\pi^{(1,0)}_{ijk}&=\exp(W_{ijk})/(\exp(Z_{ijk})+\exp(W_{ijk})+\exp(H_{ijk})+1)\\
\pi^{(0,1)}_{ijk}&=\exp(H_{ijk})/(\exp(Z_{ijk})+\exp(W_{ijk})+\exp(H_{ijk})+1)
\end{split}
\end{align}
and $\pi^{(0,0)}_{ijk}=1-\pi^{(1,1)}_{ijk}-\pi^{(1,0)}_{ijk}-\pi^{(0,1)}_{ijk}$, where superscripts correspond to strata probabilities.

We target three different proportions of harmed-patients, $2.5\%$, $5\%$, and $10\%$ by setting $\alpha_h$ to $-2.7$, $-2$, and $-1.05$ respectively. The true percentage at three decimal places of harmed-patients (HP), always-survivors (AS), protected-patients (PS), and never-survivors (NS) under these three targets ranged from:
\begin{enumerate}
    \item \textbf{Target 2.5\%} -- HP: 2.5\% to 2.6\%; AS: 40.9\% to 41.0\%; PP: 23.5\% to 23.9\%;  NS: 32.6\% to 33.1\%
    \item \textbf{Target 5\%} -- HP: 4.8\% to 5.0\%; AS: 40.0\%; PP: 22.9\% to 23.2\%; NS 31.7\% to 32.3\%
    \item \textbf{Target 10\%} -- HP: 10.9\% to 11.3\%; AS: 37.6\%; PP: 21.0\% to 21.4\%; NS: 29.7\% to 30.4\%
\end{enumerate}

For generating harmed-patients outcomes, we shift the intercept and covariate parameters (less the period indicator) of the always-survivors under treatment 0. Since the harmed-patients are falsely classified as always-survivors under treatment 0 in our estimation procedure, we aim to assess the effect of increasing how separate these two groups are on small-sample performance of our SACE estimation approach. Recall that always-survivor outcomes for treatment 0 are generated as (\ref{eq:dgmout11}):
\begin{equation*}
\mathcal{Y}_{ijk}^{(1,1)}(0)=0.9+ 0.3X_{ijk,1}-0.15X_{ijk,2}+0.1X_{ijk,3}+0.05I(j=2)+\xi_{i,(1,1)}+\gamma_{ij,(1,1)}+\epsilon_{ijk,(1,1)}
\end{equation*}
we then generate harmed-patients under treatment 0 as, 
\begin{align*}
\mathcal{Y}_{ijk}^{(0,1)}(0)&=(0.9+\delta)+(0.3+\delta)X_{ijk,1}+(-0.15+\delta)X_{ijk,2}+(0.1+\delta)X_{ijk,3}+0.05I(j=2)\\&\quad +\xi_{i,(0,1)}+\gamma_{ij,(0,1)}+\epsilon_{ijk,(0,1)}
\end{align*}
where we vary $\delta \in \{0.1,0.3,0.5,0.7\}$, increasing the average log outcome for harmed-patients relative to always-survivors by $1.25\delta$. We set the variance of $\epsilon_{ijk,(0,1)}$ to be 0.75, with the variances of the random intercepts selected to match the various BPC and WPC values. All remaining parameters are as in the main data generating process of Section \ref{sec:dgp}. All independence assumptions (including for the harmed-patient model) are the same as well.

Results for 2.5\%, 5\%, and 10\% (target) harmed-patients are in Tables \ref{tab:simresultsharm2p5}, \ref{tab:simresultsharm5}, and \ref{tab:simresultsharm10}, respectively. 
The small-sample performance of our method under this data generating mechanism worsens as the percentage of harmed-patients increases. However, under reasonable survival monotonicity departures of 2.5\% and 5\% harmed-patients, our method performs rather adequately for both SACE contrasts. The absolute value of relative bias does not exceed 2.8\% across all scenarios with 2.5\% harmed-patients and does not exceed (and often well below) 5\% across all scenarios with 5\% harmed-patients. As anticipated, when $\delta$ increases within scenarios, there is generally more empirical bias, with the biggest jump in bias tending to be from $\delta=0.5$ to $\delta=0.7$. Coverage is generally permissible for 2.5\% and for 5\% harmed-patients, but the method for the 5\% setting leads to overall worse coverage, suffering from noticeable undercoverage for certain ratio of means scenarios. Patterns regarding $\delta$ and coverage are not as straightforward as they were for bias because coverage depends on how well the modeling accounts for the artificial increase of the number of always-survivors by harmed-patients under treatment 0, whose similarity in log outcome is governed by the magnitude of $\delta$. We would not recommend applying our method if there are suspected to be at least 10\% harmed-patients, which may be unrealistic under restrictions regarding toxicity or safety for an active treatment. Under the proposed data generating mechanism for 10\% harmed-patients, our estimation procedure does not maintain reasonable performance across most scenarios (there are a few scenarios that are somewhat adequate for difference in means on the log scale). Critically, under all 10\% scenarios, there are some iterations when the MCMC algorithm fails, and the failure rate increases as $\delta$ grows, reaching $14.6\%$ to $15.5\%$ for $\delta=0.7$.}

\begin{table}[t]
    \centering
    \captionsetup{labelfont=bf,justification=raggedright,singlelinecheck=false}
    \caption{{Simulation study for a two-period CRXO trial with 18 clusters and cluster-period sizes drawn uniformly from 50 to 150. Empirical bias and root mean square error (RMSE) for Bayesian estimation of the SACE expressed in both difference in means on the log-scale and the ratio of means are included. Empirical coverage of nominal $95\%$ highest posterior density credible intervals is provided. Proposed method is assessed under violation of survival monotonicity with a target $2.5\%$ harmed-patients and varying difference, $\delta$, between non-mortality outcomes for harmed-patients and from always-survivors under treatment 0. Results are from 1,000 simulated data sets with Bayesian inference on 10,000 MCMC iterations and 2,500 burn-ins. 0.000 is used to indicate less than 0.001}}
{
\begin{tabular}{ccccc|cccc}
    \toprule
    & \multicolumn{4}{c|}{\textbf{SACE Difference in means (Log)}} & \multicolumn{4}{c}{\textbf{SACE Ratio of means}}\\
     $\delta$ & Truth & Bias & RMSE & Coverage & Truth & Bias & RMSE & Coverage\\
     \midrule
    \multicolumn{2}{l}{\textbf{Scenario 1}} & \multicolumn{7}{c}{$\text{BPC}_{\text{Out}}=0.010$ \hspace{.1cm} $\text{WPC}_{\text{Out}}=0.020$ \hspace{.1cm} $\text{ICC}_{\text{PS}}=0.020$}\\
    \midrule 
    0.1 & -1.180 & 0.021 & 0.096  & 93.5\% & 0.507 & -0.006 & 0.053 & 92.7\% \\
    0.3 & -1.180 & 0.023 & 0.093 & 92.9\% & 0.507 & -0.008 & 0.050 & 93.1\% \\
    0.5 & -1.180 & 0.026 & 0.094  & 94.0\% & 0.507 & -0.008 & 0.051 & 93.1\% \\
    0.7 & -1.180 & 0.033 & 0.101  & 92.0\% & 0.507 & -0.009 & 0.048 & 93.7\% \\
    \midrule
    \multicolumn{2}{l}{\textbf{Scenario 2}} & \multicolumn{7}{c}{$\text{BPC}_{\text{Out}}=0.030$ \hspace{.1cm} $\text{WPC}_{\text{Out}}=0.035$ \hspace{.1cm} $\text{ICC}_{\text{PS}}=0.035$}\\
    \midrule
    0.1 & -1.180 & 0.018 & 0.091 & 93.8\% & 0.509 & -0.010 & 0.051 & 93.8\%  \\
    0.3 & -1.180 & 0.016 & 0.086 & 95.0\% & 0.509 & -0.011 & 0.052 & 92.5\% \\
    0.5 & -1.180 & 0.024 & 0.095 & 94.0\% & 0.509 & -0.013 & 0.051 & 93.3\%\\
    0.7 & -1.180 & 0.033 & 0.091 & 94.6\% & 0.509 & -0.012 & 0.049 & 94.2\% \\
    \midrule
    \multicolumn{2}{l}{\textbf{Scenario 3}} & \multicolumn{7}{c}{$\text{BPC}_{\text{Out}}=0.050$ \hspace{.1cm} $\text{WPC}_{\text{Out}}=0.100$ \hspace{.1cm} $\text{ICC}_{\text{PS}}=0.100$}\\
    \midrule
    0.1 & -1.184 & 0.015 & 0.122 & 92.3\% & 0.509 & -0.006 & 0.065 & 92.3\% \\
    0.3 & -1.184 & 0.018 & 0.108 & 94.7\% & 0.509 & -0.007 & 0.062 & 93.8\% \\
    0.5 & -1.184 & 0.020 & 0.118 & 92.4\% & 0.509 & -0.010 & 0.063 & 92.1\% \\
    0.7 & -1.184 & 0.033 & 0.113 & 94.2\% & 0.509 & -0.009 & 0.061 & 94.2\% \\
    \bottomrule
\end{tabular}}
\caption*{{We use the following abbreviations: BPC-within-cluster between-period correlation, WPC-within-cluster within-period correlation, ICC-intracluster correlation coefficient, Out-outcome, PS-principal strata.}}
\label{tab:simresultsharm2p5}
\end{table}

\begin{table}[t]
    \centering
    \captionsetup{labelfont=bf,justification=raggedright,singlelinecheck=false}
    \caption{{Simulation study for a two-period CRXO trial with 18 clusters and cluster-period sizes drawn uniformly from 50 to 150. Empirical bias and root mean square error (RMSE) for Bayesian estimation of the SACE expressed in both difference in means on the log-scale and the ratio of means are included. Empirical coverage of nominal $95\%$ highest posterior density credible intervals is provided. Proposed method is assessed under violation of survival monotonicity with a target $5\%$ harmed-patients and varying difference, $\delta$, between non-mortality outcomes for harmed-patients and from always-survivors under treatment 0. Results are from 1,000 simulated data sets with Bayesian inference on 10,000 MCMC iterations and 2,500 burn-ins. 0.000 is used to indicate less than 0.001}}
{
\begin{tabular}{ccccc|cccc}
    \toprule
    & \multicolumn{4}{c|}{\textbf{SACE Difference in means (Log)}} & \multicolumn{4}{c}{\textbf{SACE Ratio of means}}\\
     $\delta$ & Truth & Bias & RMSE & Coverage & Truth & Bias & RMSE & Coverage\\
     \midrule
    \multicolumn{2}{l}{\textbf{Scenario 1}} & \multicolumn{7}{c}{$\text{BPC}_{\text{Out}}=0.010$ \hspace{.1cm} $\text{WPC}_{\text{Out}}=0.020$ \hspace{.1cm} $\text{ICC}_{\text{PS}}=0.020$}\\
    \midrule 
    0.1 & -1.177 & 0.029 & 0.100 & 91.0\% & 0.511 & -0.018 & 0.052 & 88.7\% \\
    0.3 & -1.177 & 0.031 & 0.094 & 92.2\% & 0.511 & -0.020 & 0.051 & 91.4\% \\
    0.5 & -1.177 & 0.043 & 0.103 & 90.6\% & 0.511 & -0.022 & 0.051 & 90.6\% \\
    0.7 & -1.177 & 0.052 & 0.107 & 90.4\% & 0.511 & -0.025 & 0.054 & 88.7\% \\
    \midrule
    \multicolumn{2}{l}{\textbf{Scenario 2}} & \multicolumn{7}{c}{$\text{BPC}_{\text{Out}}=0.030$ \hspace{.1cm} $\text{WPC}_{\text{Out}}=0.035$ \hspace{.1cm} $\text{ICC}_{\text{PS}}=0.035$}\\
    \midrule
    0.1 & -1.182 & 0.026 & 0.091 & 93.9\% & 0.503 & -0.011 & 0.048 & 91.9\% \\
    0.3 & -1.182 & 0.027 & 0.087 & 94.0\% & 0.503 & -0.014 & 0.049 & 92.3\% \\
    0.5 & -1.182 & 0.039 & 0.094 & 92.7\% & 0.503 & -0.016 & 0.048 & 92.7\% \\
    0.7 & -1.182 & 0.053 & 0.099 & 92.1\% & 0.503 & -0.019 & 0.048 & 90.3\% \\
    \midrule
    \multicolumn{2}{l}{\textbf{Scenario 3}} & \multicolumn{7}{c}{$\text{BPC}_{\text{Out}}=0.050$ \hspace{.1cm} $\text{WPC}_{\text{Out}}=0.100$ \hspace{.1cm} $\text{ICC}_{\text{PS}}=0.100$}\\
    \midrule
    0.1 & -1.181 & 0.020 & 0.109 & 93.9\% & 0.508 & -0.013 & 0.061 & 92.2\% \\
    0.3 & -1.181 & 0.027 & 0.113 & 93.6\% & 0.508 & -0.014 & 0.063 & 91.4\% \\
    0.5 & -1.181 & 0.036 & 0.114 & 94.6\% & 0.508 & -0.018 & 0.061 & 91.4\% \\
    0.7 & -1.181 & 0.048 & 0.118 & 91.9\% & 0.508 & -0.022 & 0.063 & 91.2\% \\
    \bottomrule
\end{tabular}}
\caption*{{We use the following abbreviations: BPC-within-cluster between-period correlation, WPC-within-cluster within-period correlation, ICC-intracluster correlation coefficient, Out-outcome, PS-principal strata.}}
\label{tab:simresultsharm5}
\end{table}
 
\begin{table}[t]
    \centering
    \captionsetup{labelfont=bf,justification=raggedright,singlelinecheck=false}
    \caption{{Simulation study for a two-period CRXO trial with 18 clusters and cluster-period sizes drawn uniformly from 50 to 150. Empirical bias and root mean square error (RMSE) for Bayesian estimation of the SACE expressed in both difference in means on the log-scale and the ratio of means are included. Empirical coverage of nominal $95\%$ highest posterior density credible intervals is provided. Proposed method is assessed under violation of survival monotonicity with a target $10\%$ harmed-patients and varying difference, $\delta$, between non-mortality outcomes for harmed-patients and from always-survivors under treatment 0. Results are from 1,000 simulated data sets with Bayesian inference on 10,000 MCMC iterations and 2,500 burn-ins. Failed runs by percentage are denoted as Failure Rate. 0.000 is used to indicate less than 0.001}}
{
\begin{tabular}{ccccc|cccc|c}
    \toprule
    & \multicolumn{4}{c|}{\textbf{SACE Difference in means (Log)}} & \multicolumn{4}{c|}{\textbf{SACE Ratio of means}} & \textbf{Failure} \\
     $\delta$ & Truth & Bias & RMSE & Coverage & Truth & Bias & RMSE & Coverage & \textbf{Rate} \\
     \midrule
    \multicolumn{2}{l}{\textbf{Scenario 1}} & \multicolumn{7}{c}{$\text{BPC}_{\text{Out}}=0.010$ \hspace{.1cm} $\text{WPC}_{\text{Out}}=0.020$ \hspace{.1cm} $\text{ICC}_{\text{PS}}=0.020$} & \\
    \midrule 
    0.1 & -1.162 & 0.040 & 0.094 & 89.0\% & 0.515 & -0.042 & 0.060 & 77.1\% & 5.2\% \\
    0.3 & -1.162 & 0.046 & 0.097 & 87.0\% & 0.515 & -0.050 & 0.065 & 70.4\% & 6.3\% \\
    0.5 & -1.162 & 0.065 & 0.110 & 83.8\% & 0.515 & -0.060 & 0.072 & 63.0\% & 11.1\% \\
    0.7 & -1.162 & 0.079 & 0.118 & 78.1\% & 0.515 & -0.072 & 0.082 & 51.2\% & 15.5\% \\
    \midrule
    \multicolumn{2}{l}{\textbf{Scenario 2}} & \multicolumn{7}{c}{$\text{BPC}_{\text{Out}}=0.030$ \hspace{.1cm} $\text{WPC}_{\text{Out}}=0.035$ \hspace{.1cm} $\text{ICC}_{\text{PS}}=0.035$}\\
    \midrule
    0.1 & -1.169 & 0.041 & 0.092 & 90.1\% & 0.512 & -0.042 & 0.058 & 79.0\% & 5.6\% \\
    0.3 & -1.169 & 0.044 & 0.095 & 88.9\% & 0.512 & -0.049 & 0.064 & 73.6\% & 6.7\% \\
    0.5 & -1.169 & 0.063 & 0.103 & 87.5\% & 0.512 & -0.057 & 0.069 & 64.4\% & 11.7\% \\
    0.7 & -1.169 & 0.077 & 0.110 & 83.3\% & 0.512 & -0.069 & 0.078 & 54.8\% & 14.6\% \\
    \midrule
    \multicolumn{2}{l}{\textbf{Scenario 3}} & \multicolumn{7}{c}{$\text{BPC}_{\text{Out}}=0.050$ \hspace{.1cm} $\text{WPC}_{\text{Out}}=0.100$ \hspace{.1cm} $\text{ICC}_{\text{PS}}=0.100$}\\
    \midrule
    0.1 & -1.170 & 0.039 & 0.112 & 92.3\% & 0.512 & -0.034 & 0.064 & 85.8\% & 3.8\% \\
    0.3 & -1.170 & 0.040 & 0.112 & 91.3\% & 0.512 & -0.045 & 0.069 & 80.5\% & 4.7\% \\
    0.5 & -1.170 & 0.046 & 0.118 & 90.9\% & 0.512 & -0.056 & 0.076 & 73.8\% & 9.8\% \\
    0.7 & -1.170 & 0.056 & 0.118 & 90.1\% & 0.512 & -0.071 & 0.087 & 65.4\% & 15.1\% \\
    \bottomrule
\end{tabular}}
\caption*{{We use the following abbreviations: BPC-within-cluster between-period correlation, WPC-within-cluster within-period correlation, ICC-intracluster correlation coefficient, Out-outcome, PS-principal strata.}}
\label{tab:simresultsharm10}
\end{table}

\FloatBarrier

\subsection{Trace plots PEPTIC}

\begin{figure}[ht]
\captionsetup{labelfont=bf,justification=raggedright,singlelinecheck=false}
\centering
\includegraphics[width=\textwidth]{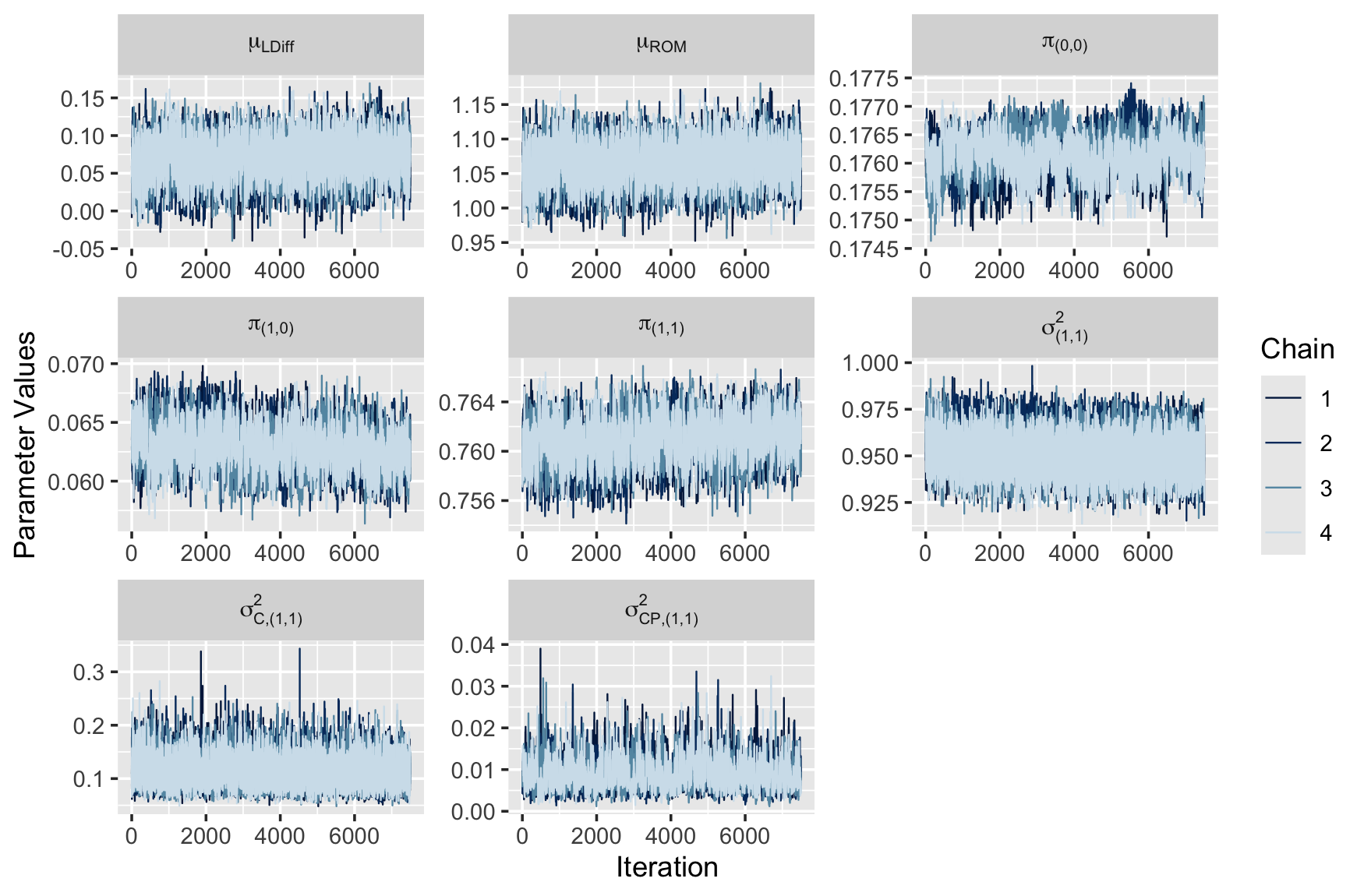}
\caption{Trace plots to assess mixing of key parameters for data application PEPTIC, a two-period cross-sectional CRXO trial. 4 chains with 10,000 iterations with 2,500 burn-ins with random initials are plotted. Run time was about 2.7 hours, parallelized over 4 chains.}
\label{fig:trace}
\end{figure}

\FloatBarrier
\begingroup

\hypersetup{linkcolor=blue,citecolor=blue,urlcolor=blue}
\section{Extensions to cross-sectional multi-period parallel-arm and stepped-wedge cluster-randomized trials}
\label{sec:otherdesigns}

The main text focuses on CRXO trials because the motivating PEPTIC trial used a crossover design, and it presents a general $J$-period formulation with the two-period design as a special case. In considering the relationship between CRXO and stepped-wedge trials, we recognized that the same principal stratification likelihood and Bayesian modeling approach can be extended to cross-sectional multi-period parallel-arm and stepped-wedge CRTs. Such an extension, however, requires more than simply reframing the CRXO design. The assignment conditions and regression design matrices must be modified to reflect the allocation mechanism of the target design. Throughout this appendix, each individual belongs to one cluster-period, so sampling over multiple periods is cross-sectional rather than repeated measurement of the same individuals (closed-cohort). The adaptations described below largely retain the outcome and principal strata model formulation used in the main analysis, with the changes to treatment assignment and the associated design matrices made explicit.

In both extensions, we retain the no-interference, consistency, between-cluster independence, and survival monotonicity conditions, with the same two well-defined treatment strategies across cluster-periods. Enrollment and the covariates used for adjustment are assumed not to be affected by the assigned schedule. Principal strata remain defined by $G_{ijk}=(S_{ijk}(1),S_{ijk}(0))$. Conditional on treatment assignment, covariates, and model parameters, including random effects, the individual observed-data likelihood for survival status and any defined non-mortality outcome retains the same mixture structure. The outcome models retain their cluster and cluster-period random intercepts, and the principal strata model retains its cluster random intercepts. No additional level of random effects is introduced because the cross-sectional clustering structure is unchanged.

The changes to the fixed-effects design matrices can be described using their individual rows. In the main analysis, the always-survivor outcome row contains $(1,a_{ij},\boldsymbol{x}_{ijk}^{T},\boldsymbol{R}_j^{T},a_{ij}\boldsymbol{x}_{ijk}^{T})$, where $\boldsymbol{R}_j$ contains $J-1$ period indicators. The principal strata row contains $(1,\boldsymbol{x}_{ijk}^{T},\boldsymbol{R}_j^{T})$, and the protected-patient outcome model uses the same non-treatment regressors, with period coding adapted to treatment-1 support where needed. The principal strata model includes all individuals and does not include realized treatment assignment as a predictor. Outcome design matrices are restricted to observed survivors in the corresponding sampled stratum; protected-patient outcomes are observed only under treatment 1. Thus, changing the treatment schedule changes the treatment and treatment-by-covariate columns of the always-survivor outcome design matrix and the observations available to the protected-patient outcome model, but not the form of the principal strata design matrix.

For both extensions, posterior SACE contrasts are obtained by standardizing both treatment conditions over the same target distribution of covariates and periods among always-survivors, as in Section~\ref{sec:augment}. In each posterior draw, the covariates and period indicators of the sampled always-survivors in the target population are held fixed while treatment is set to 1 and 0, using the marginal mean expressions in that section. For LDiff, the contrast is the difference between the standardized means of the log outcomes; for ROM, the standardized means on the original outcome scale are calculated before taking their ratio.

\subsection{Cross-sectional multi-period parallel-arm CRTs}
\label{sec:otherdesigns_parallel}

In a multi-period parallel-arm CRT, each cluster retains its assigned treatment over all $J\geq2$ periods, with different individuals sampled in each period. Assumption 1 in the main text is replaced by the following condition.

\renewcommand{\theassumption}{1P}
\begin{assumption}
\label{ass:otherdesigns_parallel}
\textit{Multi-period parallel-arm treatment assignment}: \textup{For each cluster, $a_{ij}=a_i\in\{0,1\}$ for every $j=1,\ldots,J$, so that $\boldsymbol{a}_i\in\{\boldsymbol{0}_J,\boldsymbol{1}_J\}$.}
\end{assumption}

Assumption 5 is correspondingly stated for the randomized arm $A_i$: it is independent of the cluster's preassignment covariates and potential outcomes, with $0<P(A_i=1)<1$. The always-survivor outcome design matrix is formed by replacing $a_{ij}$ with $a_i$, giving the row
\begin{equation}
\label{eq:parallel_design_row}
\left(1,a_i,\boldsymbol{x}_{ijk}^{T},\boldsymbol{R}_j^{T},a_i\boldsymbol{x}_{ijk}^{T}\right).
\end{equation}
The treatment indicator is constant within a cluster, whereas the period indicators vary across periods. Both the outcome and principal strata models retain the same period adjustment as in the main analysis. Protected-patient outcome observations come from the treatment-1 clusters throughout follow-up. The treatment contrast is informed by comparisons between randomized clusters rather than by crossover within clusters. Figure~\ref{fig:design_matrices_extensions} compares the treatment schedules and design-matrix blocks for the CRXO, multi-period parallel-arm, and stepped-wedge designs.

\subsection{Cross-sectional stepped-wedge CRTs}
\label{sec:otherdesigns_sw}

In a stepped-wedge CRT, clusters switch from control to treatment at randomized adoption periods and remain under treatment thereafter.
Although variants such as staircase designs have been proposed \citep{grantham2024staircase}, we focus here on conventional complete stepped-wedge designs, in which data collection is planned in every cluster-period.
For a design with an initial all-control period, a final all-treated period, and $J\geq3$, Assumption 1 is replaced by the following condition.

\renewcommand{\theassumption}{1S}
\begin{assumption}
\label{ass:otherdesigns_sw}
\textit{Stepped-wedge treatment assignment}: \textup{For each cluster, there is a treatment-adoption period $t_i\in\{2,\ldots,J\}$ such that $a_{ij}=0$ for $j<t_i$ and $a_{ij}=1$ for $j\geq t_i$; equivalently, $a_{ij}=I(j\geq t_i)$.}
\end{assumption}

Assumption 5 now applies to the randomized adoption period $T_i$, which is independent of the cluster's preassignment covariates and potential outcomes. Clusters are randomized among at least two admissible adoption periods, each with positive assignment probability. The always-survivor outcome design row becomes
\begin{equation}
\label{eq:sw_design_row}
\left(1,I(j\geq t_i),\boldsymbol{x}_{ijk}^{T},\boldsymbol{R}_j^{T},I(j\geq t_i)\boldsymbol{x}_{ijk}^{T}\right).
\end{equation}
In the always-survivor outcome model, the period indicators are unchanged, but the treatment indicator and treatment-by-covariate columns follow the unidirectional rollout rather than an alternating sequence. The principal strata model retains its covariate and period columns, and protected-patient outcomes contribute only after treatment adoption. The cluster and cluster-period outcome random intercepts are retained, as in standard cross-sectional stepped-wedge mixed models \citep{li2021mixed}. This adaptation assumes that current potential survival and non-mortality outcomes do not depend on earlier treatment assignments or time since treatment adoption; it does not accommodate exposure-duration or carryover effects merely by changing the design matrix.

Because the all-control baseline period contains no treatment-1 observations, it contributes no observed outcomes to the protected-patient outcome model. To avoid structural redundancy, the period indicators in that model are reparameterized using a reference period with treatment-1 support, with redundant columns omitted. Baseline survival observations still contribute to the principal strata model, and baseline outcomes among survivors contribute to the always-survivor outcome model.

Unlike the parallel-arm design, the stepped-wedge design does not provide randomized treatment variation in every period. Periods with $0<P(A_{ij}=1)<1$ provide such variation, whereas the all-control baseline and all-treated final periods do not \citep{chen2023model}. Including these endpoint periods in the target SACE therefore requires model-based extrapolation for the treatment-1 outcome at baseline and the treatment-0 outcome in the final period. The periods included in the target SACE should be specified explicitly; restricting the target periods does not require endpoint observations to be excluded from model fitting.

%

\newcommand{\crtTreatment}[3]{%
  \ifcase#1\relax
  \or
    \pgfmathtruncatemacro{\crtA}{ifthenelse(#2<=3,mod(#3+1,2),mod(#3,2))}%
  \or
    \pgfmathtruncatemacro{\crtA}{ifthenelse(#2<=3,0,1)}%
  \or
    \pgfmathtruncatemacro{\crtA}{ifthenelse(#3>ceil(#2/2),1,0)}%
  \fi
}

\newcommand{\crtDesignPanel}[2]{%
\begin{tikzpicture}[x=0.62cm,y=0.30cm,font=\scriptsize,text=black]

  \node[
    font=\small\bfseries,
    align=center,
    text width=3.6cm,
    minimum height=0.8cm
  ] at (2.5,2.6) {#2};

  \node at (2.5,0.9) {Period $j$};
  \node[anchor=east] at (0.35,0) {$i$};

  \foreach \j in {1,...,4}{
    \node at (\j,0) {$\j$};
  }

  \foreach \i in {1,...,6}{
    \node[anchor=east] at (0.35,-\i) {$\i$};

    \foreach \j in {1,...,4}{
      \crtTreatment{#1}{\i}{\j}

      \ifnum\crtA=1\relax
        \fill[gray!25]
          (\j-0.5,-\i-0.5)
          rectangle
          (\j+0.5,-\i+0.5);
      \fi

      \draw[gray!65,line width=0.3pt]
        (\j-0.5,-\i-0.5)
        rectangle
        (\j+0.5,-\i+0.5);

      \node at (\j,-\i) {$\crtA$};
    }
  }

  \draw[->,line width=0.45pt]
    (2.5,-6.8) -- (2.5,-7.6);

  \node at (2.6,-8.2) {Fixed-effect design block};

  \node[anchor=east] at (0.3,-9) {$(i,j)$};
  \node at (1,-9) {$1$};
  \node at (2,-9) {$a_{ij}$};
  \node[font=\tiny] at (3,-9) {$j{=}2$};
  \node[font=\tiny] at (4,-9) {$j{=}3$};
  \node[font=\tiny] at (5,-9) {$j{=}4$};

  \fill[gray!15]
    (1.5,-33.5) rectangle (2.5,-9.5);

  \draw[line width=0.55pt]
    (0.65,-9.5) -- (0.5,-9.5)
    -- (0.5,-33.5) -- (0.65,-33.5);

  \draw[line width=0.55pt]
    (5.35,-9.5) -- (5.5,-9.5)
    -- (5.5,-33.5) -- (5.35,-33.5);

  \foreach \i in {1,...,6}{
    \foreach \j in {1,...,4}{

      \pgfmathtruncatemacro{\crtRow}{4*(\i-1)+\j}
      \crtTreatment{#1}{\i}{\j}

      \node[anchor=east]
        at (0.3,{-9-\crtRow}) {$(\i,\j)$};

      \node at (1,{-9-\crtRow}) {$1$};
      \node at (2,{-9-\crtRow}) {$\crtA$};

      \foreach \ell in {2,3,4}{
        \pgfmathtruncatemacro{\crtIndicator}{ifthenelse(\j==\ell,1,0)}
        \node at ({\ell+1},{-9-\crtRow}) {$\crtIndicator$};
      }
    }
  }

  \foreach \y in {-13.5,-17.5,-21.5,-25.5,-29.5}{
    \draw[gray!60,densely dashed,line width=0.3pt]
      (0.55,\y) -- (5.45,\y);
  }

  \pgfresetboundingbox
  \path[use as bounding box]
    (-1.1,4.1) rectangle (5.7,-33.8);

\end{tikzpicture}%
}

\begin{figure}[p]
\centering

\captionsetup{labelfont=bf,justification=raggedright,singlelinecheck=false}

\resizebox{0.32\textwidth}{!}{%
  \crtDesignPanel{1}{(a) CRXO}%
}%
\hfill
\resizebox{0.32\textwidth}{!}{%
  \crtDesignPanel{2}{(b) Multi-period\\parallel-arm}%
}%
\hfill
\resizebox{0.32\textwidth}{!}{%
  \crtDesignPanel{3}{(c) Stepped-wedge}%
}%

\caption{\textcolor{black}{Illustrative treatment schedules and fixed-effect design blocks for three cross-sectional cluster-randomized designs with $I=6$ clusters and $J=4$ periods. Each upper matrix shows one realized allocation, with treatment-1 cells shaded. Three clusters receive each CRXO sequence or parallel-arm assignment; two clusters switch to treatment at each adoption period in the stepped-wedge design. Each lower matrix has one row per cluster-period $(i,j)$ and columns $(1,a_{ij},I(j=2),I(j=3),I(j=4))$, with the treatment column shaded and dashed lines separating clusters. Covariate and treatment-by-covariate columns are omitted. For the individual-level outcome design matrix, the treatment and period entries are repeated within cluster-periods and combined with individual covariates before selecting observed survivors in the sampled always-survivor stratum. Deleting the treatment column gives the intercept-and-period block of the principal strata design matrix, which is identical across the panels. The protected-patient outcome design uses the non-treatment columns for treatment-1 survivors in the sampled stratum, with period coding adjusted for treatment-1 support.}}
\label{fig:design_matrices_extensions}

\end{figure}

\FloatBarrier
\endgroup

%% file: Bibliography_revised.bib
@article{murray2004design,
  title={Design and analysis of group-randomized trials: a review of recent methodological developments},
  author={Murray, David M and Varnell, Sherri P and Blitstein, Jonathan L},
  journal={American Journal of Public Health},
  volume={94},
  number={3},
  pages={423--432},
  year={2004},
  publisher={American Public Health Association}
}

@article{maleyeff2023assessing,
  title={Assessing exposure-time treatment effect heterogeneity in stepped-wedge cluster randomized trials},
  author={Maleyeff, Lara and Li, Fan and Haneuse, Sebastien and Wang, Rui},
  journal={Biometrics},
  volume={79},
  number={3},
  pages={2551--2564},
  year={2023},
  publisher={Wiley Online Library}
}

@article{wang2024achieve,
  title={How to achieve model-robust inference in stepped wedge trials with model-based methods?},
  author={Wang, Bingkai and Wang, Xueqi and Li, Fan},
  journal={Biometrics},
  volume={80},
  number={4},
  pages={ujae123},
  year={2024},
  publisher={Oxford University Press}
}

@book{hayes2017cluster,
  title={Cluster Randomised Trials},
  author={Hayes, Richard J and Moulton, Lawrence H},
  year={2017},
  publisher={CRC Press}
}

@article{turner2017review,
  title={Review of recent methodological developments in group-randomized trials: part 1---design},
  author={Turner, Elizabeth L and Li, Fan and Gallis, John A and Prague, Melanie and Murray, David M},
  journal={American Journal of Public Health},
  volume={107},
  number={6},
  pages={907--915},
  year={2017},
  publisher={American Public Health Association}
}

@article{tong2025semiparametric,
  title={Semiparametric principal stratification analysis beyond monotonicity},
  author={Tong, Jiaqi and Kahan, Brennan and Harhay, Michael O and Li, Fan},
  journal={Statistica Sinica},
  year={in press}
}

@article{arnup2017understanding,
  title={Understanding the cluster randomised crossover design: a graphical illustration of the components of variation and a sample size tutorial},
  author={Arnup, Sarah J and McKenzie, Joanne E and Hemming, Karla and Pilcher, David and Forbes, Andrew B},
  journal={Trials},
  volume={18},
  number={1},
  pages={381},
  year={2017},
  publisher={BioMed Central}
}

@article{arnup2016appropriate,
  title={Appropriate statistical methods were infrequently used in cluster-randomized crossover trials},
  author={Arnup, Sarah J and Forbes, Andrew B and Kahan, Brennan C and Morgan, Katy E and McKenzie, Joanne E},
  journal={Journal of Clinical Epidemiology},
  volume={74},
  pages={40--50},
  year={2016},
  publisher={Elsevier}
}

@article{hooper2025reporting,
  title={Reporting of cluster randomised crossover trials: extension of the {CONSORT} 2010 statement with explanation and elaboration},
  author={McKenzie, Joanne E and Taljaard, Monica and Hemming, Karla and Arnup, Sarah J and Giraudeau, Bruno and Eldridge, Sandra and Hooper, Richard and Kahan, Brennan C and Li, Tianjing and Moher, David and Turner, Elizabeth L and Grimshaw, Jeremy M and Forbes, Andrew B},
  journal={BMJ},
  volume={388},
  pages={e080472},
  year={2025},
  publisher={BMJ Publishing Group}
}

@article{forbes2015cluster,
  title={Cluster randomised crossover trials with binary data and unbalanced cluster sizes: application to studies of near-universal interventions in intensive care},
  author={Forbes, Andrew B and Akram, Muhammad and Pilcher, David and Cooper, Jamie and Bellomo, Rinaldo},
  journal={Clinical Trials},
  volume={12},
  number={1},
  pages={34--44},
  year={2015},
  publisher={SAGE Publications Sage UK: London, England}
}

@article{donner1981randomization,
  title={Randomization by cluster: sample size requirements and analysis},
  author={Donner, Allan and Birkett, Nicholas and Buck, Carol},
  journal={American Journal of Epidemiology},
  volume={114},
  number={6},
  pages={906--914},
  year={1981},
  publisher={Oxford University Press}
}

@article{cook2021rationale,
  title={Rationale, methodological quality, and reporting of cluster-randomized controlled trials in critical care medicine: a systematic review},
  author={Cook, David J and Rutherford, William B and Scales, Damon C and Adhikari, Neill K J and Cuthbertson, Brian H},
  journal={Critical Care Medicine},
  volume={49},
  number={6},
  pages={977--987},
  year={2021},
  publisher={LWW}
}

@article{morgan2017choosing,
  title={Choosing appropriate analysis methods for cluster randomised cross-over trials with a binary outcome},
  author={Morgan, Katy E and Forbes, Andrew B and Keogh, Ruth H and Jairath, Vipul and Kahan, Brennan C},
  journal={Statistics in Medicine},
  volume={36},
  number={2},
  pages={318--333},
  year={2017},
  publisher={Wiley Online Library}
}

@article{young2018cluster,
  title={A cluster randomised, crossover, registry-embedded clinical trial of proton pump inhibitors versus histamine-2 receptor blockers for ulcer prophylaxis therapy in the intensive care unit ({PEPTIC} study): study protocol},
  author={Young, Paul J and Bagshaw, Sean M and Forbes, Andrew and Nichol, Alistair and Wright, Stephen E and Bellomo, Rinaldo and Bailey, Michael J and Beasley, Richard W and Eastwood, Glenn M and Festa, Marino and others},
  journal={Critical Care and Resuscitation},
  volume={20},
  number={3},
  pages={182--189},
  year={2018},
  publisher={Elsevier}
}

@article{young2020effect,
  title={Effect of stress ulcer prophylaxis with proton pump inhibitors vs histamine-2 receptor blockers on in-hospital mortality among {ICU} patients receiving invasive mechanical ventilation: the {PEPTIC} randomized clinical trial},
  author={Young, Paul J and Bagshaw, Sean M and Forbes, Andrew B and Nichol, Alistair D and Wright, Stephen E and Bailey, Michael and Bellomo, Rinaldo and Beasley, Richard and Brickell, Kathy and Eastwood, Glenn M and others},
  journal={JAMA},
  volume={323},
  number={7},
  pages={616--626},
  year={2020},
  publisher={American Medical Association}
}

@article{knaus1985apache,
  title={{APACHE II}: a severity of disease classification system},
  author={Knaus, William A and Draper, Elizabeth A and Wagner, Douglas P and Zimmerman, Jack E},
  journal={Critical Care Medicine},
  volume={13},
  number={10},
  pages={818--829},
  year={1985},
  publisher={LWW}
}

@article{hill2017long,
  title={Long-Term Outcomes and Health Care Utilization After Prolonged Mechanical Ventilation},
  author={Hill, Andrea D and Fowler, Robert A and Burns, Karen E A and Rose, Louise and Pinto, Ruxandra L and Scales, Damon C},
  journal={Annals of the American Thoracic Society},
  volume={14},
  number={3},
  pages={355--362},
  year={2017}
}

@article{alhazzani2013proton,
  title={Proton pump inhibitors versus histamine 2 receptor antagonists for stress ulcer prophylaxis in critically ill patients: a systematic review and meta-analysis},
  author={Alhazzani, Waleed and Alenezi, Farhan and Jaeschke, Roman Z and Moayyedi, Paul and Cook, Deborah J},
  journal={Critical Care Medicine},
  volume={41},
  number={3},
  pages={693--705},
  year={2013},
  publisher={LWW}
}

@article{field2007bootstrapping,
  title={Bootstrapping clustered data},
  author={Field, Christopher A and Welsh, Alan H},
  journal={Journal of the Royal Statistical Society: Series B, Statistical Methodology},
  volume={69},
  number={3},
  pages={369--390},
  year={2007},
  publisher={Oxford University Press}
}

@article{chen2024bayesian,
  title={A {Bayesian} machine learning approach for estimating heterogeneous survivor causal effects: Applications to a critical care trial},
  author={Chen, Xinyuan and Harhay, Michael O and Tong, Guangyu and Li, Fan},
  journal={The Annals of Applied Statistics},
  volume={18},
  number={1},
  pages={350--374},
  year={2024},
  publisher={Institute of Mathematical Statistics}
}

@article{tong2023bayesian,
  title={A {Bayesian} approach for estimating the survivor average causal effect when outcomes are truncated by death in cluster-randomized trials},
  author={Tong, Guangyu and Li, Fan and Chen, Xinyuan and Hirani, Shashivadan P and Newman, Stanton P and Wang, Wei and Harhay, Michael O},
  journal={American Journal of Epidemiology},
  volume={192},
  number={6},
  pages={1006--1015},
  year={2023},
  publisher={Oxford University Press}
}

@article{li2026uncovering,
  title={Uncovering treatment effect heterogeneity in pragmatic gerontology trials},
  author={Li, Changjun and Allore, Heather and Harhay, Michael O and Li, Fan and Tong, Guangyu},
  journal={Experimental Gerontology},
  volume={215},
  pages={113055},
  year={2026},
  publisher={Elsevier}
}

@article{hirano2000assessing,
  title={Assessing the effect of an influenza vaccine in an encouragement design},
  author={Hirano, Keisuke and Imbens, Guido W and Rubin, Donald B and Zhou, Xiao-Hua},
  journal={Biostatistics},
  volume={1},
  number={1},
  pages={69--88},
  year={2000},
  publisher={Oxford University Press}
}

@article{imbens1997bayesian,
  title={{B}ayesian inference for causal effects in randomized experiments with noncompliance},
  author={Imbens, Guido W and Rubin, Donald B},
  journal={The Annals of Statistics},
  volume={25},
  number={1},
  pages={305--327},
  year={1997},
  publisher={Institute of Mathematical Statistics}
}

@article{wang2024mixed,
  title={A mixed model approach to estimate the survivor average causal effect in cluster-randomized trials},
  author={Wang, Wei and Tong, Guangyu and Hirani, Shashivadan P and Newman, Stanton P and Halpern, Scott D and Small, Dylan S and Li, Fan and Harhay, Michael O},
  journal={Statistics in Medicine},
  volume={43},
  number={1},
  pages={16--33},
  year={2024},
  publisher={Wiley Online Library}
}

@article{zhang2009likelihood,
  title={Likelihood-based analysis of causal effects of job-training programs using principal stratification},
  author={Zhang, Junni L and Rubin, Donald B and Mealli, Fabrizia},
  journal={Journal of the American Statistical Association},
  volume={104},
  number={485},
  pages={166--176},
  year={2009},
  publisher={Taylor \& Francis}
}

@article{isenberg2025weighting,
  title={Weighting methods for truncation by death in cluster-randomized trials},
  author={Isenberg, Dane and Harhay, Michael O and Mitra, Nandita and Li, Fan},
  journal={Statistical Methods in Medical Research},
  volume={34},
  number={3},
  pages={473--489},
  year={2025},
  publisher={SAGE Publications Sage UK: London, England}
}

@article{kahan2024estimands,
  title={The estimands framework: a primer on the {ICH E9 (R1)} addendum},
  author={Kahan, Brennan C and Hindley, Joanna and Edwards, Mark and Cro, Suzie and Morris, Tim P},
  journal={BMJ},
  volume={384},
  year={2024},
  pages={e076316},
  publisher={British Medical Journal Publishing Group}
}

@misc{ICH_E9R1_2019,
  author = {{International Council for Harmonisation of Technical Requirements for Pharmaceuticals for Human Use}},
  title = {E9(R1) Statistical Principles for Clinical Trials: Addendum: Estimands and Sensitivity Analysis in Clinical Trials},
  year = {2019},
  note = {Accessed: 2025-05-08},
  url = {https://www.ich.org/page/efficacy-guidelines}
}

@article{kahan2023estimands,
  title={Estimands in cluster-randomized trials: choosing analyses that answer the right question},
  author={Kahan, Brennan C and Li, Fan and Copas, Andrew J and Harhay, Michael O},
  journal={International Journal of Epidemiology},
  volume={52},
  number={1},
  pages={107--118},
  year={2023},
  publisher={Oxford University Press}
}

@article{vanderweele2011principal,
  title={Principal stratification--uses and limitations},
  author={VanderWeele, Tyler J},
  journal={The International Journal of Biostatistics},
  volume={7},
  number={1},
  pages={28},
  year={2011},
  publisher={De Gruyter}
}

@article{rubin2000causal,
  title={Causal Inference without Counterfactuals: Comment},
  author={Rubin, Donald B},
  journal={Journal of the American Statistical Association},
  volume={95},
  number={450},
  pages={435--438},
  year={2000}
}

@article{zhang2003estimation,
  title={Estimation of causal effects via principal stratification when some outcomes are truncated by ``death''},
  author={Zhang, Junni L and Rubin, Donald B},
  journal={Journal of Educational and Behavioral Statistics},
  volume={28},
  number={4},
  pages={353--368},
  year={2003},
  publisher={Sage Publications Sage CA: Los Angeles, CA}
}

@article{hayden2005estimator,
  title={An estimator for treatment comparisons among survivors in randomized trials},
  author={Hayden, Douglas and Pauler, Donna K and Schoenfeld, David},
  journal={Biometrics},
  volume={61},
  number={1},
  pages={305--310},
  year={2005},
  publisher={Wiley Online Library}
}

@article{ding2017principal,
  title={Principal stratification analysis using principal scores},
  author={Ding, Peng and Lu, Jiannan},
  journal={Journal of the Royal Statistical Society: Series B, Statistical Methodology},
  volume  = {79},
  number  = {3},
  pages={757--777},
  year={2017},
  publisher={Oxford University Press}
}

@article{jiang2022multiply,
  title={Multiply robust estimation of causal effects under principal ignorability},
  author={Jiang, Zhichao and Yang, Shu and Ding, Peng},
  journal={Journal of the Royal Statistical Society: Series B, Statistical Methodology},
  volume={84},
  number={4},
  pages={1423--1445},
  year={2022},
  publisher={Oxford University Press}
}

@article{zehavi2023match,
  title={Matching methods for truncation by death problems},
  author={Zehavi, Tamir and Nevo, Daniel},
  journal={Journal of the Royal Statistical Society: Series A, Statistics in Society},
  volume={186},
  number={4},
  pages={659--681},
  year={2023},
  publisher={Oxford University Press US}
}

@article{oganisian2021practical,
  title={A practical introduction to {Bayesian} estimation of causal effects: Parametric and nonparametric approaches},
  author={Oganisian, Arman and Roy, Jason A},
  journal={Statistics in Medicine},
  volume={40},
  number={2},
  pages={518--551},
  year={2021},
  publisher={Wiley Online Library}
}

@article{oganisian2024hierarchical,
  title={Hierarchical {Bayesian} bootstrap for heterogeneous treatment effect estimation},
  author={Oganisian, Arman and Mitra, Nandita and Roy, Jason A},
  journal={The International Journal of Biostatistics},
  volume={20},
  number={1},
  pages={93--106},
  year={2024},
  publisher={De Gruyter}
}

@book{gelman2013bayesian,
  author    = {Andrew Gelman and John B Carlin and Hal S Stern and David B Dunson and Aki Vehtari and Donald B Rubin},
  title     = {{Bayesian} Data Analysis},
  edition   = {3rd},
  year      = {2013},
  publisher = {CRC Press},
  address   = {Boca Raton, FL}
}

@article{van2019shrinkage,
  title={Shrinkage priors for {Bayesian} penalized regression},
  author={van Erp, Sara and Oberski, Daniel L and Mulder, Joris},
  journal={Journal of Mathematical Psychology},
  volume={89},
  pages={31--50},
  year={2019},
  publisher={Elsevier}
}

@article{rubin1981bayesian,
  title   = {The {Bayesian} Bootstrap},
  author  = {Rubin, Donald B},
  journal = {The Annals of Statistics},
  volume  = {9},
  number  = {1},
  pages   = {130--134},
  year    = {1981}
}

@article{mercatanti2015improving,
  title={Improving inference of {Gaussian} mixtures using auxiliary variables},
  author={Mercatanti, Andrea and Li, Fan and Mealli, Fabrizia},
  journal={Statistical Analysis and Data Mining: The ASA Data Science Journal},
  volume={8},
  number={1},
  pages={34--48},
  year={2015},
  publisher={Wiley Online Library}
}

@incollection{teh2010hierarchical,
  title     = {Hierarchical {Bayesian} Nonparametric Models with Applications},
  author    = {Teh, Yee Whye and Jordan, Michael I.},
  booktitle = {Bayesian Nonparametrics},
  pages     = {158--207},
  year      = {2010},
  publisher = {Cambridge University Press}
}

@article{ding2018causal,
  title={Causal inference: A missing data perspective},
  author={Ding, Peng and Li, Fan},
  journal={Statistical Science},
  volume={33},
  number={2},
  pages={214--237},
  year={2018},
  publisher={IMS}
}

@article{he2023bayesian,
  title={{Bayesian} Framework for Causal Inference with Principal Stratification and Clusters},
  author={He, Li and Wang, Yu-Bo and Bridges, Jr., William C and He, Zhulin and Che, S Megan},
  journal={Statistics in Biosciences},
  volume={15},
  number={1},
  pages={114--140},
  year={2023},
  publisher={Springer}
}

@article{tanner1987calculation,
  title={The calculation of posterior distributions by data augmentation},
  author={Tanner, Martin A and Wong, Wing Hung},
  journal={Journal of the American Statistical Association},
  volume={82},
  number={398},
  pages={528--540},
  year={1987},
  publisher={Taylor \& Francis}
}

@article{diebolt1994estimation,
  title={Estimation of finite mixture distributions through {Bayesian} sampling},
  author={Diebolt, Jean and Robert, Christian P},
  journal={Journal of the Royal Statistical Society: Series B, Statistical Methodology},
  volume={56},
  number={2},
  pages={363--375},
  year={1994},
  publisher={Wiley Online Library}
}

@article{dempster1977maximum,
  title={Maximum likelihood from incomplete data via the {EM} algorithm},
  author={Dempster, Arthur P and Laird, Nan M and Rubin, Donald B},
  journal={Journal of the Royal Statistical Society: Series B, Statistical Methodology},
  volume={39},
  number={1},
  pages={1--22},
  year={1977},
  publisher={Wiley Online Library}
}

@article{frangakis2002principal,
  title={Principal stratification in causal inference},
  author={Frangakis, Constantine E and Rubin, Donald B},
  journal={Biometrics},
  volume={58},
  number={1},
  pages={21--29},
  year={2002},
  publisher={Wiley Online Library}
}

@article{chen2023model,
  title={Model-assisted analysis of covariance estimators for stepped wedge cluster randomized experiments},
  author={Chen, Xinyuan and Li, Fan},
  journal={Scandinavian Journal of Statistics},
  volume={52},
  number={1},
  pages={416--446},
  year={2025}
}

@article{turner2007analysis,
  title={Analysis of cluster randomized cross-over trial data: a comparison of methods},
  author={Turner, Rebecca M and White, Ian R and Croudace, Tim},
  journal={Statistics in Medicine},
  volume={26},
  number={2},
  pages={274--289},
  year={2007},
  publisher={Wiley Online Library}
}

@article{goldstein2002partitioning,
  title={Partitioning variation in multilevel models},
  author={Goldstein, Harvey and Browne, William and Rasbash, Jon},
  journal={Understanding Statistics: Statistical Issues in Psychology, Education, and the Social Sciences},
  volume={1},
  number={4},
  pages={223--231},
  year={2002},
  publisher={Taylor \& Francis}
}

@article{hedeker2003mixed,
  title={A mixed-effects multinomial logistic regression model},
  author={Hedeker, Donald},
  journal={Statistics in Medicine},
  volume={22},
  number={9},
  pages={1433--1446},
  year={2003},
  publisher={Wiley Online Library}
}

@article{jo2022handling,
  title={Handling parametric assumptions in principal causal effect estimation using {Gaussian} mixtures},
  author={Jo, Booil},
  journal={Statistics in Medicine},
  volume={41},
  number={16},
  pages={3039--3056},
  year={2022},
  publisher={Wiley Online Library}
}

@article{polson2013bayesian,
  title={{Bayesian} inference for logistic models using {P}{\'o}lya--{G}amma latent variables},
  author={Polson, Nicholas G and Scott, James G and Windle, Jesse},
  journal={Journal of the American Statistical Association},
  volume={108},
  number={504},
  pages={1339--1349},
  year={2013},
  publisher={Taylor \& Francis}
}

@article{allen2023bayesian,
  title={A {Bayesian} multivariate mixture model for high throughput spatial transcriptomics},
  author={Allen, Carter and Chang, Yuzhou and Neelon, Brian and Chang, Won and Kim, Hang J and Li, Zihai and Ma, Qin and Chung, Dongjun},
  journal={Biometrics},
  volume={79},
  number={3},
  pages={1775--1787},
  year={2023},
  publisher={Oxford University Press}
}

@article{choi2013polya,
  title={The {P}{\'o}lya-{G}amma {Gibbs} sampler for {Bayesian} logistic regression is uniformly ergodic},
  author={Choi, Hee Min and Hobert, James P},
  journal={Electronic Journal of Statistics},
  volume={7},
  pages={2054--2064},
  year={2013},
  publisher={Institute of Mathematical Statistics and Bernoulli Society}
}

@article{zhang2019lognormal,
  title={Lognormal-based mixture models for robust fitting of hospital length of stay distributions},
  author={Zhang, Xu and Barnes, Sean and Golden, Bruce and Myers, Miranda and Smith, Paul},
  journal={Operations Research for Health Care},
  volume={22},
  pages={100184},
  year={2019},
  publisher={Elsevier}
}

@article{faddy2009modeling,
  title={Modeling length of stay in hospital and other right skewed data: comparison of phase-type, gamma and log-normal distributions},
  author={Faddy, Malcolm and Graves, Nicholas and Pettitt, Anthony},
  journal={Value in Health},
  volume={12},
  number={2},
  pages={309--314},
  year={2009},
  publisher={Elsevier}
}

@article{marazzi1998fitting,
  title={Fitting the distributions of length of stay by parametric models},
  author={Marazzi, Alfio and Paccaud, Fred and Ruffieux, Christiane and Beguin, Claire},
  journal={Medical care},
  volume={36},
  number={6},
  pages={915--927},
  year={1998}
}

@inproceedings{feller2016principal,
  title={Weak separation in mixture models and implications for principal stratification},
  author={Ho, Nhat and Feller, Avi and Greif, Evan and Miratrix, Luke and Pillai, Natesh},
  booktitle={Proceedings of the 25th International Conference on Artificial Intelligence and Statistics},
  volume={151},
  series={Proceedings of Machine Learning Research},
  pages={5416--5458},
  year={2022},
  publisher={PMLR}
}

@article{plummer2006coda,
  title={{CODA}: convergence diagnosis and output analysis for {MCMC}},
  author={Plummer, Martyn and Best, Nicky and Cowles, Kate and Vines, Karen},
  journal={R News},
  volume={6},
  number={1},
  pages={7--11},
  year={2006}
}

@article{xu2022bayesian,
  title={A {Bayesian} nonparametric approach for evaluating the causal effect of treatment in randomized trials with semi-competing risks},
  author={Xu, Yanxun and Scharfstein, Daniel and M{\"u}ller, Peter and Daniels, Michael},
  journal={Biostatistics},
  volume={23},
  number={1},
  pages={34--49},
  year={2022},
  publisher={Oxford University Press}
}

@article{comment2025survivor,
  title={Survivor average causal effects for continuous time: a principal stratification approach to causal inference with semicompeting risks},
  author={Comment, Leah and Mealli, Fabrizia and Haneuse, Sebastien and Zigler, Corwin M},
  journal={Biometrical Journal},
  volume={67},
  number={2},
  pages={e70041},
  year={2025},
  publisher={Wiley Online Library}
}

@article{nevo2022causal,
  title={Causal inference for semi-competing risks data},
  author={Nevo, Daniel and Gorfine, Malka},
  journal={Biostatistics},
  volume={23},
  number={4},
  pages={1115--1132},
  year={2022},
  publisher={Oxford University Press}
}

@article{kennedy2019survivor,
  title={Survivor-complier effects in the presence of selection on treatment, with application to a study of prompt {ICU} admission},
  author={Kennedy, Edward H and Harris, Steve and Keele, Luke J},
  journal={Journal of the American Statistical Association},
  volume={114},
  number={525},
  pages={93--104},
  year={2019},
  publisher={Taylor \& Francis}
}

@article{vanderweele2013causal,
  title={Causal inference under multiple versions of treatment},
  author={VanderWeele, Tyler J and Hern{\'a}n, Miguel A},
  journal={Journal of Causal Inference},
  volume={1},
  number={1},
  pages={1--20},
  year={2013},
  publisher={De Gruyter}
}

@article{kenny2022analysis,
  title={Analysis of stepped wedge cluster randomized trials in the presence of a time-varying treatment effect},
  author={Kenny, Avi and Voldal, Emily C and Xia, Fan and Heagerty, Patrick J and Hughes, James P},
  journal={Statistics in Medicine},
  volume={41},
  number={22},
  pages={4311--4339},
  year={2022},
  publisher={Wiley Online Library}
}

@article{li2021mixed,
  title={Mixed-effects models for the design and analysis of stepped wedge cluster randomized trials: an overview},
  author={Li, Fan and Hughes, James P and Hemming, Karla and Taljaard, Monica and Melnick, Edward R and Heagerty, Patrick J},
  journal={Statistical Methods in Medical Research},
  volume={30},
  number={2},
  pages={612--639},
  year={2021},
  publisher={SAGE Publications Sage UK: London, England}
}

@article{hemming2020use,
  title={Use of multiple period, cluster randomised, crossover trial designs for comparative effectiveness research},
  author={Hemming, Karla and Taljaard, Monica and Weijer, Charles and Forbes, Andrew B},
  journal={BMJ},
  volume={371},
  pages={m3800},
  year={2020},
  publisher={British Medical Journal Publishing Group}
}

@article{grantham2024staircase,
  title={The staircase cluster randomised trial design: a pragmatic alternative to the stepped wedge},
  author={Grantham, Kelsey L and Forbes, Andrew B and Hooper, Richard and Kasza, Jessica},
  journal={Statistical Methods in Medical Research},
  volume={33},
  number={1},
  pages={24--41},
  year={2024},
  publisher={SAGE Publications Sage UK: London, England}
}
